\documentclass[aps,prb,twocolumn,superscriptaddress,floatfix,showpacm]{revtex4}
\usepackage{graphicx}
\usepackage{amssymb}
\usepackage{amsmath}
\usepackage{verbatim}

\ifx\pdftexversion\undefined
\usepackage[dvips]{hyperref}
\else
\usepackage{hyperref}
\fi
\hypersetup{colorlinks = true, linkcolor = blue}

\begin{document}

\title{Competing structures in two dimensions: square-to-hexagonal transition}

\author{Barbara Gr\"anz}
\affiliation{Theoretische Physik, ETH Zurich, CH-8093
Zurich, Switzerland}
\author{Sergey E.\ Korshunov}
\affiliation{L.D.\ Landau Institute for Theoretical Physics, 
142432 Chernogolovka, Russia}
\author{Vadim B.\ Geshkenbein}
\affiliation{Theoretische Physik, ETH Zurich, CH-8093
Zurich, Switzerland}
\author{Gianni Blatter}
\affiliation{Theoretische Physik, ETH Zurich, CH-8093
Zurich, Switzerland}

\date{\today}

\begin{abstract}
We study a system of particles in two dimensions interacting via a dipolar
long-range potential $D/r^3$ and subject to a square-lattice substrate
potential $V({\bf r})$ with amplitude $V$ and lattice constant $b$.  The
isotropic interaction favors a hexagonal arrangement of the particles with
lattice constant $a$, which competes against the square symmetry of the
underlying substrate lattice.  We determine the minimal-energy states at fixed
external pressure $p$ generating the commensurate density $n = 1/b^2 =
(4/3)^{1/2}/a^2$ in the absence of thermal and quantum fluctuations, using
both analytical techniques based on the harmonic- and continuum elastic
approximations as well as numerical relaxation of particle configurations. At
large substrate amplitude $V > 0.2\, e_{\scriptscriptstyle D}$, with
$e_{\scriptscriptstyle D} = D/b^3$ the dipolar energy scale, the particles
reside in the substrate minima and hence arrange in a square lattice. Upon
decreasing $V$, the square lattice turns unstable with respect to a
zone-boundary shear-mode and deforms into a period-doubled zig-zag lattice.
Analytic and numerical results show that this period-doubled phase in turn
becomes unstable at $V \approx 0.074\, e_{\scriptscriptstyle D}$ towards a
non-uniform phase developing an array of domain walls or solitons; as the
density of solitons increases, the particle arrangement approaches that of a
rhombic (or isosceles triangular) lattice. At a yet smaller substrate value
estimated as $V \approx 0.046\, e_{\scriptscriptstyle D}$, a further solitonic
transition establishes a second non-uniform phase which smoothly approaches
the hexagonal (or equilateral triangular) lattice phase with vanishing
amplitude $V$. At small but finite amplitude $V$, the hexagonal phase is
distorted and hexatically locked at an angle of $\varphi \approx 3.8^\circ$
with respect to the substrate lattice.  The square-to-hexagonal transformation
in this two-dimensional commensurate-incommensurate system thus involves a
complex pathway with various non-trivial lattice- and modulated phases.
\end{abstract}


\maketitle

\section{Introduction}\label{sec:intro}

Interacting particles exhibit complex structural changes when subjected to an
external modulated potential. A periodic modulation which is out of registry
with the particle lattice can induce lattice distortions, favor alternative
lattices, or generate non-uniform phases in commensurate-incommensurate
transitions \cite{Bak_82}, where solitons separate different regions of locked
phases. On the other hand, a random potential typically leads to the loss of
long-range order \cite{Larkin_70}. Besides such structural changes, both,
periodic and random potentials lead to pinning of the particles to the
substrate, thereby changing their dynamics under the action of an applied
force, a phenomenon which is of particular importance in the context of
dissipation-free transport in type II superconductors. In this paper, we focus
on the structural change of a two-dimensional particle system when it is
subjected to a periodic potential of different symmetry. We analyze the
specific case where the isotropic interaction between particles favors a
hexagonal (or equilateral triangular) lattice and subject it to a square
substrate lattice with lattice constant $b$ and of varying strength $V$,
fixing the pressure $p$ such as to generate a commensurate density $n = 1/b^2
= (4/3)^{1/2}/a^2$ for the free hexagonal phase.  Increasing the substrate
potential $V$, we find the complete pathway of transformations that takes the
freely floating hexagonal phase at $V=0$ to the fully locked square phase at
large $V$.

Effects of discommensuration in periodic external potentials appear in
numerous physical systems, prominent examples being atoms on surfaces, e.g.,
Krypton on graphite \cite{Specht_84}, vortices in modulated superconducting
films \cite{Daldini_74} and in periodic pinning arrays \cite{Tonomura_96},
flux quanta in Josephson junction arrays \cite{Webb_83}, or colloidal
monolayers on periodic substrates \cite{Mangold_03}. Recent applications of
such ideas in cold gases are the theoretical analysis of vortex pinning in a
Bose-Einstein Condensate subject to an optical lattice
\cite{Reijnders_04,Tung_06} or the proposal \cite{Graenz_14} to realize this
physics in a system of dipolar molecules \cite{pol_mol} subject to a square
optical lattice \cite{Hazzard_14}; the high tunability of these cold-gas
systems \cite{Buechler_07} then can be used to explore the various structural
phases. Another recent example is the study of graphene on boron nitride
\cite{Woods_14} where novel electronic and optical properties can be found.

The original `misfit problem' has been formulated in one dimension (1D) and
dealt with a particle lattice with lattice constant $a$ subject to a periodic
substrate with incommensurate periodicity $b\neq a$. As shown by Frenkel and
Kontorova \cite{FrenkelKontorova_39} and by Frank and Van der Merve
\cite{FrankVdMerwe_49}, the locked system at large potential $V$ (with particle
separation $b$) transforms into the free lattice at $V=0$ (with separation $a$
between particles) via a smooth commensurate-incommensurate transition. The
intermediate non-uniform phase involves solitons with cores approximating the
free phase and separating regions of locked phase; the dense soliton array
then approaches the free phase with lattice constant $a$. 

\begin{figure}[h]
\begin{center}
\includegraphics[width=8.5cm]{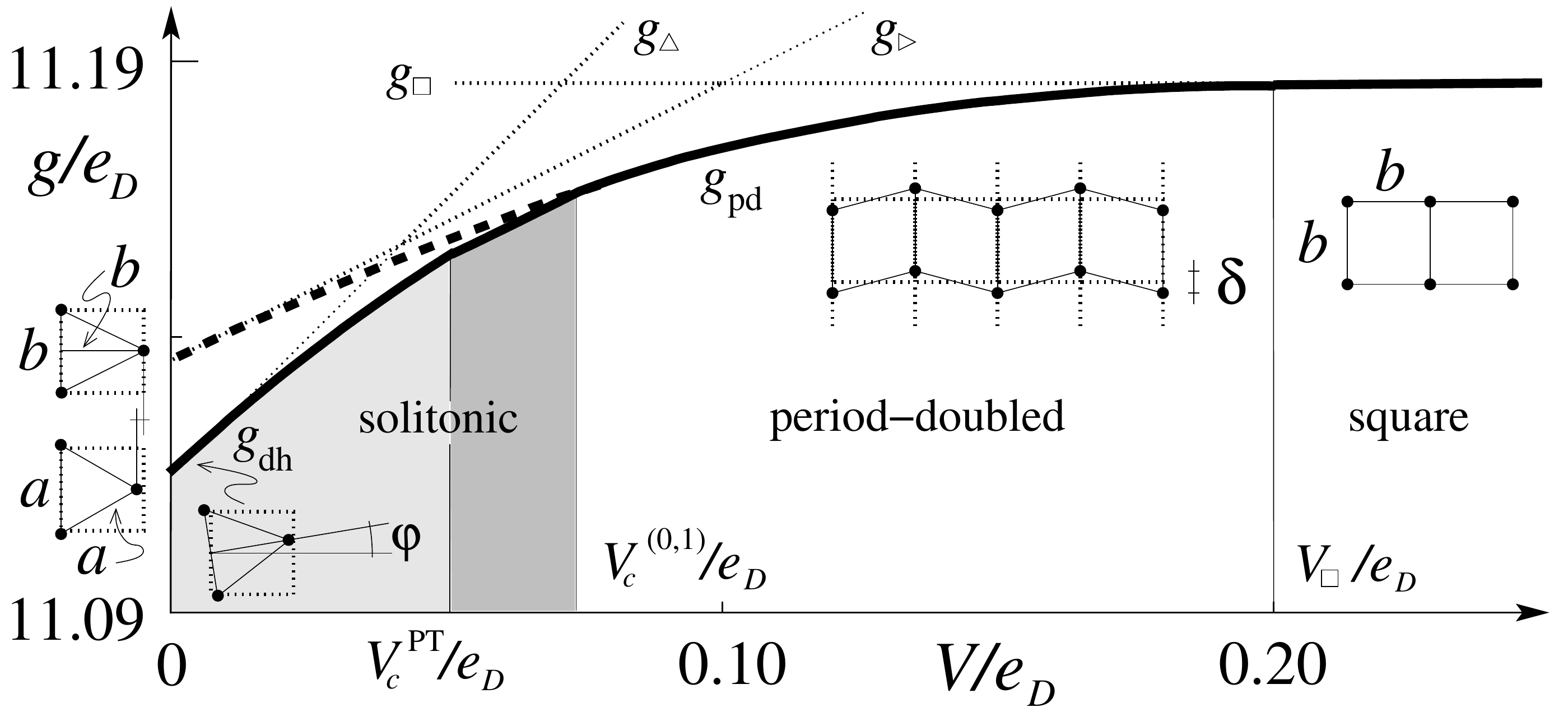}
\end{center}
\caption{\label{fig:phase_dia} Gibbs free energy of optimal states (thick
line), hexagonal at $V=0$, distorted and rotated hexagonal ($g_\mathrm{dh}$)
at small $V$, solitonic and period-doubled ($g_\mathrm{pd}$) at intermediate
$V$, and square for $V > V_{\scriptscriptstyle \square}$. Below the critical
potential $V_c^{\scriptscriptstyle (0,1)}$, the period-doubled phase smoothly
transforms into the hexagonal lattice via two soliton transitions involving
different soliton arrays. The dashed line extrapolates the energy
$g_\mathrm{pd}$ of the period-doubled phase. Dotted lines are energies of
rigid hexagonal ($\triangle$), rhombic ($\rhd$), and square ($\square$)
configurations.}
\end{figure}

The distortion of a two-dimensional particle lattice due to a weak substrate
potential has been analyzed by McTague and Novaco \cite{McTague_79} within a
perturbative (in small $V$) approach (for lattices with equal symmetry); the
distorted lattice becomes orientationally locked to the substrate at a
non-trivial angle $\varphi$ that depends on the elastic properties of the
lattice and on the misfit parameter. Adapting this analysis to our situation,
we find a locking angle $\varphi \approx 3.8^\circ$ between the height of the
equilateral triangular unit cell and one of the main axes of the substrate
potential, see Fig.\ \ref{fig:phase_dia}. The commensurate--incommensurate
transition in two dimensions (2D) has been first addressed by Pokrovsky and
Talapov \cite{PokrovskyTalapov_79,PT_pap,PT_book} within the so-called
`resonance approximation' where only the leading harmonic of the substrate
potential is accounted for. In this situation, the problem reduces to a 1D one
and the system develops a secondary structure in the form of an array of
soliton-lines.  However, with only one substrate harmonic present, the
commensurate phase at large potential $V$ is always a rhombic lattice with
base $b' > b$ (we call it the $bb'$ rhombic lattice)---in order to obtain the
complete pathway connecting the hexagonal and square lattices, both harmonics
have to be accounted for.

Rather than starting from the free hexagonal phase and increasing the
substrate potential $V$, it is then more opportune to start from the locked
square phase and decrease $V$. As the square phase is merely stabilized by the
large substrate potential, decreasing  $V$ naturally generates an instability.
It turns out that the leading instability appears at $V_{\scriptscriptstyle
\square} \approx 0.2\, e_{\scriptscriptstyle D}$ and is given by a shear
distortion ${\bf u}_\mathrm{pd} = (0,\delta/2)$ with amplitude $\delta$
parallel to the $y$-axis and a wave vector ${\bf q}_\mathrm{pd} = (\pi/b,0)$
along $x$ residing at the Brillouin zone boundary (alternatively, the
spontaneous symmetry breaking involves a distortion ${\bf u}_\mathrm{pd} =
(\delta/2,0)$ along $x$ with a wavevector ${\bf q}_\mathrm{pd} = (0,\pi/b)$
along $y$).  The resulting zig-zag lattice, see Fig.\ \ref{fig:phase_dia},
exhibits a doubled unit cell and has been found before in the context of
vortex pinning by a square-lattice pinning potential \cite{Zhuralev_03}.
Together with the symmetry breaking defining the direction of period-doubling,
a second spontaneous symmetry breaking fixes the sign of the amplitude
$\delta$, that defines two twin versions of the zig-zag phase.  Upon
decreasing $V$ further, the amplitude $\delta$ of the zig-zag distortion
increases, assuming the value $\delta = \pm b/2$, and hence resulting in a
rhombic (or isosceles triangular) lattice, at $V=0$. Although this lattice is
close to the hexagonal one, it has the wrong symmetry and hence further
transitions are needed to reach the free hexagonal phase.

These transitions are of the commensurate-incommensurate type and the task is
to find the most favorable soliton-line appearing first upon decreasing $V$.
While in one dimension only two types of point-solitons either diluting or
compressing the particle chain by $\pm b$ are possible, in two dimensions
soliton-lines with different `topological vector charge' ${\bf d}_{j,k} =
(-jb, kb/2)$, $j$ and $k$  mutually prime integers, can be conceived, where
the vector charge defines the translation of the lattice on itself or on a
twin after the passage of the soliton-line.  Thereby, shift vectors ${\bf
d}_{j,k}$ with odd values of $j+k$ are domain walls connecting unequal twins
with $\pm \delta$, while even values of $j+k$ belong to solitons connecting
same twins. Out of the many candidate defects, we find that a dilution-type
domain wall (connecting two period-doubled twins) with displacement vector
${\bf d}_{01} = (0,b/2)$ is the most favorable defect appearing at the highest
value $V_c^{\scriptscriptstyle (0,1)} \approx 0.074 \, e_{\scriptscriptstyle
D}$. Quite surprisingly, the domain wall array does not follow one of the
symmetry axes of the parent crystal, although such symmetric arrangements have
been predicted in the literature \cite{Chaikin_95}. When domain walls are
flooding the lattice upon further decrease in $V$, they wash out the substrate
mode along the $y$-axis and the particle lattice approaches the $bb'$ rhombic
phase resulting from the resonance approximation that neglects just this mode
and has been encountered in the discussion further above. The further decrease
in $V$ then follows the path described before, with the first
Pokrovsky-Talapov type soliton appearing at $V_c^{\rm\scriptscriptstyle PT}
\approx 0.046 \, e_{\scriptscriptstyle D}$ and developing into the distorted
and rotated hexagonal phase as the soliton density increases at small $V$.
Note that the vector charge of the Pokrovsky-Talapov soliton-line has only one
of its components quantized, ${\bf d}^{\rm\scriptscriptstyle PT} = (-b, \delta
y)$ with the shift $\delta y$ along $y$ fixed by the elastic properties of the
particle lattice but assuming any (non-quantized) value. The above described
rather complex pathway for the square-to-hexagonal transformation involving
distorted lattice- as well as non-uniform soliton phases is illustrated in
Fig. \ref{fig:phase_dia} and is the main result of this paper.

In the following, we define the model and discuss the rigid lattice
approximation (Sec.\ \ref{sec:model}) in order to obtain a rough layout of the
possible phases and associated energies. We then focus on large values of the
substrate potential $V$: in section \ref{sec:per-doub}, we find the shear
instability in the square lattice and derive a simplified but very accurate
model free-energy determining the amplitude $\delta$ of the distortion and the
energy density $g_\mathrm{pd}$ of the period-doubled phase, see Fig.\
\ref{fig:phase_dia}.  In section \ref{sec:trian}, we analyze the situation at
small substrate potential $V$ \cite{McTague_79}: using the harmonic- and
continuum elastic approximations for the hexagonal lattice, we minimize the
system's Gibbs free energy as a function of angle $\varphi$ between the
particle- and substrate lattices and determine the optimal value
$\varphi_\mathrm{min}$ as well as the associated energy density
$g_\mathrm{dh}(V)$ of the distorted hexagonal phase.  Non-uniform phases are
first introduced in section \ref{sec:sol_ra}: increasing $V$ further, the
distorted hexagonal phase develops into a soliton phase which we describe in
the resonance approximation\cite{PokrovskyTalapov_79,PT_pap,PT_book}, dropping
the subdominant mode in the substrate potential.  We will see, how the soliton
array generates a (locally modulated) distortion and rotation that transforms
the hexagonal lattice into a $bb'$ rhombic (or isosceles triangular) lattice.
Besides understanding the functionality of the soliton phase in transforming
the hexagonal lattice to the $bb'$ rhombic one, we are particularly interested
in the value of the critical potential $V_c$ for the
commensurate-incommensurate transition. The latter can be more easily found by
starting from the commensurate phase at higher $V$ and determining the
instability for the first soliton entry. In section \ref{sec:s_sol}, we
calculate the energy of such an individual soliton using the elasticity theory
for the $bb'$ rhombic lattice and find the critical substrate potential---the
result agrees quite well with the one obtained from the elastic theory for the
hexagonal lattice. However, other characteristics such as the soliton
direction or soliton amplitude turn out quite different, which we attribute to
anharmonicities becoming important in our problem due to the rather large
misfit between the hexagonal- and square phases. In order to determine an
accurate and reliable value for the critical substrate potential, we solve the
particle problem numerically, using the analytic solution as a variational
starting point and relaxing the particle positions to find the optimal soliton
shape.  Section \ref{sec:sol_dw} deals with the full two-dimensional problem
(beyond the resonance approximation): Several candidate solitons and domain
walls with different shift vectors ${\bf d}_{j,k} = (-jb,kb/2)$ then have to
be tested for their critical substrate potential $V_c^{\scriptscriptstyle
(j,k)}$---the soliton or domain-wall with the highest $V_c$ will then trigger
the transformation away from the period-doubled phase to a non-uniform
soliton phase.  In order to reach the required precision to separate the
critical substrate amplitudes $V_c^{\scriptscriptstyle (j,k)}$, the latter
have to be determined numerically. Finally, we will establish the
transformation pathway in Sec.\ \ref{sec:pathway} and conclude in Sec.\
\ref{sec:con}

\section{Model and Rigid Lattice Approximation}\label{sec:model}

In two dimensions (we consider the ${\bf r} = (x,y)$-plane), particles
interacting via a repulsive isotropic two-body potential $\Phi(r)$ arrange in
a hexagonal lattice. In this paper, we consider the case of particles with
long-range repulsive dipolar interactions $\Phi({\bf r}) = D/r^3$, where $D =
d^2$ derives from (electric/magnetic) dipoles ${\bf d}$ aligned parallel to
the $z$-axis; this situation describes the case of atoms physisorbed on a
surface \cite{Bolshov_77}, colloidal monolayers \cite{Mangold_03}, or dipolar
molecules in a 2D flat trap \cite{Buechler_07}.  In other cases, e.g., vortex
systems \cite{Tonomura_96, Reijnders_04}, the interaction falls off
logarithmically, changing the numerical values of the results in the analysis
described below---however, we expect that such systems exhibit a similar
overall behavior.

We submit this particle-system to a periodic lattice, in our case a square
periodic lattice with lattice constant $b$ and amplitude $V$. Such a substrate
lattice (possibly of another, e.g., hexagonal, symmetry) appears naturally in
the case of physisorption on a surface \cite{Specht_84} or is artificially
imposed with the help of optical tweezers \cite{Mangold_03}, optical lattices
\cite{Buechler_07}, or pinning arrays \cite{Tonomura_96}. We consider the case
where thermal and quantum fluctuations are negligible; this situation is
described through the Hamiltonian or total energy $E$ for $N$ particles
confined within an area $A$,
\begin{eqnarray}\label{eq:E}
   E(A,N) &=& E^\mathrm{int} + E^\mathrm{sub} \\ \nonumber
          &=& \frac{1}{2}\sum_{i\neq j} \frac{D}{r_{ij}^3} +
            \frac{V}{2}\sum_{i,\alpha} \bigl[1-\cos({\bf q}_\alpha\!\cdot\!
                                                    {\bf r}_i)\bigr],
\end{eqnarray}
where particles are located at positions ${\bf r}_i$ with distances $r_{ij}
\equiv |{\bf r}_i - {\bf r}_j|$; the substrate potential involves the two modes
${\bf q}_1 = (q,0)$ and ${\bf q}_2 = (0,q)$ with $q=2\pi/b$ along the $x$- and
$y$-axis. 

A crucial parameter is the particle density $n= N/A$ determining the lattice
constant $a = (4/3n^2)^{1/4}$ of the hexagonal lattice. We choose to work at
fixed pressure $p$, which is arranged in such a way as to define a
commensurate density $n = 1/b^2$ at $V=0$, i.e., for the free hexagonal phase;
the change to a situation with a fixed chemical potential $\mu$ is
straightforward. For a large potential $V$, the particles fit the minima of
the substrate potential (hence again $n = 1/b^2$), however, the density will
change, in fact decrease, at intermediate values of $V$. The misfit parameter
$s$ between the hexagonal and square lattices is determined by the distance
between rows in the hexagonal (the height $h = \sqrt{3/4}a$) and in the square
lattice (the lattice constant $b$),
\begin{equation}\label{eq:s}
   s = \frac{b}{h}-1 \approx 0.0746,
\end{equation}
and corresponds to a lattice constant $a$ of the hexagonal lattice that is
slightly larger than that of the substrate lattice, $a = (4/3)^{1/4} b > b$.
Note that large misfits $s$ potentially create large lattice distortions, what may
turn out problematic in the use of the harmonic approximation; we will see below
that our misfit parameter of order $\sim 0.1$ is quite large in this respect.

Working at fixed pressure $p$, the appropriate potential to minimize is the
Gibbs free energy $G(p,N)$: starting with the system's energy \eqref{eq:E} for
$N$ particles trapped within the area $A$, the Legendre transform with
$\partial_A E =-p$ provides us with the Gibbs free energy per particle
\begin{equation}\label{eq:gpN}
   g(p) = G(p,N)/N = [E(A,N) + pA]/N,
\end{equation}
where the thermodynamic limit $N,~A \to \infty$, $n=N/A = \mathrm{const.}$ is
implied.

In the rigid lattice approximation, we fix lattice sites ${\bf R}_j$ and
determine the Gibbs free energy density $g$ via straightforward summation. The
long-range potential in the interaction energy density $e^\mathrm{int}$ is
conveniently handled with an Ewald\cite{Ewald_21} summation technique (see
appendix \ref{app:ewald}), splitting the sum in Eq.\ \eqref{eq:E} into two
terms describing near and distant particles through real- and reciprocal space
contributions,
\begin{align} \nonumber
   e^\mathrm{int} &=\pi e_{\scriptscriptstyle D} \Bigl\{\frac{4}{3}
   +\sum_{j\neq0}\bigr[\Psi_{\frac{1}{2}}(\epsilon {R_j}^2)
                      +\Psi_{-\frac{3}{2}}({K_j}^2/4\epsilon)\bigl]\!\Bigr\}\\
   &= \pi e_{\scriptscriptstyle D} \Bigl\{\frac{4}{3}
   +\sum_{j\neq0}\bigr[\Psi_{\frac{1}{2}}(\pi n R_j^2)
                      +\Psi_{-\frac{3}{2}}(\pi n R_j^2)\bigl]\!\Bigr\},
   \label{eq:eint}
\end{align}
where we have chosen an Ewald parameter\cite{Ewald_21,Bonsall_77,RK} $\epsilon
= \pi n$ and have used the property $K_j^2 = (2\pi n)^2 R_j^2$ in the last
equation.  Here, $\Psi_{x}(\beta)=\beta^{-(x+1)} \Gamma(x+1,\beta)$ with
$\Gamma(x,\beta)$ the incomplete Gamma function, ${\bf K}_j$ are reciprocal
lattice sites, and $e_{\scriptscriptstyle D} = D n^{3/2} = D/b^3$ is the
dipolar energy scale.  
\begin{figure}[h]
\begin{center}
\includegraphics[width=8.5cm]{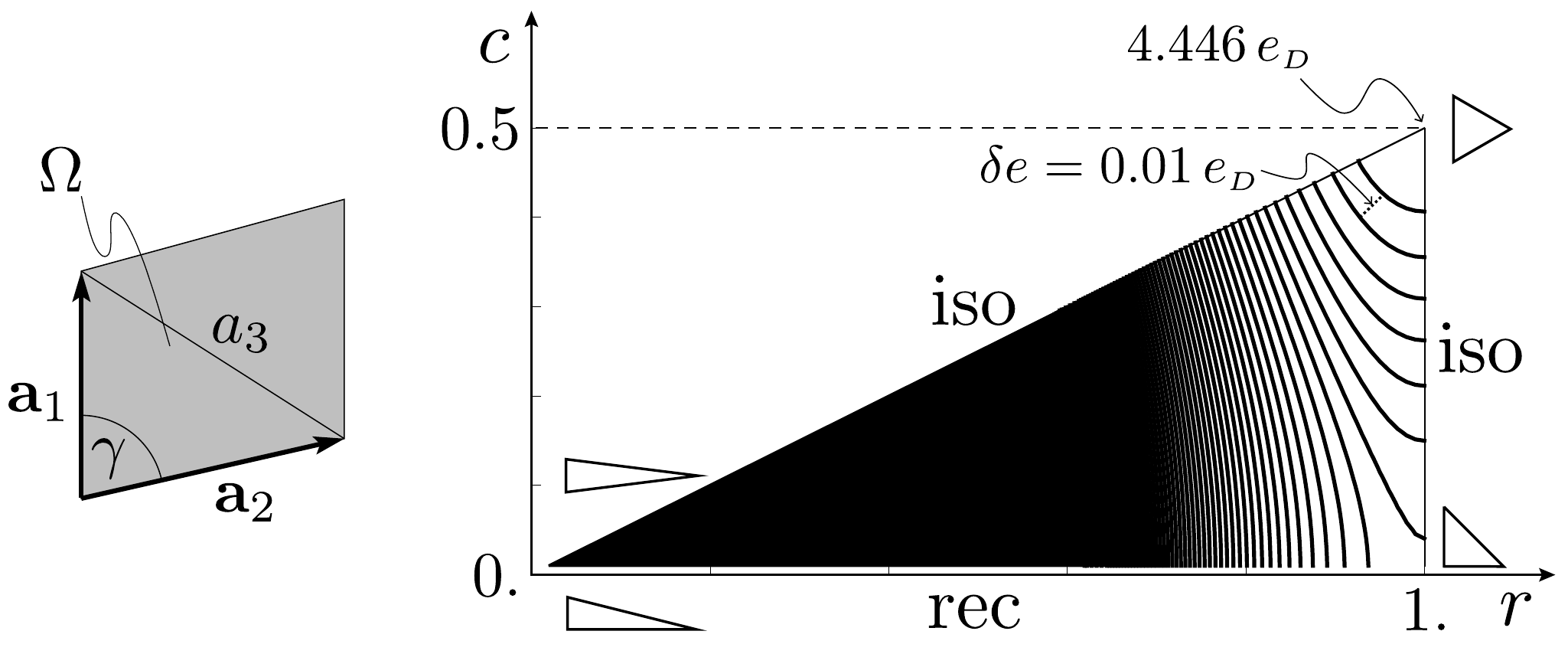}
\end{center}
\caption{\label{fig:BL} 
   Left: Parametrization of 2D Bravais lattices with given unit cell area
   $\Omega = b^2$. The lattice vectors ${\bf a}_1$ and ${\bf a}_2$ are choosen
   such that $a_1 < a_2 < a_3$ and $\gamma$ is the angle enclosed by ${\bf
   a}_1$ and ${\bf a}_2$. Right: The region $0 <r=a_1/a_2 < 1$, $0 \leq
   c=\cos\gamma \leq r/2$ uniquely covers all possible Bravais lattices.
   Boundaries correspond to rectangular unit cells and isosceles triangles
   (rhombic lattices), respectively. The contours mark constant energy lines,
   with the lowest energy $e_{\scriptscriptstyle \triangle} \approx 4.446 \,
   e_{\scriptscriptstyle D}$ attained for the hexagonal or equilateral
   triangular lattice.  The square lattice assumes a saddle-point
   configuration. Lines start at $e_{\scriptscriptstyle \triangle}$ and are
   separated by $0.01\, e_{\scriptscriptstyle D}$.}
\end{figure}

Parametrizing the 2D Bravais lattices by the lattice vectors ${\bf a}_1$ and
${\bf a}_2$ with $a_1 \leq a_2 \leq a_3$ via the length ratio $r = a_1/a_2$
and the angle $\gamma$ enclosed between the two (shortest) vectors ${\bf a}_1$
and ${\bf a}_2$, we can describe all lattices by choosing values $0 < r \leq
1$ and $0 \leq c = \cos\gamma \leq r/2$, see Fig.\ \ref{fig:BL}. The latter
boundary derives from the condition $a_3^2 = a_1^2 + a_2^2 -2 a_1 a_2 \cos
\gamma \geq a_2^2$ with $\cos\gamma >0$ (a situation with $\gamma > \pi/2$ can
be reduced to the triangular region in the $c$-$r$ diagram by choosing another
lattice vector). The interaction energy $e^\mathrm{int}$ per particle for the
different lattices is shown in the contour plot of Fig.\ \ref{fig:BL}. As
expected, the minimal energy is assumed by the hexagonal lattice with
$e_{\scriptscriptstyle \triangle} \approx 4.446\, e_{\scriptscriptstyle D}$,
but other configurations such a the rhombic lattice with height $b$
($e_{\scriptscriptstyle \rhd} \approx 4.467\, e_{\scriptscriptstyle D}$) or
the square lattice (with $e_{\scriptscriptstyle \square} \approx 4.517\,
e_{\scriptscriptstyle D}$) are close-by in energy.

Minimizing the Gibbs free energy $g(p,n) = e_{\scriptscriptstyle \triangle}(n)
+p/n$ at $V=0$ with respect to the density $n$ provides us with the expression
for the pressure\cite{remark_mu}
\begin{equation}\label{eq:p}
   p = \frac{3}{2} n e_{\scriptscriptstyle \triangle}(n).
\end{equation}
In the following, we will fix the pressure $p$ to generate the density $n =
1/b^2$ at $V=0$; when increasing the substrate potential $V$ at this fixed
value of $p$, the density will change and return back to the value $n = 1/b^2$
at larger $V$ in the period-doubled and square phases. Note that with a purely
repulsive interaction it is the pressure $p$ which determines the lattice
constant $a$, i.e., the hexagonal lattice has no own `generic' lattice
constant, see also the discussion in appendix \ref{app:el_const}.

When calculating the substrate energy in a rigid lattice approximation, we
encounter three classes with the following substrate energies per particle,
depending whether none, one, or two substrate modes line up with the particle
positions,
\begin{equation}
\label{eq:classes}
   e^\textrm{sub}(V)=\left\{\begin{array}{ll}
   V & {\bf q}_1,{\bf q}_2\notin\{{\bf K}_j\},\\
   \noalign{\vspace{5 pt}}
   V/2 & {\bf q}_1\textrm{ or }{\bf q}_2\in\{{\bf K}_j\},\\
   \noalign{\vspace{5 pt}}
   0 & {\bf q}_1,{\bf q}_2\in\{{\bf K}_j\},
\end{array}\right.
\end{equation}
where $\{{\bf K}_j\}$ denotes the set of reciprocal lattice vectors of the
particle lattice with sites $\{{\bf R}_i\}$. This follows from the sum
\begin{align} \label{eq:periodic}
   e^\textrm{sub}({\bf d})
   &=\frac{1}{N}\sum_{i=1}^N V^\textrm{sub}({\bf R}_i+{\bf d}) \\ \nonumber
   &=\frac{V}{2}\sum_{\alpha=1,2}
   \Bigl[1-\cos({\bf q}_\alpha\cdot{\bf d})
   \Bigl(\sum_j\delta_{{\bf K}_j,{\bf q}_\alpha}\Bigr)\Bigr],
\end{align}
where changes in the translational shift ${\bf d}$ quantify the locking energy
of the rigid lattice.

Combining the results for the interaction and substrate energies as well as
the pressure $p$, we find three favorable configurations, hexagonal
(unlocked), rhombic with base $b$ and height $b$ (single
locked\cite{Pogosov_03}, below called the $bb$-lattice), and square (double
locked) with Gibbs free energies $g(V) = e^\mathrm{int} + p/n +
e^\mathrm{sub}(V)$,
\begin{align} \nonumber
    g_{\scriptscriptstyle \triangle}(V) &= g_{\scriptscriptstyle \triangle} + V
    \approx 11.115 \, e_{\scriptscriptstyle D} + V, \\ \nonumber
    g_{\scriptscriptstyle \rhd}(V) &= g_{\scriptscriptstyle \rhd} + V/2
    \approx 11.136\, e_{\scriptscriptstyle D}+V/2, \\ 
    g_{\scriptscriptstyle \square}
    &\approx 11.186\, e_{\scriptscriptstyle D}.
    \label{eq:ri_lat}
\end{align}
Note that in choosing the orientation of the $bb$-lattice we have to break the
symmetry of the system spontaneously---this symmetry breaking is triggered by
the shear instability of the square phase leading to the period-doubled phase
as described in Sec.\ \ref{sec:per-doub} below.  The above expressions for
$g_{\scriptscriptstyle \triangle}(V)$, $g_{\scriptscriptstyle \rhd}(V)$, and
$g_{\scriptscriptstyle \square}$ already provide a reasonable approximation to
the energy $g$ versus potential $V$ diagram as illustrated in Fig.\
\ref{fig:phase_dia} (dotted lines).

\section{Period-doubled Phase}\label{sec:per-doub}

\begin{figure}[b]
\begin{center}
\includegraphics[width=7.0cm]{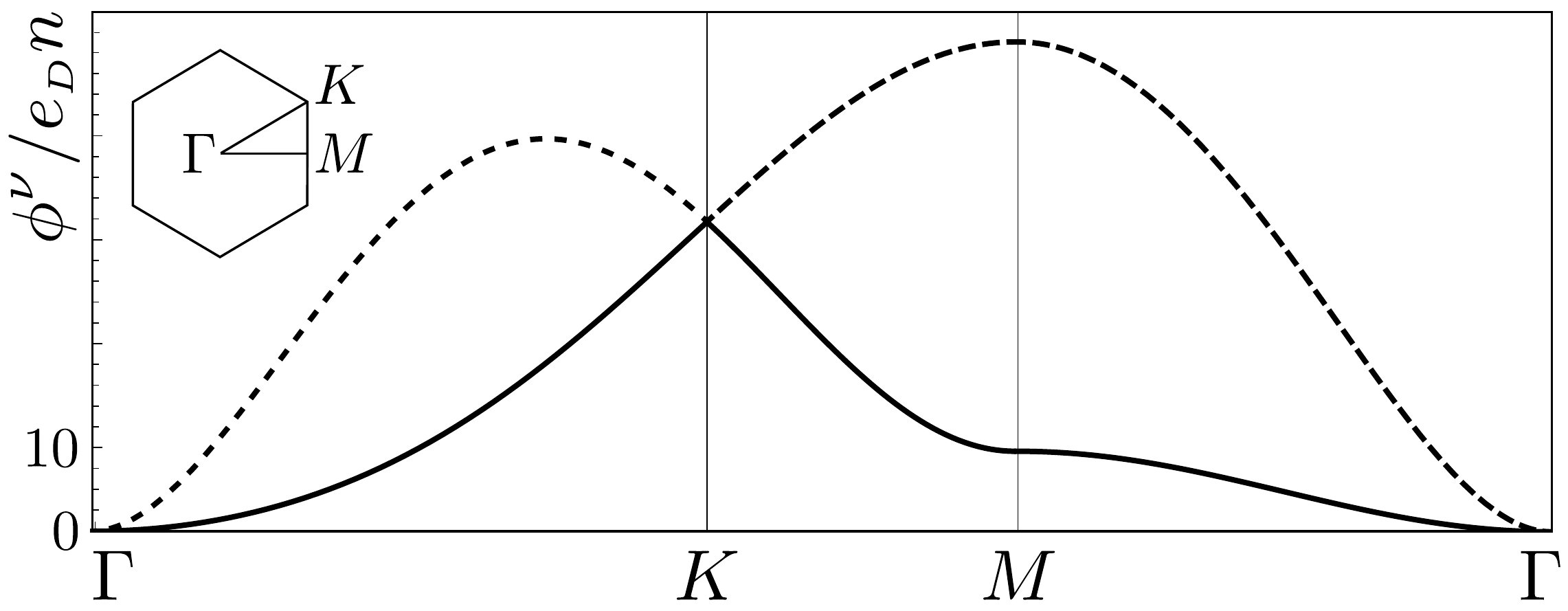}
\end{center}
\caption{\label{fig:ev3} 
   Transverse (lower branch) and longitudinal (upper branch) eigenvalues for
   the hexagonal lattice along symmetry axes as calculated using Eq.\
   \eqref{eq:dynmat}. For small wavevectors $k$ around the $\Gamma$-point,
   $\phi^\perp_k \approx (\mu/n) k^2$ and $\phi^\parallel_k
   \approx [(\kappa+\mu)/n] k^2$ describe shear and compression modes.}
\end{figure}

Going beyond the rigid lattice approximation, we account for small
deviations ${\bf u}_i$ of the particle coordinates ${\bf r}_i = {\bf
R}_i^\mathrm{latt}+{\bf u}_i$ from regular lattice positions ${\bf
R}_i^\mathrm{latt}$. The harmonic expansion of the energy \eqref{eq:E} in the
displacement field ${\bf u}_i$ provides us with corrections to the Gibbs free
energy $g = g_\mathrm{latt} +\delta g$. The interaction energy $g^\mathrm{int}
= e^\mathrm{int} +p/n$ contributes a term
\begin{equation}\label{eq:gint}
   \delta g^\mathrm{int} \approx \frac{1}{2N} \sum_{i,j} u_{i\mu} \,
   \Phi_{\mu\nu}^{\scriptscriptstyle D}
   ({\bf R}_{ij}^\mathrm{latt}) \, u_{j\nu},
\end{equation}
with the elastic matrix $\Phi_{\mu\nu}^{\scriptscriptstyle D}({\bf
R}_{ij}^\mathrm{latt})$ depending on the chosen lattice, while the substrate
potential adds a second term $\delta e^\mathrm{sub}$ to $\delta g$, $\delta g
= \delta g^\mathrm{int} + \delta e^\mathrm{sub}$. For a dipolar system, the
elastic matrix assumes the form
\begin{align} \nonumber
   \Phi^{\scriptscriptstyle D}_{\mu\nu}({\bf R}_{ij})
   &= D(1-\delta_{ij})\Bigl[3\frac{\delta_{\mu\nu}}{{R_{ij}}^5}
   -15\frac{{\bf{R}}_{ij,\mu}{\bf{R}}_{ij,\nu}}{{R_{ij}}^7}\Bigr]\\
   &-D\delta_{ij}\sum_{l,l\neq i}\Bigl[3\frac{\delta_{\mu\nu}}{R_{il}^5}
   -15\frac{{\bf{R}}_{il,\mu}{\bf{R}}_{il,\nu}}{{R_{il}}^7}\Bigr].
   \label{eq:Phi_int}
\end{align}
Its Fourier transform\cite{remark_inv_FT}
   $\Phi_{\mu\nu}^{\scriptscriptstyle D}({\bf  k})
   =\sum_j \Phi_{\mu\nu}^{\scriptscriptstyle D}({\bf R}_{ij})
   \exp{(-i{\bf  k}\cdot{\bf  R}_{ij})}$
is conveniently calculated with the help of the Ewald summation technique,
\cite{Ewald_21}
\begin{widetext}
\begin{align} \label{eq:dynmat}
   \Phi_{\mu\nu}^{\scriptscriptstyle D}({\bf {k}})
   &= 4\pi^2Dn^{5/2} \Bigl\{ \frac{k_\mu k_\nu}{2\epsilon}\, 
     \Psi_{-\frac{3}{2}}(k^2/4\epsilon)
   + \sum_{j\neq0} \Bigl[2\epsilon R_{j\mu} R_{j\nu}\, 
    \Psi_{\frac{5}{2}}(\epsilon {R_j}^2)
   -\delta_{\mu\nu} \Psi_{\frac{3}{2}}(\epsilon {R_j}^2)\Bigr]
     \bigl[1-\cos({\bf {k}}\cdot{\bf {R}}_j)\bigr] \nonumber\\
   &+\sum_{j\neq0}\Bigl[\frac{(K_{j\mu}-k_\mu)(K_{j\nu}-k_\nu)}
   {2\epsilon}\,\Psi_{-\frac{3}{2}}(\vert{\bf {K}}_j
   -{\bf \bf {k}}\vert^2/4\epsilon)
    -\frac{K_{j\mu} K_{j\nu}}{2\epsilon}\,
   \Psi_{-\frac{3}{2}}(K_j^2/4\epsilon)\Bigr]\Bigr\}
\end{align}
\end{widetext}
with ${\bf k}$ residing in the first Brillouin zone.  Determining the
eigenvectors ${\bf e}_{\bf k}^\nu$ and eigenvalues $\phi_{\bf k}^\nu$ allows
for a stability analysis of the lattice ${\bf R}_{i}^\mathrm{latt}$ under
small local distortions.  E.g., evaluating the eigenvalues along the high
symmetry directions in the Brillouin zone  for the hexagonal lattice ${\bf
R}_{i}^\mathrm{latt} = {\bf R}_{i}^{\scriptscriptstyle \triangle}$ one finds
positive transverse ($\nu = \>\perp$) and longitudinal ($\nu = \>\parallel$)
eigenvalues, see Fig.\ \ref{fig:ev3}; in the long-wavelength limit, these are
the usual compression and shear modes (see also appendix \ref{app:eltr})
\begin{align}\label{eq:LT_modes}
   {\bf  e}_{\bf  k}^\parallel &= (k_x,k_y)/k,
   \qquad \phi^\parallel_{k} \approx [(\kappa+\mu)/n]\,k^2,
   \\ \nonumber
  {\bf  e}_{\bf  k}^\perp &= (k_y,-k_x)/k,
  \quad~ \phi^\perp_{k} \approx (\mu/n)\, k^2,
\end{align}
with
\begin{align}\label{eq:mu_ka}
   \mu = \frac{3}{8}\,ne_{\scriptscriptstyle\triangle}, \qquad
   \kappa = \frac{15}{4}\,ne_{\scriptscriptstyle\triangle},
\end{align}
the shear and compression moduli of the isotropic (at small $k$) hexagonal
lattice.
\begin{figure}
\begin{center}
\includegraphics[width=7.0cm]{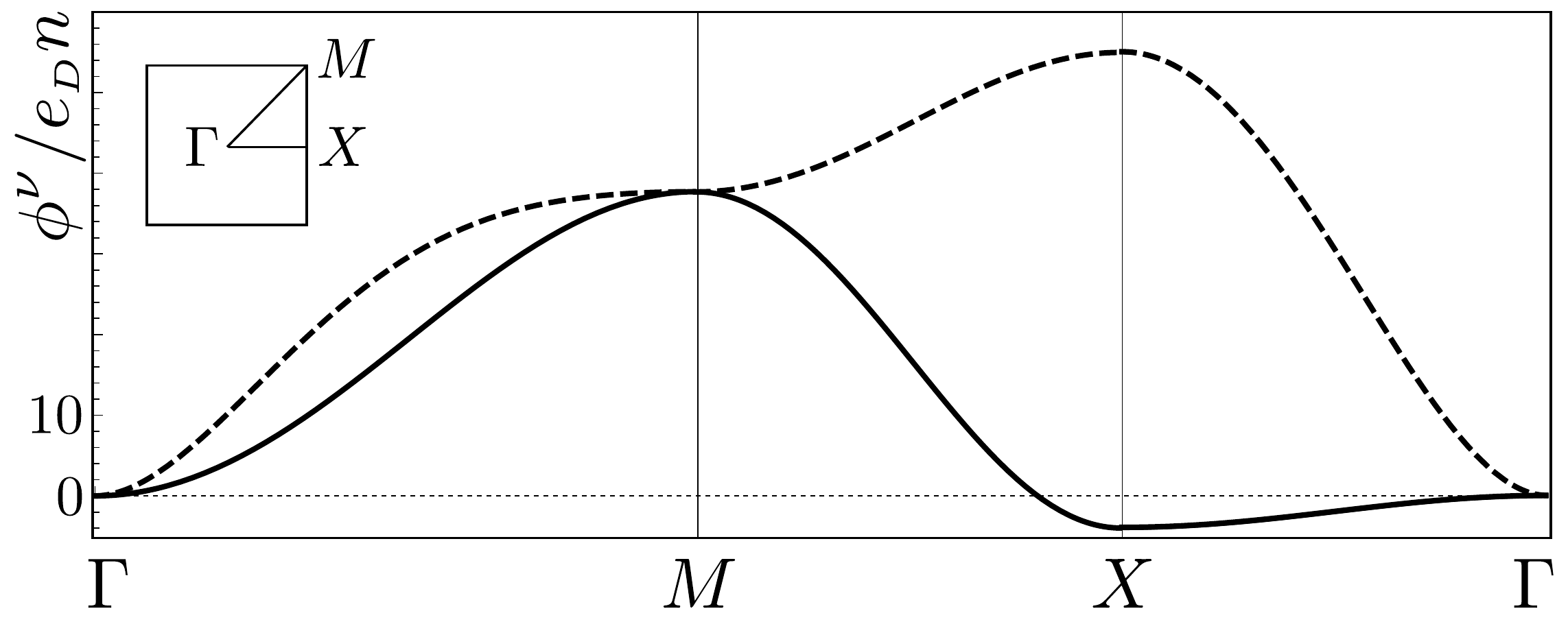}
\end{center}
\caption{\label{fig:ev4}
   Transverse (lower branch) and longitudinal (upper branch) eigenvalues for
   the square lattice along high symmetry axes as calculated using Eq.\
   \eqref{eq:dynmat}. The transverse branch becomes negative along the line
   $\Gamma-X-M$, indicating an instability of the square lattice.  The presence
   of the square substrate potential lifts these eigenvalues to stabilize them
   above a threshold $V_{\scriptscriptstyle\square}$.}
\end{figure}

\begin{figure}[h]
\begin{center}
\includegraphics[width=7cm]{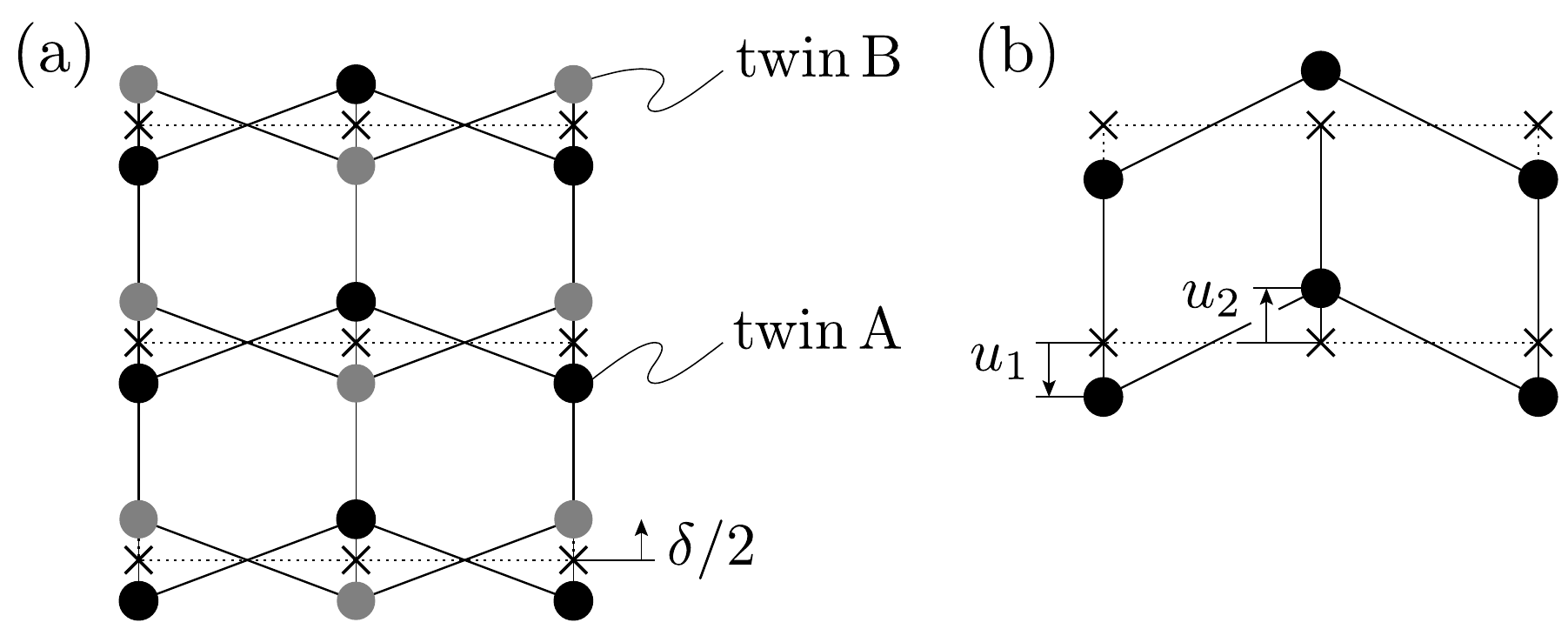}
\end{center}
\caption{\label{fig:twin}
   The two possible zig-zag structures for an instability at the $X$-point 
   $(\pi/b,0)$.  The crosses represent the undistorted square
   lattice, whereas the black and dark grey points show the twins A and B,
   respectively.  The twin structures transform into one another either by
   changing the sign of the distortion $\delta/2$ or by a shift by $b$ along
   $x$.}
\end{figure}

For a large substrate potential $V$, the substrate enforces a square lattice
with particle positions ${\bf R}_i^\mathrm{latt}={\bf R}_i^{\scriptscriptstyle
\square}$. Evaluating the elastic matrix $\Phi_{\mu\nu}^{\scriptscriptstyle
D}({\bf k})$ in Eq.\ \eqref{eq:dynmat} for ${\bf k}$ within the first
Brillouin zone and determining its eigenvalues, we find an unstable branch
with the largest negative eigenvalue appearing at the $X$-point assuming a
numerical value $\phi^\perp(0,\pi/b) =-3.958\, e_{\scriptscriptstyle D} n$,
see Fig.\ \ref{fig:ev4}. On the other hand, the substrate potential
contributes a term $\delta e^\mathrm{sub} \approx V q^2 |u_{\bf k}|^2/2$ to
the free energy correction $\delta g$, pushing up the entire spectrum.
Including this upward shift, all modes remain stable for $V >
V_{\scriptscriptstyle\square}$ with the critical value for the potential
$V_{\scriptscriptstyle\square}$ defined by the equation
\begin{equation} \label{eq:vstab}
   \frac{q^2}{2}V_{\scriptscriptstyle\square}=|\phi^\perp(0,\pi/b)|
   =3.958\, e_{\scriptscriptstyle D} n,
\end{equation}
hence
\begin{equation} \label{eq:V_sq}
   V_{\scriptscriptstyle\square}=0.201\, e_{\scriptscriptstyle D}.
\end{equation}
At $V=V_{\scriptscriptstyle\square}$ the lowest eigenvalue  of
$\Phi_{\mu\nu}^{\scriptscriptstyle D}({\bf k})$ touches zero at the $X$-points
$(\pi/b,0)$ and $(0,\pi/b)$ and the lattice deforms, with a shear mode
doubling the unit cell in one of the two principal directions $x$ or $y$; the
two possible choices for this zig-zag distortion correspond to a
$\mathbb{Z}_2$-symmetry breaking.  Furthermore, for each choice of principal
direction $x$ or $y$, the sign of the distortion $\delta/2$ can be reversed,
defining two twins as shown in Fig.\ \ref{fig:twin}(a) (alternatively, the
sign change in the distortion can be viewed as a shift by $b$).

\subsection{Period-doubled phase relative to the square lattice}
\label{ssec:pd_square}

Next, we determine the amplitude $\delta$ of the zig-zag distortion in the
period-doubled phase for $V < V_{\scriptscriptstyle \square}$ assuming an
instability realized at $X = (\pi/b,0)$. We start from the square lattice and
consider a rectangular unit cell with lattice vectors ${\bf a}_1^{\rm
\scriptscriptstyle R} = (2b,0)$ and ${\bf a}_2^{\rm \scriptscriptstyle R} =
(0,b)$ holding two particles at positions ${\bf c}_1 = (0,u_1)$ and ${\bf c}_2
= (b,u_2)$, see Fig.\ \ref{fig:twin}(b).  Defining center-of-mass and
difference coordinates
\begin{equation} \label{eq:sig_del}
   \sigma = (u_1+u_2)/2,\qquad \delta = u_1-u_2,
\end{equation}
we determine the interaction energy of the period-doubled phase choosing
a reference frame shifted by $\delta/2$,
\begin{align} \label{eq:eintdist}
   e^\mathrm{int}_\mathrm{pd}(\delta)
   &=\frac{1}{2}\sum_{j=1}^{N/2}\frac{D}{(R^{\,\rm \scriptscriptstyle R}_j)^3}
   +\frac{1}{2}\sum_{j=1}^{N/2}\frac{D}{\vert{\bf R}^{\rm \scriptscriptstyle
   R}_j +{\bf c}\vert^3}
\end{align}
with the shift ${\bf c}=(b,\delta)$.  The first sum is the energy of the
rectangular lattice and is evaluated with an Ewald summation to provide the
energy per particle $e_{\rm\scriptscriptstyle R}^\textrm{int} =
2.025\,e_{\scriptscriptstyle D}$.  The second sum is decomposed into a sum
over columns (index $m$) and rows (index $l$); applying the Poisson summation
rule to the sum over $l$, Eq.\ \eqref{eq:eintdist} can be rewritten as
(see appendix \ref{app:eff_pot})
\begin{align}
\label{eq:eds}
   e^\mathrm{int}_\mathrm{pd}(\delta) 
   =& e_{\rm\scriptscriptstyle R}^\textrm{int}
   +\frac{\pi^2}{4}\, e_{\scriptscriptstyle D} \\ 
   &+8\pi e_{\scriptscriptstyle D}\, \sum_{m>0}\sum_{l'>0}
   \frac{l' K_1[2\pi l' (2m-1)]}{2m-1}\cos(q l' \delta),\nonumber
\end{align}
where $l'$ accounts for the particle rows in reciprocal space; the term $l'=0$
has been treated separately and contributes the energy $(\pi^2/4)\,
e_{\scriptscriptstyle D}$. The modified Bessel function of the second kind
$K_1(z)$ decays rapidly, $K_1(z)\propto e^{-z}$, such that we can 
discard terms with $l' > 1$ and $m > 1$; the interaction energy per
particle then takes the simple form 
\begin{align} \label{eq:edsapp}
   e^\textrm{int}_\mathrm{pd}(\delta)&\approx e_{\rm\scriptscriptstyle R}^\textrm{int}
   +\frac{\pi^2}{4} e_{\scriptscriptstyle D}
   +8\pi e_{\scriptscriptstyle D} K_1[2\pi]\cos(q \delta)\\
   &\equiv C_1+C_2\cos(q \delta)\nonumber
\end{align}
with the constants $C_1=4.492\,e_{\scriptscriptstyle D}$ and $C_2=0.0248\,
e_{\scriptscriptstyle D}$. Going over to the Gibbs energy by adding the
pressure term $p/n =3e_{\scriptscriptstyle\triangle}/2$
and rearranging terms, we obtain
\begin{align} \label{eq:edsapp2}
   g^\textrm{int}_\mathrm{pd}(\delta)&=g_{\scriptscriptstyle\rhd}
   +\Delta\bigl[1+\cos{(q\delta)}\bigr]
   \intertext{with}
   \Delta&=\frac{g_{\scriptscriptstyle\square}-g_{\scriptscriptstyle\rhd}}{2}
   = 0.0248\, e_{\scriptscriptstyle D}.
\end{align}
It is easily seen that the `asymptotic cases' are in agreement with our
expectations, i.e., $g^\textrm{int}_\mathrm{pd}(\delta =0) =
g_{\scriptscriptstyle \square}$ and $g^\textrm{int}_\mathrm{pd}(\delta = \pm
b/2) = g_{\scriptscriptstyle\rhd}$.  The interaction energy
\eqref{eq:edsapp2} differs from the exact result obtained by the Ewald
method by far less than a per mill such that the approximation made in the
step going from \eqref{eq:eds} to \eqref{eq:edsapp} is well-justified.

\begin{figure}[h]
\begin{center}
\includegraphics[width=6cm]{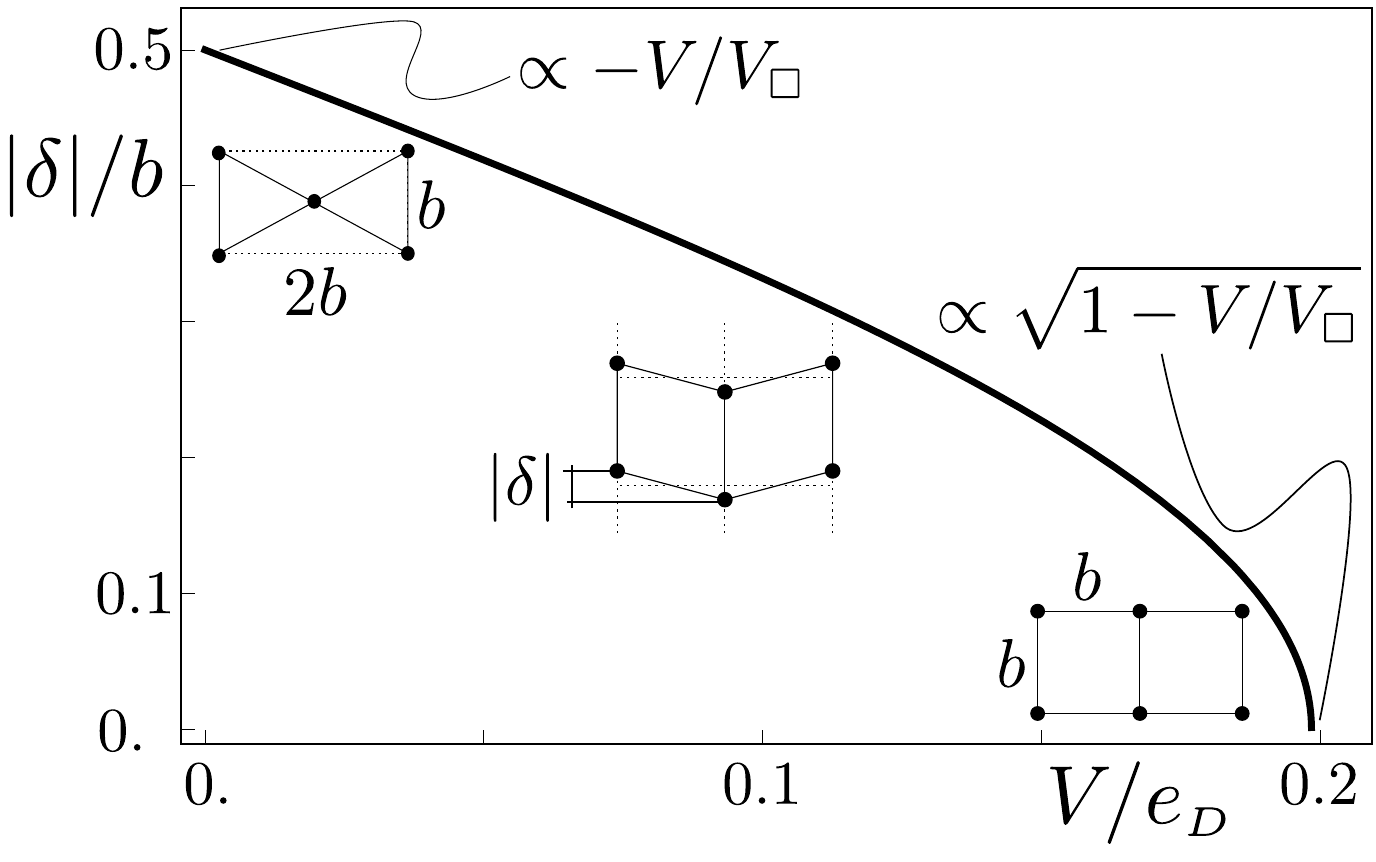}
\end{center}
\caption{\label{fig:rel_dist} 
   The relative distortion $|\delta|$ assumes its maximal value $b/2$ in the
   $bb$ rhombic phase at $V=0$, decreases $\propto V/V_{\scriptscriptstyle
   \square}$ for small substrate amplitudes, and goes to zero
   $\propto\sqrt{1-V/V_{\scriptscriptstyle \square}}$ as $V$ approaches
   $V_{\scriptscriptstyle \square}$.}
\end{figure}

The substrate potential $e^\mathrm{sub}$ contributes a term 
\begin{align} \label{eq:esubdist}
   e^\mathrm{sub}_\mathrm{pd}(V,\sigma,\delta)&=\frac{V}{2N}
   \sum_j\bigl[2-\cos(q u_1)-\cos(q u_2)\bigr]\nonumber\\
   &=\frac{V}{2}[1-\cos(q\sigma)\cos(q \delta/2)\bigr],
\end{align}
where the sum over $j$ goes over $N/2$ particles. Minimizing the Gibbs free
energy $g_\mathrm{pd} = g^\mathrm{int}_\mathrm{pd} +
e^\mathrm{sub}_\mathrm{pd}$ with respect to the distortion $\delta$, we find
the latter related to the center-of-mass coordinate $\sigma$ via
\begin{equation} \label{eq:fud}
   \cos(q \delta/2)=\frac{V}{8\Delta}\cos(q\sigma)
\end{equation}
and obtain the energy of the period-doubled phase
\begin{equation} \label{eq:egz}
   g_\textrm{pd}(V,\sigma)
   = g_{\scriptscriptstyle\rhd} +\frac{V}{2}-\frac{V^2}{32\Delta}
     +\frac{V^2}{64\Delta}\bigl[1-\cos(2q\sigma)\bigr],
\end{equation}
with the distortion $\delta$ slaved to $\sigma$. This slaved distortion
generates the period-halfing $b/2$ and a small periodic energy $V^2/32\Delta$
for the motion of the particle lattice along the $y$ axis (by increasing
$\sigma$), hence the period-doubled phase is pinned to the substrate with
respect to both directions $x$ and $y$, although much weaker along the
$y$-axis. Minimal energy configurations are realized for $\sigma=nb/2$,
$n\in\mathbb{Z}$.  Choosing the solution $\sigma = 0$ (or $2\sigma$ equal to
an even multiple of $b$), Eq.\ \eqref{eq:fud} provides us with 
the distortion amplitude
\begin{align}\label{eq:dV}
   \delta(V)=\frac{b}{\pi}\arccos(V/8\Delta),
\end{align}
where both signs of the $\arccos$ are relevant; the sign of $\delta$ then
decides into which of the two degenerate zig-zag solutions $u_1 = -u_2 =
\delta/2$ the system deforms, see Fig.\ \ref{fig:twin} ($\delta < 0$ for twin
A).  The condition $\delta = 0$ provides us with an alternative result for the
critical potential $V_{\scriptscriptstyle \square} = 8\Delta \approx 0.198\,
e_{\scriptscriptstyle D}$; this value is close to the previous result
(\ref{eq:V_sq}), again confirming that terms with $m>1$ or $l'>1$ in Eq.\
(\ref{eq:eds}) are indeed small. The order parameter approaches zero as
$\delta \approx \pm (\sqrt{2}\, b/\pi) \sqrt{1-V/V_{\scriptscriptstyle
\square}}$ on approaching the square lattice, while $\delta = \pm
(b/2)[1-2V/\pi V_{\scriptscriptstyle \square}]$ near $V=0$ describes the
vicinity of the $bb$ rhombic lattice with energy $g_{\scriptscriptstyle
\rhd}$, see Fig.\ \ref{fig:rel_dist}. When approaching the state with maximal
distortion amplitude $\pm b/2$ at $V=0^+$ the particles assume the symmetric
positions between the potential maxima and minima along $y$.  The energy of
the period-doubled phase
\begin{equation} \label{eq:epd}
   g_\textrm{pd}(V)
   = g_{\scriptscriptstyle\rhd} +\frac{V}{2}-\frac{V^2}{32\Delta}
\end{equation}
undercuts that of the rigid phase approximation and smoothly interpolates
between the energy $g_{\scriptscriptstyle \rhd}$ of the $bb$ rhombic lattice
at $V=0$ and the energy $g_{\scriptscriptstyle \square} =
g_{\scriptscriptstyle \rhd} + 2\Delta$ of the square lattice at
$V_{\scriptscriptstyle \square} = 8 \Delta$, see Fig.\ \ref{fig:phase_dia}.

Equivalent solutions (involving the branch of $\arccos$ around $0$) are
obtained for $2\sigma = nb$ with even $n$ (although the displacements $u_1$
and $u_2$ are no longer antisymmetric).  Care has to be taken when choosing
$2\sigma = nb$ with an odd integer $n$; in this case, the right hand side of
Eq.\ \eqref{eq:fud} is negative and the solutions for $\delta$ involve the
branches of $\arccos$ around $\pm\pi$.

\subsection{Period-doubled phase relative to the $bb$ rhombic
lattice} \label{ssec:pd_phase}

\begin{figure}[h]
\begin{center}
\includegraphics[width=7cm]{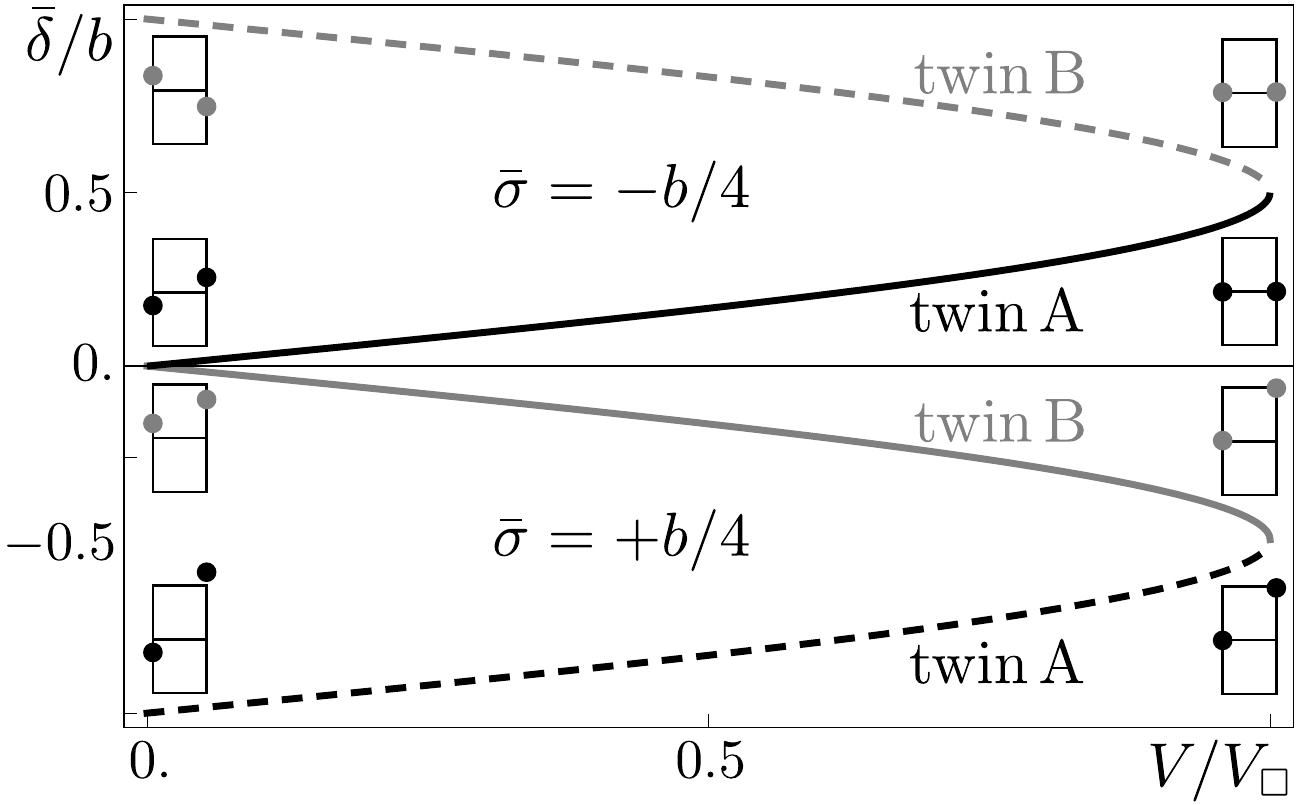}
\end{center}
\caption {\label{fig:branchArcSin}
   The twin-B phase may be reached from the twin-A phase by changing the
   center-of-mass coordinate $\bar{\sigma}$ from $\bar{\sigma}=-b/4$ to
   $\bar{\sigma}=b/4$.}
\end{figure}

In the analysis above, we have described the period-doubled phase as it
develops out of the square phase under a shear distortion that is increasing
with decreasing substrate amplitude $V$. On the other hand, when studying the
instability of the period-doubled phase towards formation of topological
defects (soliton- or domain-wall lines, see Sec.\ \ref{sec:sol_dw}) a
description with reference to the $bb$ rhombic phase is more convenient.
Defining the displacements $\bar{u}_1$ and $\bar{u}_2$ with respect to the
latter, we define the positions of the particles in the rectangular unit cell
via $\bar{\bf c}_1=(0,\bar{u}_1)$ and $\bar{\bf c}_2 =(b,b/2+\bar{u}_2)$ and
determine once more the interaction and substrate energies of the distorted
phase,
\begin{align} \label{eq:edsapp_iso}
   g^\textrm{int}_\mathrm{pd}(\bar{\delta})
   &=g_{\scriptscriptstyle\rhd} + \Delta\bigl[1-\cos{(q\bar{\delta})}\bigr]
\end{align}
and 
\begin{align} \label{eq:esub_iso}
   e^\textrm{sub}(V,\bar\sigma,\bar\delta)
   &=\frac{V}{2}\bigl\{1+\sin(q\bar{\sigma})\sin(q \bar{\delta}/2)\bigr\},
\end{align}
where $\bar\sigma = (\bar{u}_1 + \bar{u}_2)/2$ and $\bar\delta = \bar{u}_1 -
\bar{u}_2$.  Minimizing the total free energy with respect to $\bar\delta$ we
obtain
\begin{equation} \label{eq:ud}
   \sin(q \bar{\delta}/2)=-\frac{V}{8\Delta}\sin(q\bar{\sigma})
\end{equation}
and the energy
\begin{align} \label{eq:egz_iso}
   g_\mathrm{pd}(V,\bar{\sigma})
   &=g_{\scriptscriptstyle\rhd}
   +\frac{V}{2}-\frac{V^2}{32\Delta}
   +\frac{V^2}{64\Delta}\bigl[1+\cos(2q\bar{\sigma})\bigr].
\end{align}
Minima now are located at $2\bar{\sigma}= b(2n+1)/2, \,n \in \mathbb{Z}$, in
agreement with the results above as $\bar\sigma=\sigma-b/4$. Choosing $n=-1$,
$\bar{\sigma} =-b/4$ provides us with the identical particle positions as
before when starting from the square phase: The relative distortion
$\bar{\delta} = \delta + b/2$ grows from $\bar{\delta}=0$ at $V=0^+$ (twin A
solution, see Fig.\ \ref{fig:branchArcSin}) to $\bar{\delta}=b/2$ (square
lattice) as $V\to 8\Delta$,
\begin{align}\label{eq:pd}
   \bar{\delta}(V)=(b/\pi)\arcsin(V/8\Delta).
\end{align}
On returning back to $V=0^+$, we can follow the same path or choose another
branch of the $\arcsin$-function that has $\bar{\delta}$ increase further,
generating the twin B solution on returning back to $V=0^+$, see Fig.\
\ref{fig:branchArcSin}.  Note that the negative branch of the arcsin is not
compatible with Eq.\ \eqref{eq:ud} and $\bar\sigma = -b/4$. Instead, the
alternative twin phase, previously realized by changing the sign of $\delta$,
is now conveniently encoded through a change in the center-of-mass coordinate
by going over to the value $\bar\sigma = b/4$, see Fig.\
\ref{fig:branchArcSin}.
Below, we will find domain walls defined through shifts of the lattice along
$y$, i.e., by increasing $\bar\sigma$ by one period $b/2$ from one minimum in
the energy $g_\mathrm{pd}(V,\bar\sigma)$ to the next, see Eq.\
(\ref{eq:egz_iso}). When pushing the center-of-mass coordinate $\bar\sigma$
from $-b/4$ to $b/4$, the slaved distortion $\bar\delta$ will transit through
zero (where the lattice has the $bb$ rhombic geometry) and connect the twin A
with the twin B phase.

\section{Locked Hexagonal Phase}\label{sec:trian}

Next, we focus our interest on weak substrate potentials $V$. Going again
beyond the rigid lattice approximation, the lattice will deform and the
particle positions will deviate away from regular hexagonal lattice positions,
i.e., in Eq.\ (\ref{eq:E}) we choose ${\bf R}_i^\mathrm{latt} = {\bf
R}_i^{\scriptscriptstyle \triangle}$ and ${\bf r}_i = {\bf
R}_i^{\scriptscriptstyle \triangle} +{\bf u}_i$.  At $V=0$ the position and
orientation of the floating hexagonal lattice is arbitrary; without loss of
generality, we can fix the point ${\bf R}_0^{\scriptscriptstyle \triangle}$ in
a substrate minimum coming up at finite $V > 0$, e.g., ${\bf
R}_0^{\scriptscriptstyle \triangle} = (0,0)$. At finite but small $V$, the
particle lattice will relax and optimize its energy. This optimization depends
on the relative orientation $\varphi$, the angle enclosing the $x$-axis and
the height of a triangle as shown in Fig.\ \ref{fig:phase_dia}. Our task then
is to find the optimal angle providing the largest energy relaxation.
At small values of $V$, the displacements ${\bf u}_i$ remain small, $u_i \ll
a$ for all $i$, and the change in the interaction energy of Eq.\ (\ref{eq:E})
can be calculated in a harmonic approximation using $\delta
g^\mathrm{int}_{\scriptscriptstyle \triangle}$, see Eq.\ (\ref{eq:gint}). 
\begin{figure}[t]
\begin{center}
\includegraphics[width=4.5cm]{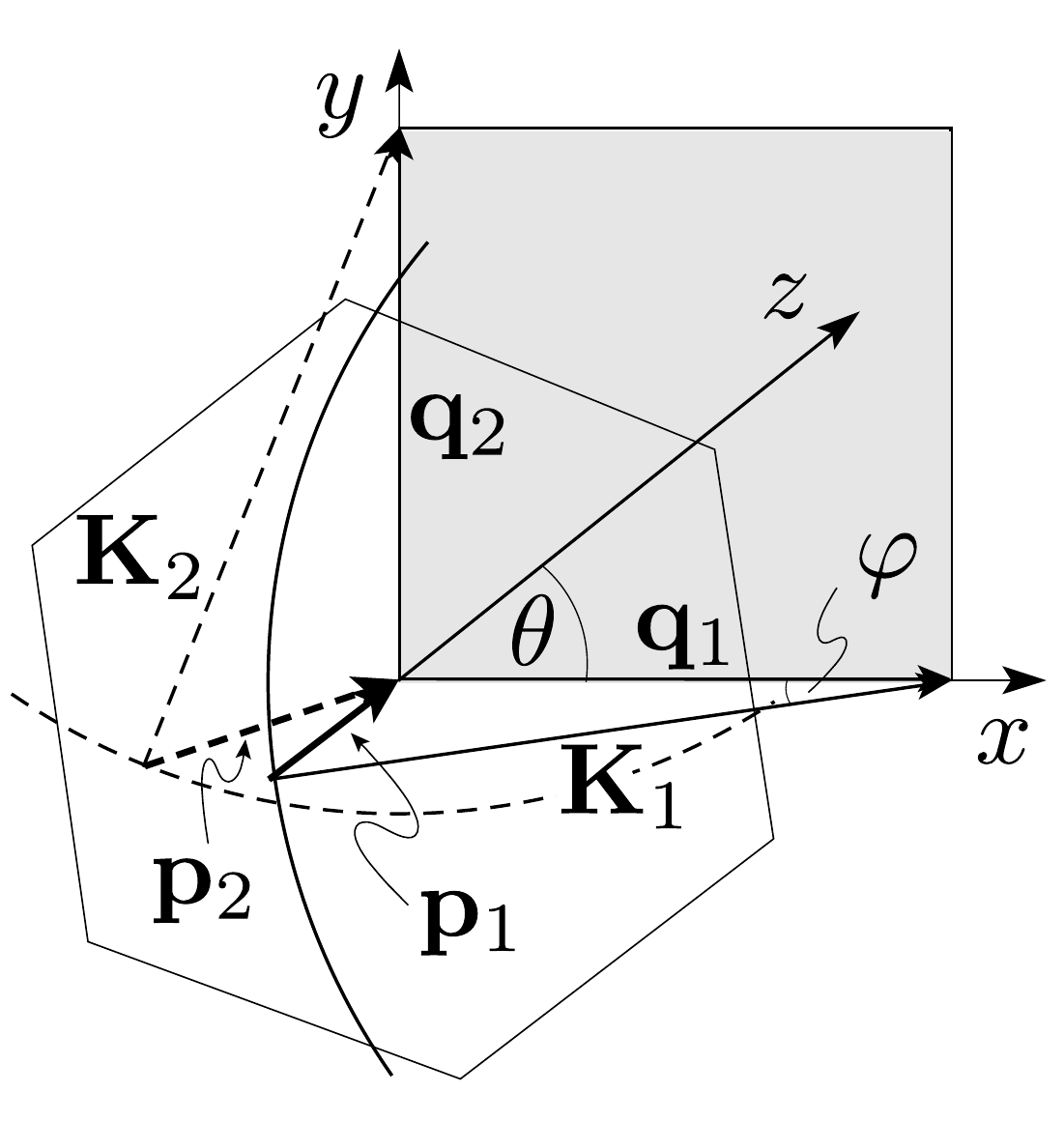}
\end{center}
\caption{\label{fig:BZ}
   Sketch of the reduction of the substrate's ${\bf q}_1$- and ${\bf
   q}_2$-vectors back to the first Brillouin zone of the (rotated hexagonal)
   particle lattice.  The back-folded ${\bf q}$-vectors ${\bf p}_1$ and ${\bf
   p}_2$ assume values on circular segments; these segments derive from
   different sectors of circles of radii $K_1$ and $K_2$ around ${\bf q}_1$
   and ${\bf q}_2$ which emerge when rotating the particle lattice against the
   fixed substrate (angle $\varphi$). A small value of ${\bf p}_1$ or ${\bf
   p}_2$ provides a large relaxation energy. For later use, the $z$-axis
   pointing along ${\bf p}_1$ and enclosing an angle $\theta$ with the
   $x$-axis is also shown.}
\end{figure}
Expanding the substrate potential to linear order in the
displacement \cite{McTague_79}, we obtain the contribution
\begin{align}
\label{eq:subrel}
   \delta e^\mathrm{sub}_{\scriptscriptstyle \triangle}
   &\approx \frac{1}{N}\sum_i {\bf u}_i\cdot{\bf f}^\mathrm{sub}_{i} 
   \quad \textrm{with}\\ 
   \label{eq:fsub}
   {\bf f}^\mathrm{sub}_{i}
   &=\frac{V}{2}\sum_{\alpha}{\bf q}_\alpha
   \sin[{\bf q}_\alpha\cdot{\bf R}^{\scriptscriptstyle\triangle}_i]
\end{align}
to the system's free energy correction $\delta g_{\scriptscriptstyle
\triangle}$.  The minimization of the Gibbs free energy $\delta
g_{\scriptscriptstyle \triangle} = \delta g^\mathrm{int}_{\scriptscriptstyle
\triangle} +\delta e^\mathrm{sub}_{\scriptscriptstyle \triangle}$ with respect
to the displacement field ${\bf u}_i$ is conveniently done in Fourier
space\cite{remark_FT} and we obtain the solution
\begin{equation} \label{eq:flinfourier}
   {\bf u}({\bf k}) =-[\hat\Phi^{\scriptscriptstyle D}]^{-1}({\bf k})\, 
      {\bf f}^\mathrm{sub}({\bf k}),
\end{equation}
with $\hat\Phi^{\scriptscriptstyle D}({\bf k})$ the Fourier transform of the
elastic matrix $\hat\Phi^{\scriptscriptstyle D}({\bf
R}_{ij}^{\scriptscriptstyle \triangle})$ and ${\bf k}$ belonging to the first
Brillouin zone of the ($\varphi$-rotated) hexagonal lattice.  The force field
\cite{remark_Kron} ${\bf f}^{\textrm{sub}}({\bf{k}}) = (VN/4i) \sum_\alpha
(\delta_{{\bf k},-{\bf p}_\alpha} - \delta_{{\bf k},{\bf p}_\alpha})\,{\bf
q}_\alpha$ involves the two modes ${\bf q}_\alpha$, $\alpha = 1,2$, of the
substrate potential, folded back to the first Brillouin cell of the particle
lattice, see Fig.\ \ref{fig:BZ}, ${\bf q}_\alpha - n_\alpha {\bf K}_1 -
m_\alpha {\bf K}_2 \equiv -{\bf p}_\alpha$,  with ${\bf K}_1$, ${\bf K}_2$ the
reciprocal lattice vectors of the (rotated hexagonal) particle lattice,
$n_\alpha,~m_\alpha$ are appropriate integers, and we have included a minus
sign in the definition of ${\bf p}_\alpha$ for convenience.

\begin{figure}[t]
\begin{center}
\includegraphics[width=7cm]{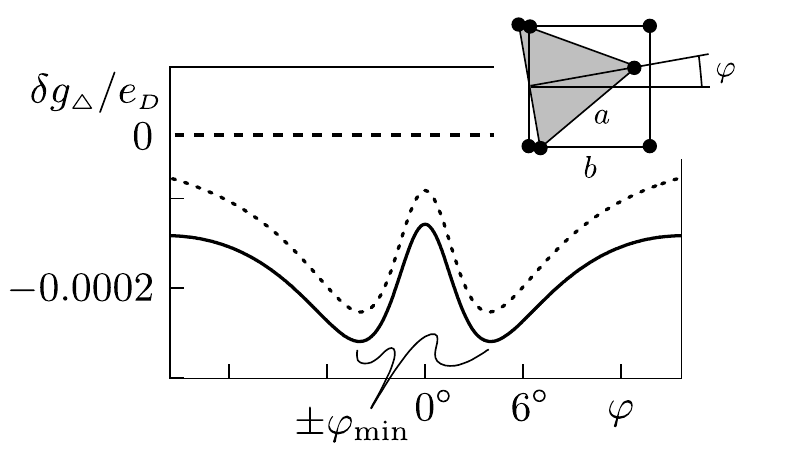}
\end{center}
\caption{\label{fig:angledep}
Lowering of the free energy due to particle relaxation as a function of
relative orientation $\varphi$ between the particle lattice and the substrate
for a small substrate amplitude $V=0.01 \, e_{\scriptscriptstyle D}$.  The
maximal energy gain is reached at $\varphi_0 \approx \pm 3.83^\circ$ and leads
to an orientational locking of the particle lattice. The dotted line is the
result of the resonance approximation, the dashed line marks the energy
without relaxation.}
\end{figure}

Inserting the solution for the displacement field back into the expression
for the free energy relaxation, we obtain the result
\begin{align} \label{eq:erel}
   \delta g_{\scriptscriptstyle \triangle}(V,\varphi)
   =-\frac{\pi^2}{4} n V^2 
    \bigl\{[{\hat\Phi}^{\scriptscriptstyle D}]^{-1}_{11}({\bf p}_1)
          +[{\hat\Phi}^{\scriptscriptstyle D}]^{-1}_{22}({\bf p}_2)\bigr\},
\end{align}
where the dependence on the angle $\varphi$ is encoded in the misfit vectors
${\bf p}_\alpha$, see Fig.\ \ref{fig:BZ}.  Calculating
$\hat\Phi^{\scriptscriptstyle D}({\bf k})$ with the help of Eq.\
(\ref{eq:dynmat}) and evaluating the energy relaxation $\delta
g_{\scriptscriptstyle \triangle}(V,\varphi)$ as a function of $\varphi$, see
Fig.\ \ref{fig:angledep}, we find the locking angle
\begin{align}\label{eq:phi0}
   \varphi_\mathrm{min} \approx \pm 3.83^\circ
\end{align}
minimizing the free energy of the distorted hexagonal lattice. Corrections
to this result are of order $V^2$ and require to go beyond the harmonic
approximation.

Instead of a numerial minimization of the free energy, one can make use of the
{\it resonance approximation} that includes only the dominant mode in the
substrate potential\cite{PT_pap,PT_book}.  Rotating the hexagonal particle
lattice with respect to the square substrate potential, the misfit vectors
${\bf p}_\alpha$ move on arcs through the Brillouin zone, see Fig.\
\ref{fig:BZ}. For a small misfit parameter $s$, one of the ${\bf p}_\alpha$
passes near zero, inducing a large relaxation (and accordingly a large energy
gain) as the elastic matrix becomes soft with small eigenvalues, see Eq.\
(\ref{eq:LT_modes}).  Within the resonance approximation
\cite{PT_pap,PT_book}, only the dominant term in the relaxation deriving from
the small misfit vector, say ${\bf p}_1 = {\bf K}_1- {\bf q}_1$, is included,
while the small correction due to the other mode is dropped; in the following,
we drop the index 1 on ${\bf q}_1$, ${\bf K}_1$, and ${\bf p}_1$.  A similar
approximation has been used by McTague and Novaco \cite{McTague_79} when
calculating the accommodation of a hexagonal lattice to a substrate with the
same (hexagonal) symmetry but with a different lattice constant.  Adopting
the long-wavelength approximation (\ref{eq:LT_modes}), the expression
(\ref{eq:erel}) for the energy relaxation simplifies considerably,
\begin{align} \label{eq:longwave}
  \delta g_{\scriptscriptstyle \triangle}(V,\varphi)
  &\approx-\frac{V^2}{16}
  \sum_{\lambda}\frac{1}{\phi^\lambda_{\bf p}}
   \bigl[{\bf e}^\lambda_{\bf p}\cdot{\bf q}\bigr]^2,
\end{align}
where ${\bf e}^\parallel_{\bf p} = {\bf p}/p$ and ${\bf e}^\perp_{\bf p} =
{\bf p}^\perp/p$.  With ${\bf e}^\parallel_{\bf p}\cdot {\bf q} = q\cos\theta$
and ${\bf e}^\perp_{\bf p}\cdot {\bf q} = q\sin\theta$, see Fig.\
\ref{fig:BZ}, and using the law of sines $K/\sin\theta=p/\sin\varphi$, we
arrive at the simple result
\begin{equation} \label{eq:longwave_res}
  \delta g_{\scriptscriptstyle \triangle}(V,\varphi)
   =-\frac{nV^2}{16(\kappa+\mu)} \frac{q^2}{p^2}
   \biggl[1+\frac{\kappa}{\mu}
   \frac{K^2}{p^2}\sin^2{\varphi}\biggr].
\end{equation}
The first term favors a minimal modulus $p$ at $\varphi = 0$, while the second
term favors a finite angle $\varphi$. Replacing $p^2 = q^2 + K^2
-2qK\cos\varphi$ and defining $r = K/q = 1+s$, this can be rewritten as
\begin{align}\label{eq:angledep}
  \delta g_{\scriptscriptstyle \triangle}(V,\varphi)
   &= -\frac{nV^2}{16(\kappa+\mu)}\biggl[
   \frac{1}{1+r^2-2r\cos\varphi}\\ \nonumber
   &\qquad\qquad + \frac{\kappa}{\mu}
   \Bigl(\frac{r\sin\varphi}{1+r^2-2r\cos\varphi}\Bigr)^2
   \biggr]
\end{align}
and the minimization of this expression with respect to $\varphi$ provides us
with the optimal angle $\varphi_\mathrm{min}$ given through
\begin{equation} \label{eq:cangle}
  \cos{\varphi_\mathrm{min}} = 1-s^2\,\frac{r-\mu/\kappa}{r(1+r^2+
  2\mu/\kappa)}.
\end{equation}
Expanding this result for small $\varphi_\mathrm{min}$ and small $s$ we obtain
the final answer
\begin{equation} \label{eq:angle_phi}
  \varphi_\mathrm{min} = s \sqrt{\nu} \approx 3.86^\circ
\end{equation}
with $\nu = (\kappa-\mu)/(\kappa+\mu)$ the Poisson ratio and we have made use
of the elastic constants in Eq.\ (\ref{eq:mu_ka}). Within the same accuracy
(i.e., to leading order in $s$), we find the misfit vector 
\begin{equation} \label{eq:mis_p}
  {\bf p} = sq \binom{1}{\sqrt{\nu}}
\end{equation}
enclosing an angle 
\begin{equation} \label{eq:angle_theta}
   \theta = \arctan\sqrt{\nu} \approx 42.13^\circ
\end{equation}
with the $x$-axis, see Fig.\ \ref{fig:BZ}. The displacement field
${\bf u}$ evolves periodically along ${\bf p}$ (or $z$, see Fig.\
\ref{fig:BZ})
\begin{equation}\label{eq:u_sin}
   {\bf u}({\bf R}) = \frac{b}{8\pi s^2} \frac{nV}{\mu \sqrt{1+\nu} }\,\hat{\bf p}_s
   \sin({\bf p}\cdot{\bf R}) ,
\end{equation}
where $\hat{\bf p}_s = (1+\nu)^{-1/2}(1, -\sqrt{\nu})$ is the vector $\hat{\bf
p}$ mirror reflected about the $x$-axis. With $\hat{\bf p}$ close to the
diagonal, the displacement field is predominantly shear-type (and a perfect
shear distortion in the incompressible limit $\kappa \to \infty$).  Finally,
the displacement (\ref{eq:u_sin}) relaxes the energy of the hexagonal lattice
to
\begin{equation}\label{eq:dgpt}
   g_\mathrm{dh} (V) = g_{\scriptscriptstyle \triangle} (V)
   - \frac{n V^2}{64 s^2 \mu} (1+\mu/\kappa).
\end{equation}
Note that the displacement ${\bf u}$ diverges $\propto s^{-2}$ on approaching
the density $n = \sqrt{3}/2 b^2$ where $h = b$ and $s = 0$ and our
approximation breaks down. Limiting the displacement $u$ to a fraction $c \sim
0.1$ of the lattice constant $a$ then restricts the validity of our analysis
to potentials $V < 8 \pi \sqrt{1+\nu}\, c (\mu/n) s^2$.  Higher order (in $V$)
corrections are of order $V^4$ in the energy relaxation $g_\mathrm{dh}$ and of
order $V^2$ in the angle $\varphi_\mathrm{min}$. Rather than studying such
corrections in $V$, we proceed with the analysis of the full non-linearity in
the force field which takes us to a non-uniform soliton phase. The precision
of this calculation then is limited by our use of the harmonic approximation
(to be improved later with a numerical analysis) and the resonance
approximation (to be abandoned when including the second mode of the substrate
potential in Sec.\ \ref{sec:sol_dw}).

\section{Soliton Phase in the Resonance Approximation}\label{sec:sol_ra}

With increasing $V$, the periodic shear-type displacement (\ref{eq:u_sin})
evolving along the misfit vector ${\bf p}$ becomes large, of order $b$, and
turns into a soliton array as first described by Pokrovsky and Talapov
\cite{PT_pap,PT_book} within the resonance approximation discussed above.  For
completeness, we will briefly sketch their analysis and present the main
results here. We describe the change in the interaction energy within the
harmonic approximation, adopting the long wave-length approximation,
introduced in Sec.\ \ref{sec:per-doub} and used in Sec.\ \ref{sec:trian},
in a continuum elastic formulation [${\bf R}_i^{\scriptscriptstyle \triangle}
\to {\bf R}$, ${\bf u}_i \to {\bf u}({\bf R}$), see also appendix
\ref{app:eltr}],
\begin{align}\label{eq:dg_int_hex}
   \delta g^\mathrm{int}_{\scriptscriptstyle \triangle}
   &= \frac{1}{N}\int_A\!d^2 R\,\Bigl[
   \frac{\kappa}{2}(\partial_x u_x+\partial_y u_y)^2\\ \nonumber
   &\qquad+\frac{\mu}{2}\bigl((\partial_x u_x-\partial_y u_y)^2
   +(\partial_y u_x+\partial_x u_y)^2\bigr)\Bigr].
\end{align}
The drive in the substrate potential derives from the misfit between the
lattice positions ${\bf R}_i^{\scriptscriptstyle \triangle}$ and the ${\bf
q}$-vector; this can be made more explicit by the transformation ${\bf q}\cdot
{\bf R}_i^{\scriptscriptstyle \triangle} \to ({\bf K}-{\bf p})\cdot {\bf
R}_i^{\scriptscriptstyle \triangle} = - {\bf p} \cdot {\bf
R}_i^{\scriptscriptstyle \triangle} + 2\pi \mathbb{Z}$. Within the continuum
approximation, the substrate potential contributes with a term (we use ${\bf
q}\cdot ({\bf R}+{\bf u}) = -{\bf p}\cdot {\bf R} + {\bf q}\cdot{\bf u}$)
\begin{equation}\label{eq:dg_sub}
   e^\mathrm{sub} = \frac{1}{N}\int_A\!d^2 R\, \frac{nV}{2}
   \Bigl[2-\cos\bigl({\bf p}\cdot {\bf R}-{\bf q}\cdot{\bf u}
   \bigr)\Bigr],
\end{equation}
where we account for the additional average energy $V/2$ of the second mode.
The task then is to minimize the total free energy $g(V) =
g_{\scriptscriptstyle \triangle} (V) + \delta
g^\mathrm{int}_{\scriptscriptstyle \triangle} + e^\mathrm{sub}$. For small
amplitudes $V$ this is achieved by the period modulation ${\bf u}({\bf R})$ in
(\ref{eq:u_sin}) of the $\varphi$-rotated hexagonal lattice. At large values
of $V$, the lowest energy will be assumed by a rhombic or isosceles triangular
lattice with height $b$ (along $x$) and base $b'$ (along $y$), the so-called
$bb'$ rhombic lattice (within the resonance approximation, we account only for
the leading substrate mode that we choose along $x$). In order to find this
lattice, we minimize the free energy $g(V)$ at large $V$ with respect to a
global displacement field
\begin{equation}\label{eq:u_g_def}
     {\bf u}_\mathrm{g}({\bf R})= \binom{{\bf w}\cdot{\bf R}}
     {{\bf t}\cdot{\bf R}},
\end{equation}
parametrized by the vectors ${\bf w}=(w_1,w_2)$ and ${\bf t}=(t_1,t_2)$.
Minimizing $\delta g^\mathrm{int}_{\scriptscriptstyle \triangle}[{\bf
u}_\mathrm{g}] = (\kappa/2n) (w_1+t_2)^2 + (\mu/2n)(w_2+t_1)^2$ with respect
to ${\bf t}$ at fixed ${\bf w}$, we find that $t_1 = -w_2$ and $t_2 = - \nu
w_1$, resulting in a displacement
\begin{equation}\label{eq:u_s_r}
   {\bf u}_\mathrm{g}({\bf R})=w_1\binom{x}{-\nu y}-w_2\binom{-y}{x}
                     = {\bf u}_\mathrm{d}+ {\bf u}_\mathrm{r}
\end{equation}
that combines a shear displacement ${\bf u}_\mathrm{d}$ (a stretching by $w_1$
along $x$ and a compression by $w_1\nu$ along $y$) and a rotation ${\bf
u}_\mathrm{r}$ (by the angle $-w_2$; note that we cannot go beyond the
linearized rotation ${\bf u}_\mathrm{r}$ as higher order terms are beyond our
accuracy and generate unphysical terms). At large $V$, the potential minima
lock the particles into a rhombic lattice with height $b$ along $x$, hence
$(1+w_1)h = b$ and $w_1 = s$; at the same time, the rotation has to align the
particle lattice back to the substrate, hence $w_2 = \varphi = s \sqrt{\nu}$
(here we drop the index and rename $\varphi = \varphi_\mathrm{min}$), hence
${\bf w} = s(1,\sqrt{\nu})$. The global displacement field \eqref{eq:u_s_r}
then can be written in the form [we define the coordinate $z= (x+\sqrt{\nu}
y)/\sqrt{1 + \nu}$]
\begin{equation}\label{eq:u_g_f}
   {\bf u}_\mathrm{g} = s \sqrt{1+\nu} \, z \binom{1}{-\sqrt{\nu}}
\end{equation}
and generates the new $bb'$ rhombic lattice out of the hexagonal one. The
lattice constant $b'$ along the $y$-axis assumes a value intermediate between
$b$ and $a$,
\begin{equation}\label{eq:b'}
   b' = a(1-\nu s) \approx 1.0090 \, b > b.
\end{equation}
The elastic energy required to generate this distortion is
\begin{equation}\label{eq:g_bb'}
   \delta g^\mathrm{int}_{\scriptscriptstyle \triangle}
   = g_{\scriptscriptstyle \rhd'} - g_{\scriptscriptstyle\triangle}
   \approx \frac{\kappa}{\kappa+\mu}\frac{2\mu}{n}s^2 = 0.0169 \,
   e_{\scriptscriptstyle D}.
\end{equation}
Note that this deformation involves a change in density or area $\delta A/A =
\nabla\cdot {\bf u}_\mathrm{d} = s(1-\nu) = 0.0136$.  A more accurate result
is obtained by minimizing the free energy $g_{\scriptscriptstyle \rhd'} (b') =
e_{\scriptscriptstyle \rhd'}^\mathrm{int}(b') + p/n'$ with respect to $b'$,
fixing the height of the rhombic lattice to $b$; here, $e_{\scriptscriptstyle
\rhd'}^\mathrm{int}$ is the true interaction energy in Eq.\ (\ref{eq:E}) (to
be calculated with the Ewald technique) and $n' = 1/bb'$. The result of such a
calculation provides the base length $b' \approx 1.0173\, b$ and
$g_{\scriptscriptstyle \rhd'} - g_{\scriptscriptstyle \triangle} = 0.0179 \,
e_{\scriptscriptstyle D}$. The relative difference $(b-b')/(a-b) \approx 0.11$
is quite large, of the order of 10 \%, indicating that the result of the
elastic theory is not very accurate.

\begin{figure}[h]
\begin{center}
\includegraphics[width=5.5cm]{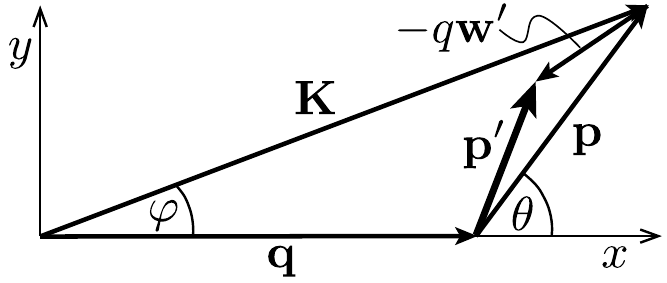}
\end{center}
\caption{\label{fig:pprime} 
   The effective mismatch ${\bf p}'$ is a combination of the true mismatch
   ${\bf p}$ and the correction $-q{\bf w}'$ due to the global displacement field
   ${\bf u}_\mathrm{g}'$.}
\end{figure}

Next, we determine the non-uniform soliton phase that interpolates between
the rotated distorted hexagonal lattice at small substrate potential $V$ and
the $bb'$ rhombic lattice at large $V$. We adopt an Ansatz ${\bf u}={\bf
u}_\mathrm{g}' + \tilde{\bf u}$ for the displacement field involving a
periodic modulation $\tilde{\bf u}$ on top of a global displacement ${\bf
u}_\mathrm{g}'$ (parametrized by ${\bf w}'$ and ${\bf t}'$).  The parameters
${\bf w}',{\bf t}'$ now depend on the amplitude $V$ of the substrate potential
with ${\bf w}'=0$ at $V=0$ and ${\bf w}' = {\bf w}$ at large $V$.  Inserting
this Ansatz into the free energy $g = g_{\scriptscriptstyle \triangle} +
\delta g_{\scriptscriptstyle \triangle} + e^\mathrm{sub}$, we first minimize
with respect to ${\bf t}'$ to find that ${\bf t}' = -(w_2',\nu w_1')$. The
free energy per particle then assumes the form
\begin{align} \label{eq:g_tr_sol}
   g =&\, g_\mathrm{bg}(V)
   + \frac{1}{N}\int_A\!d^2 R\,\Bigl\{\frac{\kappa}{2}(\partial_x \tilde{u}_x
   +\partial_y \tilde{u}_y)^2
   \\
   &\qquad\qquad+\frac{\mu}{2}\Bigl[(\partial_x \tilde{u}_x-\partial_y \tilde{u}_y)^2
   +(\partial_y \tilde{u}_x+\partial_x \tilde{u}_y)^2\Bigr]\nonumber\\
   \nonumber
   &\qquad\qquad\qquad+\frac{nV}{2}\Bigl[1-\cos\bigl({\bf p}'\cdot
   {\bf R}-q \tilde{u}_x\bigr)\Bigr] \Bigr\}.
 \end{align}
with $g_\mathrm{bg}(V) = g_{\scriptscriptstyle \triangle} + [2\mu
\kappa/n(\kappa+\mu)](s-s')^2 +V/2$ the energy of the homogeneous background
and
\begin{align} \label{eq:pw}
  {\bf p}'= {\bf p}-q{\bf w}'\equiv q \binom{s'}{\varphi'}.
\end{align}
The further minimization of (\ref{eq:g_tr_sol}) with respect to the periodic
displacement $\tilde{\bf u}$ and the effective misfit vector ${\bf p}'(V)$,
see Fig.\ \ref{fig:pprime}, will provide us with the geometry of the
non-uniform soliton phase.  The misfit vector ${\bf p}'$ starts out with
${\bf p} = qs (1,\sqrt{\nu})$ at $V=0$, see Eq.\ (\ref{eq:mis_p}), and
vanishes in the $bb'$ rhombic phase at large $V$ where ${\bf w}' = {\bf w} =
s(1,\sqrt{\nu})$; the parameters $s'$ and $\varphi'$ describe the evolution of
the global displacement as a function of $V$.

Minimizing (\ref{eq:g_tr_sol}) with respect to the periodic displacement field
$\tilde{\bf u}$, we obtain the set of differential equations
\begin{align}
   \kappa(\partial_x^2\tilde{u}_x+\partial_x\partial_y\tilde{u}_y)
   +\mu\Delta\tilde{u}_x
   &=\frac{-nV}{2}q \sin\bigl({\bf p}'\!\cdot\! {\bf R}\!-\!q\tilde{u}_x\bigr),
   \nonumber\\
   \kappa(\partial_x\partial_y\tilde{u}_x\!+\!\partial_y^2\tilde{u}_y)
   +\mu\Delta\tilde{u}_y&=0.\label{eq:ecu}
\end{align}

These equations admit a uniaxial solution $\tilde{\bf u}(x,y)=\tilde{\bf
u}(z')$ along $z'=x\cos{\theta'}+y\sin{\theta'}$ with the direction of $z'$
determined by the effective mismatch vector ${\bf p}' = p' (\cos\theta',
\sin\theta')$ and the boundary condition $\tilde{\bf u}(z') = \tilde{\bf u}
(z'+L)$ with $L = 2\pi/p'$. The second equation relates the two components of
the displacement field via
\begin{align} \label{eq:eta}
   \tilde{u}_y=-\eta(\theta')\,\tilde{u}_x\quad \textrm{with}\quad
   \eta=\frac{\kappa\sin{\theta'}
   \cos{\theta'}}{\kappa\sin^2{\theta'}+\mu},
\end{align}
where we have used the boundary condition $\tilde{u}_x(0) = \tilde{u}_x(L) =0$
and the same for $\tilde{u}_y$. Assuming $\kappa\gg\mu$, $\eta$ increases with
$\theta'$ from zero, $\eta\approx (\kappa/\mu)\theta'$, goes through a maximum
$\eta\approx\sqrt{\kappa/\mu}/2$ at $\theta' \approx\sqrt{\mu/\kappa}$,
decreases as $\eta \approx \cot\theta'$, and approaches zero at $\pi/2$
linearly as $\eta\approx (1-\mu/\kappa)(\pi/2-\theta')$; while for angles
close to 0 and $\pi/2$, the $y$-component $\tilde{u}_y$ is very small, it
increases to about 1.5 times the $x$-component $\tilde{u}_x$ for
$\kappa/\mu=10$.

Inserting the result (\ref{eq:eta}) back into the first equation of
(\ref{eq:ecu}) and transforming variables $\tilde{z}={\bf p}'\cdot {\bf
R}=p'z'$, $u= \tilde{z}-q\tilde{u}_x$, we obtain the Sine-Gordon equation
\begin{align} \label{eq:SG}
     \tilde{\alpha} \,\partial_{\tilde{z}}^2u=\sin{u},
\end{align}
with boundary conditions $u(0)=0$ and $u(2\pi)=2\pi$ and 
\begin{align}
   \tilde{\alpha}&=\frac{2\mu}{Vn}\frac{\kappa+\mu}{\kappa\sin^2{\theta'}+\mu}
   \Bigl(\frac{p'}{q}\Bigr)^2.\label{eq:alpha}
\end{align}
With the total displacement $q u_x = ({\bf p}\cdot{\bf R} -\tilde{z}) +
q\tilde{u}_x$, the displacement $u = \tilde{z}-q\tilde{u}_x$ contributes both
to the global and periodic parts of $u_x$. Indeed, $u$ has a stair-case shape,
while the periodic function $q \tilde{u}_x$ has a saw-tooth form. 

In the limit $p' \to 0$ we have $L \to \infty$ and it is convenient to
rewrite (\ref{eq:SG}) in the form $\alpha \,\partial_{z'}^2u=\sin{u}$ with
$\alpha = \tilde{\alpha}/{p'}^2$ and boundary conditions $u(-\infty) = 0$ and
$u(\infty) = 2\pi$. The single-soliton solution then is given by
\begin{equation}\label{eq:SG_sol}
   u(z') = 4 \arctan[\exp(z'/\sqrt{\alpha})],
\end{equation}
with a core region of width $\sqrt{\alpha} \approx (b/2\pi \sin\theta')
\sqrt{2\mu/Vn}$, where we have dropped the correction from the shear modulus
$\mu$ in $\alpha$. Within this core region, $\tilde{u}_x = -u/q$ quickly goes
from 0 to $-b$.

At finite $p'$, the solution is given by a soliton array with period $L=
2\pi/p'$ as obtained by integrating the `velocity' $\partial_{\tilde z}u =
[2(\tilde\alpha_0 - \cos u)/\tilde\alpha]^{1/2}$, with the integration
constant $\tilde\alpha_0$ (the minimal slope between subsequent solitons)
given by the implicit equation ($K$ is the complete elliptic integral of the
first kind\cite{abramowitz_72})
\begin{equation} \label{eq:alpha0}
   (\pi/2)\sqrt{2(1+\tilde{\alpha}_0)/\tilde\alpha}
   = K\bigl[\sqrt{2/(1+\tilde{\alpha}_0)}\bigr]
 \end{equation}
and assuming asymptotic values $\tilde\alpha_0 (\tilde\alpha \to 0) \to 1$
(single sharp soliton at large $V$ or small misfit $p'$) and $\tilde\alpha_0
(\tilde\alpha \to \infty) \to \tilde\alpha/2$ (smoothly modulated and steadily
increasing solution $u \approx \tilde{z}$ at small $V$ as solitons strongly
overlap). 

In order to find the parameter $p'$ (the soliton density $1/L = p'/2\pi$) and
the angle $\theta'$ of the soliton array, we have to minimize the energy
(\ref{eq:g_tr_sol}) of the solitonic solution. After the reduction to a
one-dimensional Sine-Gordon problem, we find the expression
 \begin{align}
  \label{eq:g_sol}
  &g - g_\mathrm{bg} =
   \frac{V}{2}\int_0^{2\pi}\!\frac{d\tilde{z}}{2\pi}\,
  \Bigl[\frac{\tilde{\alpha}}{2}(\partial_{\tilde{z}} u-1)^2+1-\cos{u}\Bigr]
  \\ \nonumber
  &~~= \frac{V}{2}(1\!-\!\tilde\alpha_0\!-\!\tilde\alpha/2)
        \!+\! \frac{V}{\pi}\sqrt{2\tilde{\alpha}(1\!+\!\tilde{\alpha}_0)}
          E\bigl[\sqrt{2/(1\!+\!\tilde{\alpha}_0)}\bigr]
 \end{align}
with $E$ the complete elliptic integral of the second kind, see Ref.\
\onlinecite{abramowitz_72}.  The energy Eq.\ \eqref{eq:g_sol} grows
monotonically with $\tilde{\alpha}$, starting from 0 at $\tilde{\alpha}=0$
(large $V$) and saturating at $V/2$ as $\tilde{\alpha} \to \infty$ (small $V$).
Expressing $\tilde\alpha$ through $s'$ and $\varphi'$, see Eqs.\ (\ref{eq:pw})
and (\ref{eq:alpha}), we minimize $\tilde\alpha$ with respect to $\varphi'$
(note that $\varphi'$ only enters the soliton energy, while $s'$ also appears
in $g_\mathrm{bg}$) and obtain the minimal value
\begin{align} \label{eq:alphamin}
   \tilde{\alpha}=\frac{2\, \mu}{Vn} \frac{4\kappa}{\kappa+\mu}\, s'^2
   \quad \textrm{at} \quad \varphi'=\sqrt{\nu}s'
\end{align}
and hence the direction of the (effective) misfit ${\bf p}\,'$ coincides with
that of ${\bf p}$ in Eq.\ (\ref{eq:mis_p}),
 \begin{align} \label{eq:optdir}
   {\bf p}'&=q s'\binom{1}{\sqrt{\nu}} \quad\textrm{and}\quad
   \theta' = \theta = \arctan\sqrt{\nu},
 \end{align}
i.e., the dense soliton array smoothly appears out of the perturbative
displacement modulation \eqref{eq:u_sin} of the locked phase found in Sec.\
\ref{sec:trian}.  The soliton density is given by $1/L = s'\sqrt{1+\nu}/b$ and
the global displacement field in Eq.\ (\ref{eq:u_s_r}) which takes the
hexagonal phase smoothly into the $bb'$ rhombic lattice reads
\begin{align} \label{eq:ug3}
   {\bf u}_\mathrm{g}'({\bf R})&=
   (s-s')\sqrt{1+\nu}\, z\, \binom{1}{-\sqrt{\nu}},
\end{align}
where we have used that ${\bf w}' =(s-s')(1,\sqrt{\nu})$ and $\sqrt{1+\nu}\, z
= x + \sqrt{\nu}\, y$, see Eq.\ \eqref{eq:optdir}. At small $V$, $s'= s$, the
density of solitons is high, ${\bf u}_\mathrm{g}' = 0$, and the lattice is
close to the hexagonal one. For a large substrate potential $V$, $s' = 0$, the
density of solitons vanishes, ${\bf u}_\mathrm{g}' = {\bf u}_\mathrm{g}$, and
the particles are arranged in the $bb'$ rhombic lattice. An alternative---and
actually the conventional---view is to start from the $bb'$ rhombic lattice at
large $V$, the commensurate phase, and then have solitons deform the lattice
until the dense soliton array describes the hexagonal phase. The shape of the
individual soliton (along $x$) is given by Eq.\ \eqref{eq:SG_sol} and making
use of the result \eqref{eq:optdir} for the angle $\theta$ and Eq.\
(\ref{eq:eta}), we find the ratio $\eta = \sqrt{\nu}$, i.e., the displacement
field of one soliton is ${\bf d}^{\rm\scriptscriptstyle PT} = b
(-1,\sqrt{\nu}) \approx b(1, -0.905)$.  As the solitons become denser with
decreasing $V$, their shift vectors add up to produce the global displacement
field ${\bf u}_\mathrm{g}' - {\bf u}_\mathrm{g}$ on top of the rhombic lattice
until the latter has transformed into the hexagonal lattice at vanishing $V$
(where ${\bf u}_\mathrm{g}'=0$). The periodic part $\tilde{\bf u}$ of the
displacement field coincides with the result (\ref{eq:u_sin}) of the
perturbative analysis at small $V$ and turns into a saw-tooth shape with sharp
shifts $\sim b(-1,\sqrt{\nu})$ in the core regions and a small slope
$\tilde{\bf u} \approx ({\bf p}'\cdot {\bf R}/q) (1,-\sqrt{\nu})$ in between
two solitons.

It remains to calculate the critical substrate potential $V_c^{\rm
\scriptscriptstyle PT}$ for the first soliton entry on decreasing $V$ and the
dependence $s'(V)$ determining the density $1/L$ of solitons. This last step
involves the minimization of $g(s';V)$ with respect to $s'$ at fixed $V$. 

At small $V$, where $\tilde\alpha$ is large, we set $\tilde{\alpha}_0 \approx
\tilde{\alpha}/2$ and expand the energy \eqref{eq:g_sol} to order
$1/\tilde{\alpha}$,
\begin{align} \label{eq:g_Vsmall}
   g &\approx g_{\scriptscriptstyle\triangle} \!\!+\! V  
   \!+\!\frac{\kappa}{\kappa+\mu}\frac{2\mu}{n}\;(s-s')^2
   \!-\!\frac{n V^2}{64{s'}^2\mu}(1\!+\!\mu/\kappa).
\end{align}
The optimal $s'$ then satisfies the equation $\partial_{s'}g=0$, i.e.,
\begin{align} \label{eq:sp_Vsmall}
  (s-s')\approx \frac{n^2V^2(1+\mu/\kappa)^2}{128\mu^2 \, {s'}^3}
\end{align}
and we find that $(s-s') \propto V^2$. Hence, we can set $s=s'$ in Eq.\
\eqref{eq:g_Vsmall} and the free energy assumes the form
\begin{align}
\label{eq:gPT}
  g \approx g_{\scriptscriptstyle\triangle}(V) 
  -\frac{n V^2}{64 s^2 \mu }(1+\mu/\kappa)
\end{align}
in agreement with the Eq.\ \eqref{eq:dgpt}. Furthermore, the angle $\varphi'
\approx \varphi$ up to corrections of order $V^2$.

At large values of $V$, we can approximate the complete elliptic
integrals\cite{remark:KE} $K$ and $E$ to arrive at the free energy in the form
\begin{align} \label{eq:g_Vlarge}
   g\approx &\, g_{\scriptscriptstyle\triangle} +\frac{V}{2}
    +\frac{\gamma}{2}s^2
    +\bigl(\epsilon-\gamma s\bigr)s'+4\epsilon\, s'\,e^{-4V/\epsilon s'},
\end{align}
where we have defined the elastic and soliton energies
\begin{equation}\label{eq:singlesol}
   \gamma =\frac{\kappa}{\kappa+\mu} \frac{4 \mu}{n},
   \quad
    \epsilon=\frac{4}{\pi}\sqrt{\frac{\kappa}{\kappa+\mu}
    \frac{V}{2} \frac{4 \mu}{n}}.
\end{equation}
The first three terms of Eq.\ \eqref{eq:g_Vlarge} represent the energy
$g_{\scriptscriptstyle \rhd'}$ of the $bb'$ rhombic structure.  The term
$(\epsilon-\gamma s)s'$ turns negative when the soliton energy $\epsilon$ is
balanced against the drive (or chemical potential for solitons) $\gamma s$.
Finally, the last term describes the exponential interaction between solitons
and stabilizes $s'$ at a finite value, i.e., a finite soliton density.  The
transition from the $bb'$ rhombic phase to the non-uniform soliton phase
then takes place when $\epsilon= \gamma s$, corresponding to the critical
substrate strength
\begin{align} \label{eq:V0crit}
   V_c^{\rm\scriptscriptstyle PT} =\frac{\pi^2}{2} \frac{\kappa}{\kappa+\mu} 
   \frac{\mu}{n} \, s^2.
\end{align}
For particles interacting via a $1/r^3$-potential, the compression and shear
moduli fulfill the relation $\kappa=10\,\mu$, see Eq.\ \eqref{eq:mu_ka}, such
that at commensurate density one finds the critical substrate
amplitude\cite{mistake}
\begin{align} \label{eq:VcPThex}
   V_c^{\rm\scriptscriptstyle PT} 
   = \frac{5 \pi^2}{11}\frac{\mu}{n}\,s^2 =0.0416\, e_{\scriptscriptstyle D}
   \quad \textrm{at}\quad \theta = 42.13^\circ.
\end{align}
We find the effective misfit parameter $s'(V)$ by minimizing the free energy
(\ref{eq:g_Vlarge}), $\partial_{s'} g = 0$, providing us with the relation
\begin{align} \label{eq:solitongrow}
   (\epsilon-\gamma s)+4\epsilon\, e^{-4V/\epsilon s'}
   \bigl(1+{4V}/{\epsilon s'}\bigr)=0.
\end{align}
The last factor is dominated by the term $4V/\epsilon s'$. Close 
to $V_c^{\rm\scriptscriptstyle PT}$, we write $V=V_c^{\rm\scriptscriptstyle PT}(1-\delta)$
with $0<\delta\ll 1$ and find
\begin{align} \label{eq:solitongrow4}
  s'\approx-\frac{4V_c^{\rm\scriptscriptstyle PT}}{\gamma s}\frac{1}{\log(\delta/8)}
  =\frac{\pi^2}{2|\log[(1-V/V_c^{\rm\scriptscriptstyle PT})/8]|}\,s,
\end{align}
where we have used that $\epsilon_\mathrm{c}=\gamma s$ in the last step.
Combining Eqs.\ (\ref{eq:g_Vlarge}) and (\ref{eq:solitongrow4}), we obtain the
free energy near the transition
\begin{align} \label{eq:edimV0large4}
   g =g_{\scriptscriptstyle \triangle}
     \! +\! \frac{V}{2}\!+\!\frac{\kappa}{\kappa+\mu} \frac{2\mu}{n}\,s^2
      \!-\! \frac{2(V_c^{\rm\scriptscriptstyle PT}\!-\! V)}
           {\log{\bigl[8V_c^{\rm\scriptscriptstyle PT}\!/
                (V_c^{\rm\scriptscriptstyle PT}\!-\!V)\bigr]}}.
\end{align}
Hence, we find that decreasing $V$ below $V_c^{\rm\scriptscriptstyle PT}$, the
particle system is rapidly flooded with solitons, $n_\mathrm{sol} \propto 1/
|\log[(1-V/V_c^{\rm\scriptscriptstyle PT})]|$, similar to the rapid entry of
flux lines in a type II superconductor when the field $H$ is increased above
the lower critical fields $H_{c1}$. This result, is changed to an algebraic
behavior $n_\mathrm{sol} \propto \sqrt{1-V/V_c^{\rm\scriptscriptstyle PT}}$
for $V$ very close to $V_c^{\rm\scriptscriptstyle PT}$, a consequence of the
long-range interaction $\propto 1/R^3$ between particles. The latter generates
an algebraic repulsion $\propto (b/L)^2$ between solitons, see Eq.\
(\ref{eq:e_2s_asym}) below, replacing the exponential law $\propto \exp [-\pi
(L/b)\sqrt{Vn/\mu}]$ in Eq.\ (\ref{eq:g_Vlarge}) at large
distances\cite{HaldaneVillain_81}.
%

\section{Solitons with one Substrate Mode}\label{sec:s_sol}

Having understood the appearance and evolution of the non-uniform soliton
phase in the 2D hexagonal particle system with increasing substrate potential
$V$, we now focus on the first appearance of the (PT or Pokrovsky-Talapov)
soliton when decreasing the substrate potential $V$ in the $bb'$ rhombic
phase. Using the elastic theory of the hexagonal lattice, we expect to find
accurate results for the distorted hexagonal phase and the dense vortex array
at small substrate potential $V$.  On the other hand, the first PT soliton
appears out of the commensurate phase at large substrate amplitudes $V$ and
accordingly, we expect more accurate results for $V_c^{\rm\scriptscriptstyle
PT}$ when using the elastic theory for the $bb'$-lattice. Furthermore, the
comparison of the results provided by these different starting points will
tell us about the relevance of anharmonicities and guide us when evaluating
the critical potentials $V_c$ for the first soliton entry in the presence of
both lattice modes, see Sec.\ \ref{sec:sol_dw}. We first find an analytical
result based on elasticity theory and then compare with numerical results
using direct summation of the interaction and substrate potential energies in
(\ref{eq:E}).

\subsection{Continuum elastic approach}\label{ssec:ana_1M}

We define the displacement field ${\bf v}({\bf R})$ with respect to the
rhombic lattice ${\bf R}^{\scriptscriptstyle\rhd'}_{m,n}= (mb,
(n-{m}/{2})b') \equiv {\bf R}$ and assume a uniaxial defect ${\bf v}(z)$
evolving along $z$, shifting the lattice by ${\bf v}(\infty) = (-b,
v_{y,\infty})$ with $v_{y,\infty}$ to be determined (the soliton starts at
${\bf v}(-\infty) = (0,0)$). We then have to minimize the soliton line energy
\begin{align} \label{eq:singlesolbb'}
  \varepsilon = \int_{-\infty}^{\infty}\!dz\,
  \Bigl\{g^\mathrm{el}_{\scriptscriptstyle \rhd '}(\mathbf{v}) +\frac{n'V}{2}
  \bigl[1-\cos{(-qv_x)}\bigr]\Bigr\}
\end{align}
with the rhombic lattice density $n' = 1/bb'$. Furthermore, we have used
that $\mathbf{q}\cdot\mathbf{R} = 2\pi \mathbb{Z}$ as the undisturbed $bb'$
lattice is in registry with the substrate potential along the $x$-axis.  The
above soliton line energy relates to the usual free energy density via
\begin{align}\label{eq:eps_line}
  \varepsilon & \approx \!\!\!
   \lim_{L,L_\perp \to \infty}\!\!\!
   L_\perp^{-1}\!\! \int_{L \times L_\perp} \!\!\!\!\!\!\!\!\!\!\!
   dz\, dz_\perp \bigl[g^\mathrm{el}_{\scriptscriptstyle \rhd '}(\mathbf{v})
   +n' e^\mathrm{sub}(\mathbf{R}+\mathbf{v})\bigr].
\end{align}
The elastic theory of the $bb'$-lattice is described by the energy density
$g^\mathrm{el}_{\scriptscriptstyle\rhd'} = g_p + g_\kappa + g_\mu$ with
the linear term
\begin{align}\label{eq:g_p_bb'}
   g_p = (\gamma_x'+p)(\partial_x v_x) + (\gamma_y'+p) (\partial_y v_y) 
\end{align}
driving the system towards the hexagonal phase and the usual compression- and
shear-type energy densities 
\begin{align}\label{eq:g_k_m_bb'}
   g_\kappa &= \frac{\kappa_x'}{2}(\partial_x v_x)^2 
        + \frac{\kappa_y'}{2} (\partial_y v_y)^2 
        + \kappa_{xy}' (\partial_x v_x)(\partial_y v_y), \\ \nonumber
   g_\mu &= \frac{\mu_x'}{2} (\partial_y v_x)^2 
        + \frac{\mu_y'}{2} (\partial_x v_y)^2 
        + \mu_{xy}' (\partial_y v_x)(\partial_x v_y).
\end{align}
The linear contribution (\ref{eq:g_p_bb'}) is due to the purely repulsive
dipolar interaction that is balanced only by the external pressure term $p
\delta A /A$ and has been included in $g^\mathrm{el}_{\scriptscriptstyle
\rhd'}$; for the hexagonal lattice this pressure term generates a stable
minimum relating to the density via Eq.\ (\ref{eq:p}) and balances the
$\gamma$ terms, $\gamma_x = \gamma_y = -p$.  Deforming the hexagonal lattice
into the $bb'$ rhombic lattice (in our case via the underlying substrate
potential) this term attempts to drive the particle lattice back to the
rhombic shape as the $\gamma'$-terms are not compensated by the pressure.  The
various coefficients $\gamma_{x,y}'$, $\kappa_{x,y,xy}'$, and $\mu_{x,y,xy}'$
are determined with the help of the Ewald summation technique\cite{Ewald_21}
as described in the appendix \ref{app:el_const}.

\begin{figure}
\begin{center}
\includegraphics[width=7cm]{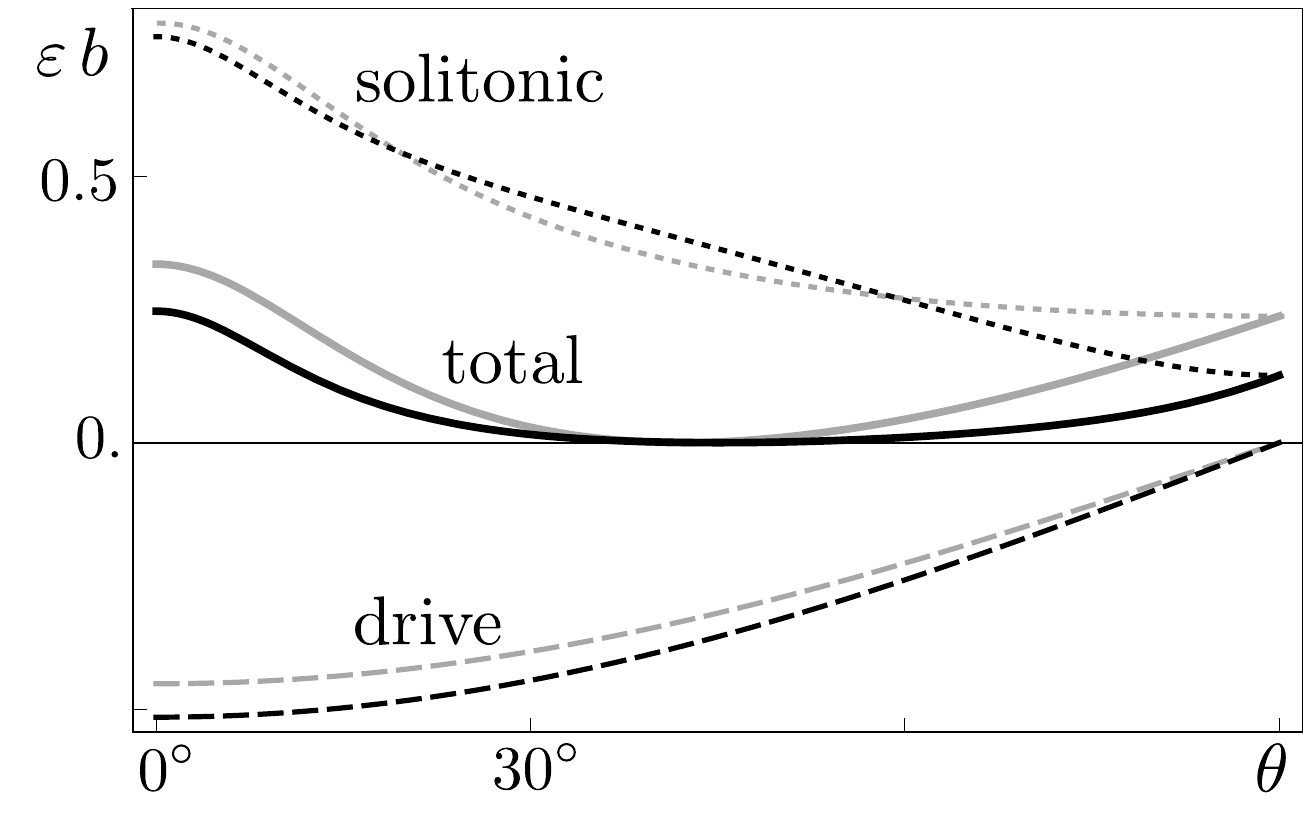}
\end{center}
\caption {\label{fig:sol_es_ed} Solitonic energy $\varepsilon_\mathrm{s} b$
(dotted) and drive $\varepsilon_\mathrm{d}b$ (dashed) calculated with the
elastic theory for the rhombic lattice (black) and the one for the
hexagonal lattice (grey) at the critical potential $V_c^{\rm\scriptscriptstyle
PT} \approx 0.0417\, e_{\scriptscriptstyle D}$ in units of
$e_{\scriptscriptstyle D}$. The solid lines represent the total energy
$\varepsilon b=\varepsilon_\mathrm{d} b+\varepsilon_\mathrm{s} b$.}
\end{figure}

Assuming a uniaxial soliton ${\bf v}(z)$ oriented along $z = x \cos \theta + y
\sin \theta$ (with $\theta$ to be determined), the expression for the total
line energy (\ref{eq:eps_line}) can be simplified and naturally splits into a
soliton part
\begin{widetext}
\begin{align}\label{eq:eps_s}
   \varepsilon_\mathrm{s} &=  \int_{-\infty}^{\infty}\!\!\!\!dz\,
   \Bigl\{\frac{\kappa'_x\cos^2\theta
   +\mu'_x\sin^2\theta}{2}(\partial_zv_x)^2
   +\frac{\kappa'_y\sin^2\theta+\mu'_y\cos^2\theta}{2}(\partial_zv_y)^2
   \\ \nonumber
   & \qquad \qquad \qquad \qquad+(\kappa'_{xy}+\mu'_{xy})\sin{\theta}
   \cos{\theta}(\partial_zv_x)
   (\partial_zv_y) +\frac{Vn'}{2}
   \bigl[1-\cos{(-qv_x)}\bigr]\Bigr\}
\end{align}
\end{widetext}
and a drive 
\begin{align}\label{eq:eps_d}
   \varepsilon_\mathrm{d} &= \! \int_{-\infty}^{\infty}\!\!\!\!dz\,
   (\gamma'_x\!+\!p)\cos{\theta}\, (\partial_zv_x)
   =-(\gamma'_x\!+\!p)\,b\cos{\theta},
\end{align}
where we have used the boundary condition $v_x(\infty) = -b$. Minimizing the
soliton energy, we obtain a Sine-Gordon equation $\alpha_{\scriptscriptstyle\rhd'}
\partial_z^2 (q v_x) = \sin (q v_x)$ with
\begin{align}\label{eq:abb'}
   \alpha_{\scriptscriptstyle\rhd'}
   \!=\! \frac{2}{Vn'q^2}\Bigl[ \kappa'_x\cos^2\theta \! + \! \mu'_x\sin^2\theta
    \! - \! \frac{(\kappa'_{xy} \! +\! \mu'_{xy})^2\cos^2\theta}
   {\kappa'_y+\mu'_y\cot^2\theta}\Bigr].
\end{align}
The displacement $v_y$ along $y$ is slaved to the displacement along $x$
via $v_y=-\eta_{\scriptscriptstyle\rhd'} v_x$ with
\begin{align}\label{eq:etabb'}
  \eta_{\scriptscriptstyle\rhd'}= \frac{(\kappa'_{xy}+\mu'_{xy})\cot\theta}
   {\kappa'_y+\mu'_y\cot^2\theta}
\end{align}
and the soliton line energy takes the form
\begin{align} \label{eq:esolbb'}
   \varepsilon_\mathrm{s}&=4n'V\sqrt{\alpha_{\scriptscriptstyle\rhd'}}.
\end{align}
The first soliton appears when the soliton and drive energies compensate one
another, $\varepsilon = \varepsilon_\mathrm{s} +\varepsilon_\mathrm{d} = 0$;
figure \ref{fig:sol_es_ed} shows these energies as a function of angle
$\theta$ at the critical potential where the minimum in $\varepsilon$ vanishes
for the first time, providing the critical substrate potential and the soliton
angle
\begin{align} \label{eq:VcPTbb'}
   V_c^{\rm \scriptscriptstyle PT}=0.0417\, e_{\scriptscriptstyle D},\quad 
   \theta=45.05^\circ.
\end{align}
The lattice displacement along $y$ associated with this soliton is determined
by $\eta{\scriptscriptstyle\rhd'} \approx 0.696$ and we obtain the overall shift vector
for the Pokrovskii-Talapov soliton ${\bf d}^{\rm\scriptscriptstyle PT} = b (-1,0.696)$.

At first sight, these results compare favorably with those obtained using the
elasticity theory for the hexagonal lattice, particularly for the critical
potential $V_c^{\rm\scriptscriptstyle PT}$, see Eqs.\ (\ref{eq:VcPThex}) and
(\ref{eq:VcPTbb'}), and to a lesser degree for $\theta$; the results for the
shift along $y$ differ quite substantially, however, $\eta \approx 0.905$
versus $\eta{\scriptscriptstyle\rhd'} \approx 0.696$. Furthermore, Fig.\
\ref{fig:sol_es_ed} shows, that the individual results for the soliton energy
$\varepsilon_\mathrm{s}$ and the drive $\varepsilon_\mathrm{d}$ again differ
quite appreciably. Overall, we have to conclude that anharmonicities are not
negligible and have the potential to change the results on the order of 10 \%.

\subsection{Numerical analysis}\label{ssec:num_ana}

In order to obtain accurate and reliable results for the first appearance (at
$V_c^{\rm\scriptscriptstyle PT}$) and the characteristic parameters ($\theta,
\eta$) of the PT soliton, we determine these quantities with the help of a
numerical analysis. Such an analysis will be even more relevant when analyzing
solitons and domain walls in the presence of two substrate modes, see Sec.\
\ref{sec:sol_dw} below. In the following, we determine the optimal shape for
the PT soliton within a variational approach and find the critical
substrate potential $V_c^{\rm\scriptscriptstyle PT}$. The latter is determined
by comparing the free energies with and without soliton on the $bb'$-lattce
background as calculated directly from the `microscopic' expressions Eqs.\
(\ref{eq:E}) and (\ref{eq:gpN}), where the geometry of the $bb'$-lattice is
determined by minimization of $g_{\scriptscriptstyle \rhd'}(b')$ with $b' =
1.0173\, b$.

\begin{figure}[h!]
\begin{center}
\includegraphics[width=8cm]{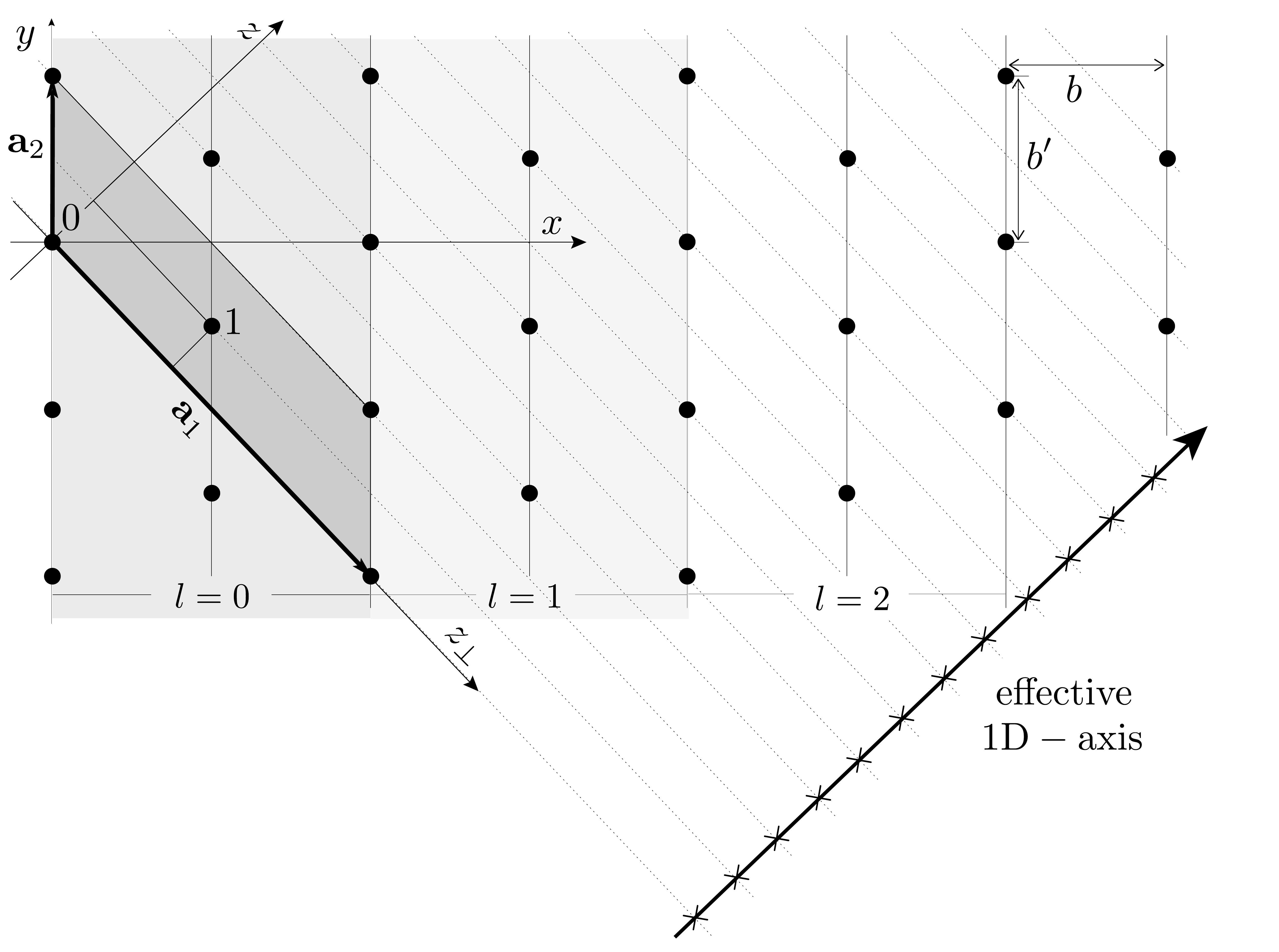}
\end{center}
\caption
{\label{fig:uc445} Coordinates $z_\perp$ and $z$ for $m=n=2$, $\theta\approx
44.5^\circ$ with a supercell containing two particles with labels 0 and 1.}
\end{figure}

Summing the long-range interaction for a two-dimensional particle system is
unpractical (note that the Ewald summation cannot be applied to the
non-uniform soliton phase). However, we can reduce the problem to a
one-dimensional one by selecting angles $\theta$ where $z_\perp$ is directed
along a particle row, see Fig.\ \ref{fig:uc445}. We then make use of
appropriate supercells with lattice vectors arranged along the $z_\perp$-axis
and along the $y$-axis, ${\bf a}_1=(mb,-nb')$ and ${\bf a}_2=(0,b')$ where $m$
and $n$ are Miller indices.  Below, we analyze configurations with small
Miller indices, $m = n = 2$ with $\theta = \arctan (mb/nb') = 44.5^\circ$ and
2 particles per supercell, $m = 2$, $n=1$ with $\theta = 63.4^\circ$, $m = 2$,
$n=3$ with $\theta = 33.2^\circ$, and $m = 2$, $n=5$ with $\theta =
21.5^\circ$ and 1 particle per cell, and $m = 4$, $n=1$ with $\theta =
75.7^\circ$ and 4 particles per supercell.  The matrix
\begin{align}\label{eq:Umn}
   U_{m,n} = \frac{1}{a_1}
   \begin{pmatrix}
             mb &  -nb' \\
             nb' & mb\\
  \end{pmatrix}
\end{align}
transforms the coordinates from the $xy$- to the $z_\perp z$-frame, in
particular, ${\bf a}_1= (a_1,0)$ and ${\bf a}_2 = (-n{b'}^2/a_1,$ $ mbb'/a_1)$.
In the following, we sketch the main steps of the analysis for the case
$m=n=2$, see Fig.\ \ref{fig:uc445}, and cite the results for the remaining cases.

The sums in the energy $E(A,N)$, Eq.\ (\ref{eq:E}), involve the particle positions
of the $bb'$ lattice
\begin{align}\label{eq:pos445_bb'}
   {\bf R}_{lq}^{{\scriptscriptstyle\rhd'},\mu} = \binom{z_{\perp,lq}^\mu}{z_q^\mu}
   =l{\bf a}_1+q{\bf a}_2+U_{2,2}{\bf c}^\mu,
\end{align}
with the basis ${\bf c}^0=(0,0)$ and ${\bf c}^1=(b,-b'/2)$ (in the $xy$-frame)
and the positions of the distorted lattice including one soliton
\begin{align}\label{eq:coord_sol445_PT}
   {\bf R}_{lq}^{\mathrm{s},\mu} = 
   \binom{z_{\perp,lq}^{\mathrm{s},\mu}}{z_{q}^{\mathrm{s},\mu}}
   ={\bf R}_{lq}^{{\scriptscriptstyle\rhd'},\mu}
   +U_{2,2}\binom{v_x(z_q^\mu)}{v_y(z_q^\mu)},
\end{align}
with the soliton displacement field (again in the $xy$-frame)
\begin{align}
   \label{eq:PTsol_vx_opt}
   v_x(z)&=-\frac{2b}{\pi}\arctan\{\exp[(z-z_\mathrm{s})/\sqrt{w_x\alpha}\,]\},\\
   \label{eq:PTsol_vy_opt}
   v_y(z)&=s_y\eta\,\frac{2b}{\pi}\arctan\{\exp[(z-z_\mathrm{s})/\sqrt{w_y\alpha}\,]\}.
\end{align}
Here, $z_\mathrm{s}$ defines the soliton position and $w_x$, $w_y$, and $s_y$ are
variational parameters for the soliton widths (along $x$ and $y$) and the
soliton shift along $y$ with respect to the values for $\alpha$ (see
(\ref{eq:abb'})) and $\eta$ (see (\ref{eq:etabb'})) obtained from the analytic
results based on the $bb'$ rhombic lattice.

A fast evaluation of the interaction energy in $E(A,N)$ is crucial for the
optimization process of the soliton shape and the evaluation of
$V_c^{\rm\scriptscriptstyle PT}$. The sum
\begin{align}\label{eq:sum_int}
   E_\mathrm{s}^\mathrm{int}=
   \frac{1}{2} \sum_{l,l',q,q',\mu,\nu} \frac{D}{|{\bf R}^{\mathrm{s},\mu}_{lq}
   -  {\bf R}^{\mathrm{s},\nu}_{l'q'}|^3}
\end{align}
is split into one along $z_\perp$ (over $l$ and $l'$) which can be resummed
analytically and the remaining sum along $z$ (over $q$ and $q'$). Using
Poisson's formula at fixed shifts $\alpha = \alpha_{qq'}^{\mathrm{s},\mu\nu} =
(z_{\perp,0q}^{\mathrm{s},\mu} - z_{\perp,0q'}^{\mathrm{s},\nu})/a_1$ along
$z_\perp$ in the supercell $l=0$ and $\beta = \beta_{qq'}^{\mathrm{s},\mu\nu}
= (z_q^{\mathrm{s},\mu} - z_{q'}^{\mathrm{s},\nu})/a_1$ along $z$, we find
that (see also appendix \ref{app:eff_pot})
\begin{align}\label{eq:sum_l}
   &\sum_l \frac{1}{[(l+\alpha)^2 + \beta^2]^{3/2}} \\
   &\qquad\qquad \nonumber
   = \frac{2}{\beta^2}
   + 8 \pi \sum_{\bar{l}>0} \bar{l} \, \cos(2\pi \bar{l} \alpha)
     \frac{K_1(2\pi \bar{l}|\beta|)}{|\beta|}
\end{align}
can be approximated by taking only few terms (of order 10) in the second sum
over $\bar{l}$, as the modified Bessel function $K_1(z)$ (of the second kind)
rapidly decreases, $K_1(z)\propto e^{-z}$ for large $z$ (see Ref.\
\onlinecite{abramowitz_72}). The remaining sums in the interaction energy
\begin{align}\label{eq:sum_int_res}
   E_\mathrm{s}^\mathrm{int} &=
   \frac{N_{\perp} D}{2 a_1^3}
   \sum_{q,q',\mu,\nu} \biggl\{
   \frac{2}{(\beta^{\mathrm{s},\mu\nu}_{qq'})^2}\\
   &\nonumber
   +8\pi\sum_{\bar{l}>0}  \bar{l} \,
   \cos(2\pi \bar{l}\alpha^{\mathrm{s},\mu\nu}_{qq'})
   \frac{K_1(|2\pi \bar{l} \beta^{\mathrm{s},\mu\nu}_{qq'}|)}
   {|\beta^{\mathrm{s},\mu\nu}_{qq'}|}
   \biggr\}
\end{align}
then have to be evaluated numerically. Here, $N_\perp$ denotes the number of
unit cells (of extension $a_1=2\sqrt{b^2+{b'}^2}$) along $z_\perp$.  The  sums
are to be taken over the particles $\mu,\nu \in \{0,1\}$ in the supercell and
$q,\,q'\in\{0,\dots, N/2-1\}$ go over the supercells in the $l=0$ strip, see
Fig.\ \ref{fig:uc445}, with $N$ the particle number in the $l=0$ strip.
Equation \eqref{eq:sum_int_res} then evaluates the interaction energy of a
one-dimensional chain of particles with an effective interaction that accounts
for the transverse dimension. Note that the terms with $q=q'$ at $\mu=\nu$ are
discarded as these are compensated by an equal term appearing in the
interaction energy $E_\mathrm{int}$ without the soliton (the two compensating
terms are easily evaluated via the direct sum $(D/2)\sum_{l\neq 0} (a_1
l)^{-3} = D\zeta(3)/2 a_1^3$ with $\zeta(s)=\sum_{n=1}^\infty n^{-s}$ the
Riemann zeta function).
\begin{table}
\caption{\label{table:1mode} Numerical results for $V_c^{\rm\scriptscriptstyle
PT}$ and optimal parameters $w_x=\alpha'_x/\alpha$, $w_y=\alpha'_y/\alpha$,
and $s_y=\eta'/\eta$ for the PT-soliton evaluated at discrete angles $\theta$
determined by small Miller indices. The corresponding analytic results for the
hexagonal- and the $bb'$ rhombic elasticity theories are $V_c^{\rm
\scriptscriptstyle PT} \approx 0.0416\, e_{\scriptscriptstyle D}$ at $\theta
\approx 42.13^\circ$ and $V_c^{\rm \scriptscriptstyle PT} \approx
0.0417\,e_{\scriptscriptstyle D}$ at $\theta \approx 45.05^\circ$.}
  \begin{center}
  \begin{tabular}{| c | c | c | c | c |}
  \hline
  $\theta$ &  $V_c^{\rm \scriptscriptstyle PT}/e_{\scriptscriptstyle D}$ &
  $w_x$ & $w_y$ & $s_y$\\
  \hline
  $21.5^\circ$ & 0.033 & 1.1 & 1.2 & 0.95\\
  $33.2^\circ$ & 0.042 & 0.9 & 0.95 & 1.05\\
  $44.5^\circ$ & 0.046 & 0.8 & 0.75 & 1.1\\
  $63.0^\circ$ & 0.031 & 1.3 & 1.15 & 1.05\\
  $75.7^\circ$ &  0.016 & 1.85 & 2.05 & 1\\
  \hline
\end{tabular}
\end{center}
\end{table}

The substrate energy $E^\mathrm{sub}$ in Eq.\ (\ref{eq:E}) is cast into a
similar form with the sum going over all basis vectors $\mu \in \{0,1\}$ in
the supercell and summation over cells $q\in\{0,\dots, N/2-1\}$ in the
$l=0$-strip,
\begin{align} \label{eq:sum_sub}
   E_\mathrm{s}^\mathrm{sub} &=
   \frac{V N_{\perp}}{2} \sum_{q,\mu}
        \bigl\{2- \cos\bigl[4\pi (\beta^{\mathrm{s},\mu}_{q}
          +\alpha^{\mathrm{s},\mu}_{\perp,q})\bigr]\bigr\}. 
\end{align}
Repeating the calculation for the particle system without soliton
($\rightarrow E^\mathrm{int}_{\scriptscriptstyle\rhd'},~
E^\mathrm{sub}_{\scriptscriptstyle\rhd'}$), the final expression for the
soliton line energy per length $a_1$ is
\begin{align}
\label{eq:linEn_PT}
    \varepsilon =(E_\mathrm{s}^\mathrm{int}-E_{\scriptscriptstyle\rhd'}^\mathrm{int}
   +E_\mathrm{s}^\mathrm{sub}-E_{\scriptscriptstyle\rhd'}^\mathrm{sub} + p \delta A)
   /N_\perp a_1,
\end{align}
where the last term $p\delta A$ represents the cost due to the area change
which comes along with the soliton deformation and depends on the direction of
$z$; for the PT-soliton with its lattice shift ${\bf d}^\mathrm{PT}=(-b,s_y \eta
b)$, we find the area change per $l$-strip (such that $\delta A=N_\perp
\delta A^{\rm\scriptscriptstyle PT}_{m,n}$)
\begin{align} \label{eq:areaPT}
   \delta A^{\rm\scriptscriptstyle PT}_{m,n} 
   = a_1({\bf d}^{\rm\scriptscriptstyle PT}\cdot\hat{\bf e}_z).
\end{align}
This area change is negative and hence the PT soliton involves a lattice
compression, in agreement with the fact that the smaller density $n'$ of the
$bb'$ lattice has to approach the larger density $n$ of the hexagonal lattice
when overlapping PT solitons approximate the distorted and rotated hexagonal
phase at small $V$. For the PT soliton at $m = n = 2$ the area change is of
order of 25 \% of the supercell area, implying that about 1 particle is added
to every two such cells along the soliton.
\begin{figure}[h!]
\begin{center}
\includegraphics[width=7cm]{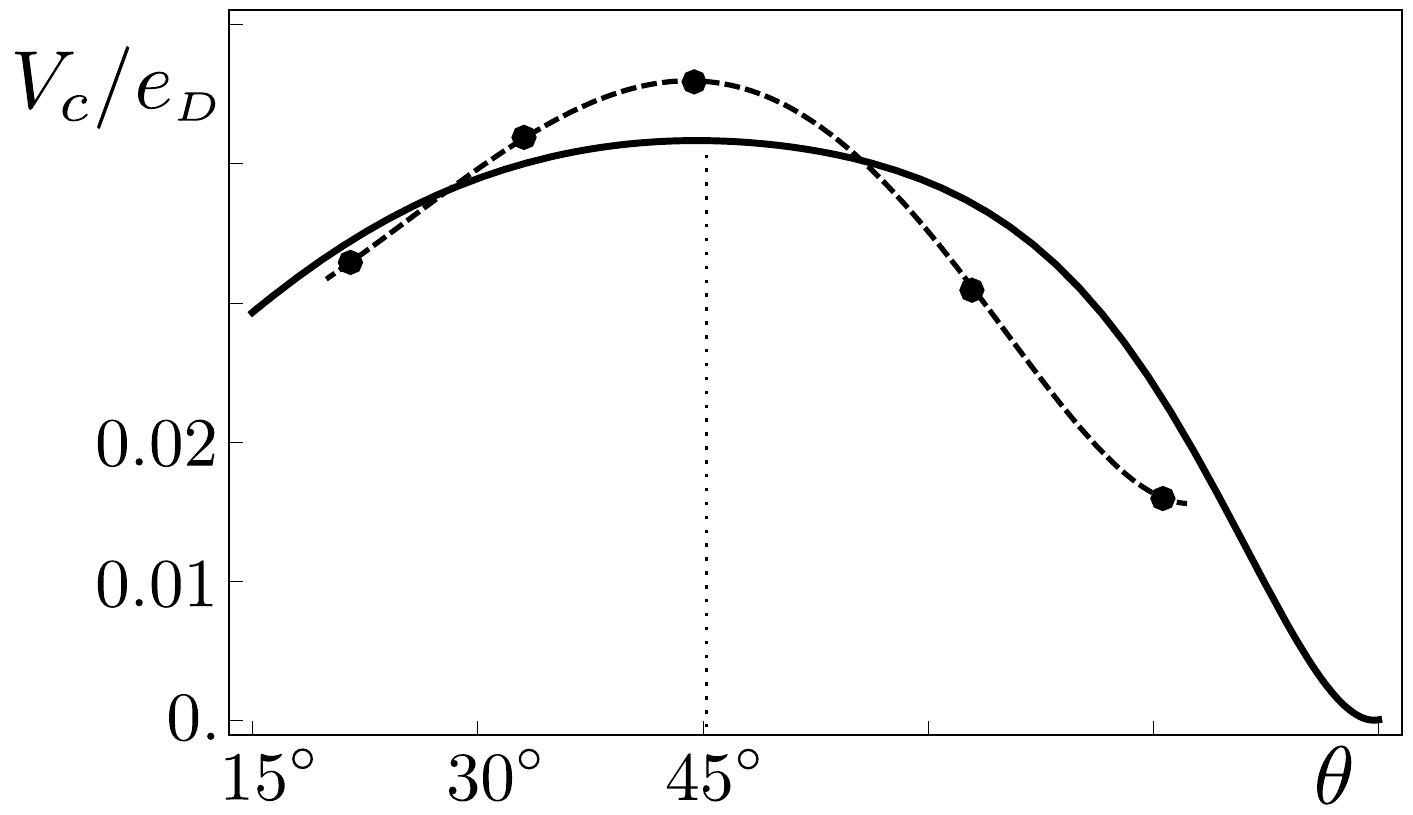}
\end{center}
\caption{\label{fig:Vc_PT} Numerical results for the critical substrate
potential $V_c^{\rm\scriptscriptstyle PT}$ of the PT-soliton at discrete
angles defined by small Miller indices (black dots); the dashed line is a
guide to the eye.  The solid line is the analytic result based on the elastic
description of the $bb'$ rhombic lattice. The thin dotted line marks the
optimal angle $\theta \approx 45.05^\circ$ where the analytic result assumes
its maximal value $V_c^{\rm \scriptscriptstyle PT}\approx 0.0417\,
e_{\scriptscriptstyle D}$.}
\end{figure}

In order to verify the numerical accuracy, we have calculated the lattice
energy $e_{\scriptscriptstyle\rhd'}$ using the uniform version of Eq.\
(\ref{eq:sum_int_res}) (without the soliton) and have compared it to the value
$e_{\scriptscriptstyle\rhd'}  = 4.3489 \, e_{\scriptscriptstyle D}$ obtained
with the help of the Ewald summation technique: Going up to $N = 25 000$
particles (where $e_{\scriptscriptstyle\rhd'} \approx 4.3465 \,
e_{\scriptscriptstyle D}$) the value obtained from Ewald summation is
approached with an error (due to boundary effects) vanishing as $1/N$. When
calculating the properties of the soliton, we go up to system sizes with $N =
5000$ particles (corresponding to a system size $Z = Nh_z/2$ along $z$ with
the height $h_z= bb'/2 \sqrt{b^2 + b'^2}$ of the unit cell along $z$). This
size is sufficently large to produce results with an accuracy in the per mill
range (note that boundary effects are less relevant in the energy differences
(\ref{eq:linEn_PT}) determining the soliton energy). Placing the soliton
midway, $z_\mathrm{s} = Z/2$, and varying the parameters $w_x$, $w_y$, and
$s_y$, we find the first soliton entry where $\varepsilon = 0$ at the critical
potential
\begin{align} \label{eq:Vc445_PT}
   V_c^{\rm \scriptscriptstyle PT} \approx 0.046\,e_{\scriptscriptstyle D},
   \quad \theta\approx 44.5^\circ,
\end{align}
appreciably larger than the result (\ref{eq:VcPTbb'}) of the analytic
calculation.  The optimal soliton parameters are $w_x \approx 0.8$, $w_y
\approx 0.75$, and $s_y \approx 1.1$, i.e., the optimized soliton shape is
narrower in both directions and the shift vector is larger along $y$, ${\bf
d}^{\rm \scriptscriptstyle PT} \approx (-b, 0.766\, b)$.

The calculation for the other Miller indices follows the same program as the
one described above and the results are summarized in Table \ref{table:1mode};
Fig.\ \ref{fig:Vc_PT} shows the critical substrate potentials
$V_c^{\rm \scriptscriptstyle PT}$ at the discrete angles for small Miller indices
in comparison with the analytic results based on the elastic theory for the
$bb'$ rhombic lattice, with the critical potential at $\theta = 44.5^\circ$
being the largest.

\section{Solitons and Domain Walls with two Substrate Modes}\label{sec:sol_dw}

The soliton array obtained within the resonance approximation transforms the
$bb'$ rhombic lattice to the hexagonal one, while our goal here is to study
the transformation of the particle system from square to hexagonal.  The
solitonic instability then should appear at small $V < V_{\scriptscriptstyle
\square}$ on the background of the period-doubled phase, which requires us to
include the second harmonic of the substrate potential into our analysis. We
treat the period-doubled phase as a $bb$ rhombic lattice distorted by the
relative shift $\bar\delta$ of the two sublattices see Sec.\
\ref{ssec:pd_phase}. The soliton is described by a smooth displacement field
${\bf v}({\bf R})$ relative to the $bb$ lattice. Inside the soliton, the
amplitude of the short-scale distortion $\bar\delta = (b/\pi)
\arcsin[V\cos(qv_y)/8\Delta]$ is slaved to the displacement ${\bf v}({\bf R})=
(b/4)\, {\bf e}_y +\bar{\boldsymbol{\sigma}}$ which is replacing the scalar
center of mass variable $\bar\sigma$ introduced above, see Eq.\ (\ref{eq:ud}).
We then have to minimize the energy\cite{terms}
\begin{eqnarray} \label{eq:g_ca2m}
   \delta g &=&\frac{1}{N}\!\int\!\! d^2R\, 
   \Bigl\{g^\mathrm{el}_{\scriptscriptstyle\rhd}({\bf v})
   +\frac{Vn}{2}[1-\cos(q v_x)]
   \\ \nonumber
   &&\qquad\qquad\qquad + \frac{nV^2}{64 \Delta}[1-\cos(2qv_y)]\Bigr\},
\end{eqnarray}
where $g^\mathrm{el}_{\scriptscriptstyle\rhd}$ is the elastic Gibbs free
energy \cite{Gfed} density of the $bb$ rhombic lattice describing the long
wave-length distortions of the period-doubled lattice and $\delta g$ denotes
the deviation from $g_\mathrm{pd}(V)$, Eq.\ (\ref{eq:epd}).

\begin{figure}
\begin{center}
\includegraphics[width=7cm]{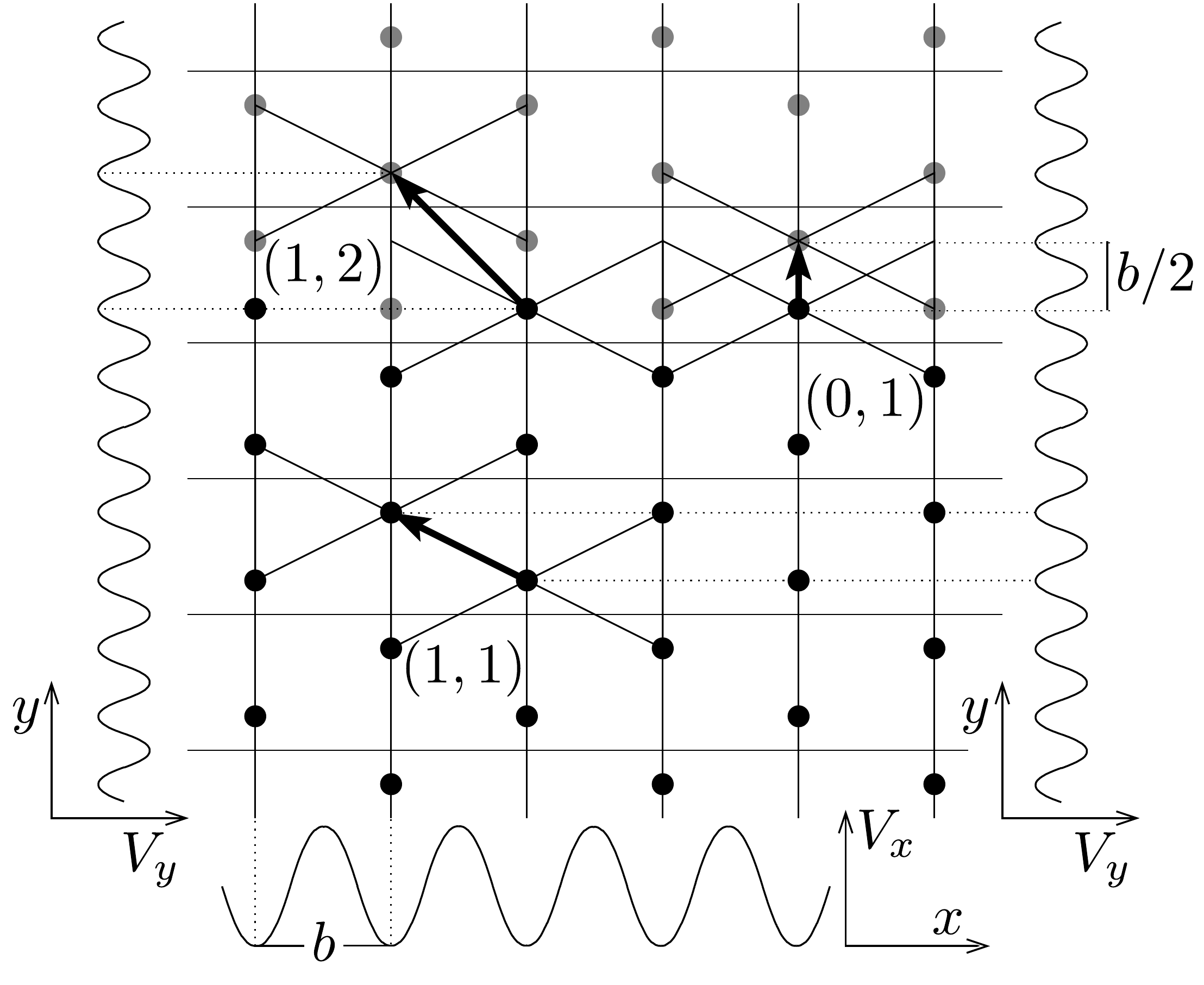}
\end{center}
\caption {\label{fig:soliton-types} Selection of low-energy solitons and
domain-walls shifting the period-doubled lattice by ${\bf d}_{j,k} =
(-jb,kb/2)$, $j,k \in \mathbb{Z}$, the topological vector-charge associated
with the defect.  The simplest defect is the $(0,1)$-domain wall crossing only
the barrier along $y$ and connecting different twins of the period-doubled
phase. The $(1,1)$-soliton crosses both barriers along $x$ and along $y$ and
connects identical twins. Finally, the domain-wall $(j,k) = (1,2)$ crosses
both potential barriers, once the barrier along $x$ and twice that along $y$.
Note that the barrier along $y$ is reduced with respect to the barrier along
$x$ by the factor $V/32\Delta = V/4 V_{\scriptscriptstyle \square}$.}
\end{figure}

While the resonance approximation admits only one low-energy soliton, the full
problem with both substrate modes present allows for several types of
line-defects with different quantized topological vector-charges ${\bf
d}_{j,k} = (-jb,kb/2)$, $j,k \in \mathbb{Z}$. The latter correspond to the
shift ${\bf d}_{j,k} = {\bf v}^{\scriptscriptstyle (j,k)}(\infty) - {\bf
v}^{\scriptscriptstyle (j,k)}(-\infty)$ of the lattice associated with the
defect, similar to the Burger's vector characterizing the displacement field
of a dislocation.  A selected set of defects with potentially low energies are
shown in Fig.\ \ref{fig:soliton-types}: promising candidates reminding about
the PT soliton are the $(j,k) = (1,k)$ defects with $k = 1,2,3$, but a simple
Ansatz with the shift ${\bf d}_{01} = (0,b/2)$ should be tried as well, since
the particles merely have to overcome the weak effective potential $\propto
V^2/64\Delta \ll V/2$ along the $y$-direction, see Eq.\ (\ref{eq:g_ca2m}).
All these line defects fall into two classes, the domain walls with odd values
$j+k$ and taking the period-doubled phase from one twin to the other, $\delta
\to -\delta$, and the genuine solitons with $j+k$ even and the same twin on
both sides, $\delta \to \delta$, see Fig.\ \ref{fig:soliton-types}.

When considering both substrate modes, the topological vector-charge ${\bf
d}_{j,k}= (-j,k/2)b$ of the soliton array is quantized in both directions $x$
and $y$. The global displacement field ${\bf u}_\mathrm{g}$ resulting from the
soliton array then must be compatible with both quantization conditions along
$x$ and $y$. The (shear) displacement field taking the $bb$ rhombic
lattice\cite{square} into the hexagonal one is given by ${\bf u}_\mathrm{g} =
(-\alpha_x x, \alpha_y y)$ with $\alpha_x = s/(1+s)$ and $\alpha_y = s$. Such
a displacement field cannot be built from a single soliton array with the
quantized geometrical constraint on ${\bf d}_{j,k}$ since $s = (4/3)^{1/4} -1$
is an irrational number.  Hence, the quantization of the topological-vector
charge provides us with a valuable input on the square-to-hexagonal transition
pathway: this pathway has to involve more than one soliton transition or
another more complex route.  As we will show below, the way the system will
deal with this problem is by undergoing two transitions: in the first
transition, involving a (0,1)-domain-wall, the mean lattice constant along $y$
smoothly changes from $b$ to $b'$ as domain-walls flood the system, implying
that the substrate potential along $y$ is washed out (we remind that, for a 1D
commensurate--incommensurate transition\cite{Bak_82}, the mean lattice constant
$\langle a \rangle$ smoothly goes from $b$ (in the locked phase) to $a$ (in
the free phase) as the substrate potenial is reduced from the critical value
$V_c$ to zero).  The resulting $bb'$ rhombic lattice then undergoes a second
soliton transition with the PT soliton array taking the rhombic lattice to the
hexagonal phase by washing out the second substrate mode along $x$. Hence the
two geometrical constraints on the locked phase are subsequently released by
two consecutive transitions.

In the following, we search for line solitons directed along an angle $\theta$
using the Ansatz ${\bf v}(x,y) = {\bf v}(z)$ with $z = x\cos\theta + y
\sin\theta$ and focus on their first appearance---the physically relevant
topological defect then is that one with the largest critical substrate
potential $V_c^{\scriptscriptstyle (j,k)}$.  Again, we first analyze the
problem within a continuum elastic theory and then refine our results with a
numerical analysis.

\subsection{Continuum elastic approach}\label{ssec:ana_2M}

Accounting for the boundary conditions, the energy (\ref{eq:g_ca2m}) can be
rewritten as the sum of a line energy $\varepsilon_\mathrm{s}$ and a drive
$\varepsilon_\mathrm{d}$. The line energy assumes the form (\ref{eq:eps_s})
with the elastic coefficients $\gamma'_{x,y},\, \kappa'_{x,y,xy}, \,
\mu'_{x,y,xy} \to \gamma_{x,y},\, \kappa_{x,y,xy}, \, \mu_{x,y,xy}$ replaced
by those for the $bb$ rhombic lattice, see appendix \ref{app:el_const}
for their evaluation. In addition, the double-periodic effective potential
$(nV^2/64\Delta)[1-\cos(2q v_y)]$ along $y$ has to be accounted for, see
(\ref{eq:g_ca2m}). The drive replacing Eq.\ (\ref{eq:eps_d}) reads
\begin{align}\label{eq:eps_dxy}
   \varepsilon_\mathrm{d} 
   =-j(\gamma_x\!+\!p)\,b\cos{\theta}
   +\frac{k}{2}(\gamma_y\!+\!p)\,b\sin{\theta}
\end{align}
and includes an additional term along $y$.

The simplest case to evaluate is the $(0,1)$ domain-wall with the displacement
field directed along $y$, ${\bf v}(z) = (0,v_y(z))$. The variation of the line
energy produces a Sine-Gordon equation with the double-periodic potential along
$y$ and inserting the standard soliton solution $v_y = (b/\pi) \arctan[\exp(z/
\sqrt{\alpha^y_{\scriptscriptstyle\rhd}})]$ back into $\varepsilon_\mathrm{s}$,
we obtain the line energy
\begin{equation}
   \varepsilon_\mathrm{s}^{\scriptscriptstyle (0,1)}
   = \frac{nV^2}{8\Delta} \sqrt{\alpha^y_{\scriptscriptstyle\rhd}} 
\end{equation}
with
\begin{equation}\label{eq:aybb}
   {\alpha}^y_{\scriptscriptstyle\rhd} = \frac{64\Delta}{nV^2}
   \frac{\kappa_y\sin^2\theta+\mu_y\cos^2\theta} {4q^2}.
\end{equation}
Balancing this energy with the drive $\varepsilon_\mathrm{d} = (\gamma_y +
p) b$ $\sin\theta/2$, we find the critical field
\begin{align} \label{eq:vc01}
   V_c^{\scriptscriptstyle (0,1)}(\theta)=-\frac{2\pi(\gamma_y+p)}{n}
  \sqrt{\frac{n\Delta}{\kappa_y+\mu_y\cot^2{\theta}}},
\end{align}
which is monotonically increasing with $\theta$. The first $(0,1)$ domain-wall
then appears at
\begin{align} \label{eq:vc01_num}
   V_c^{\scriptscriptstyle (0,1)}\approx 0.0753\, e_{\scriptscriptstyle D},
   \qquad \theta^{\scriptscriptstyle (0,1)} = 90^\circ.
\end{align}

Next, we analyze the $(1,k)$ solitons crossing the large barrier along $x$
once and $k$ times the small barrier along $y$. While in the resonance
approximation the $v_y$ displacement was slaved to the $v_x$ field (see
(\ref{eq:etabb'})), the potential $\propto V^2/\Delta$ along $y$ renders the
solution of the differential equations more difficult. Since the potential
along $y$ is small as compared to the one along $x$, $V^2/\Delta \ll V$, we
seek a perturbative solution ${\bf v} = {\bf v}^{\scriptscriptstyle (0)} +
{\bf v}^{\scriptscriptstyle(1)}$. To lowest order, we drop the potential along
$y$ and obtain the usual soliton solution
\begin{align} \label{eq:body}
   v_x^{\scriptscriptstyle (0)} &= -(2b/\pi) \arctan[\exp(z/ 
            \sqrt{\alpha^x_{\scriptscriptstyle\rhd}})], \\ \nonumber
   v_y^{\scriptscriptstyle (0)} &= -\eta_{\scriptscriptstyle\rhd} 
            v_x^{\scriptscriptstyle (0)},
\end{align}
with
\begin{align}\label{eq:axbb}
   \alpha^x_{\scriptscriptstyle\rhd} 
   &=\frac{2}{nVq^2}\Bigl[
   \kappa_x\cos^2\theta +\mu_x\sin^2\!\theta \\ \nonumber
   &\qquad \qquad
   -\frac{(\kappa_{xy}+\mu_{xy})^2\cos^2\theta}
         {\kappa_y + \mu_y\cot^2\theta}\Bigr],\\
   \eta_{\scriptscriptstyle\rhd} &=\frac{(\kappa_{xy}+\mu_{xy})\cot\theta}
   {\kappa_y+\mu_y\cot^2\theta}.\label{eq:etabb}
\end{align}

\begin{figure}[t]
\begin{center}
\includegraphics[width=7.5cm]{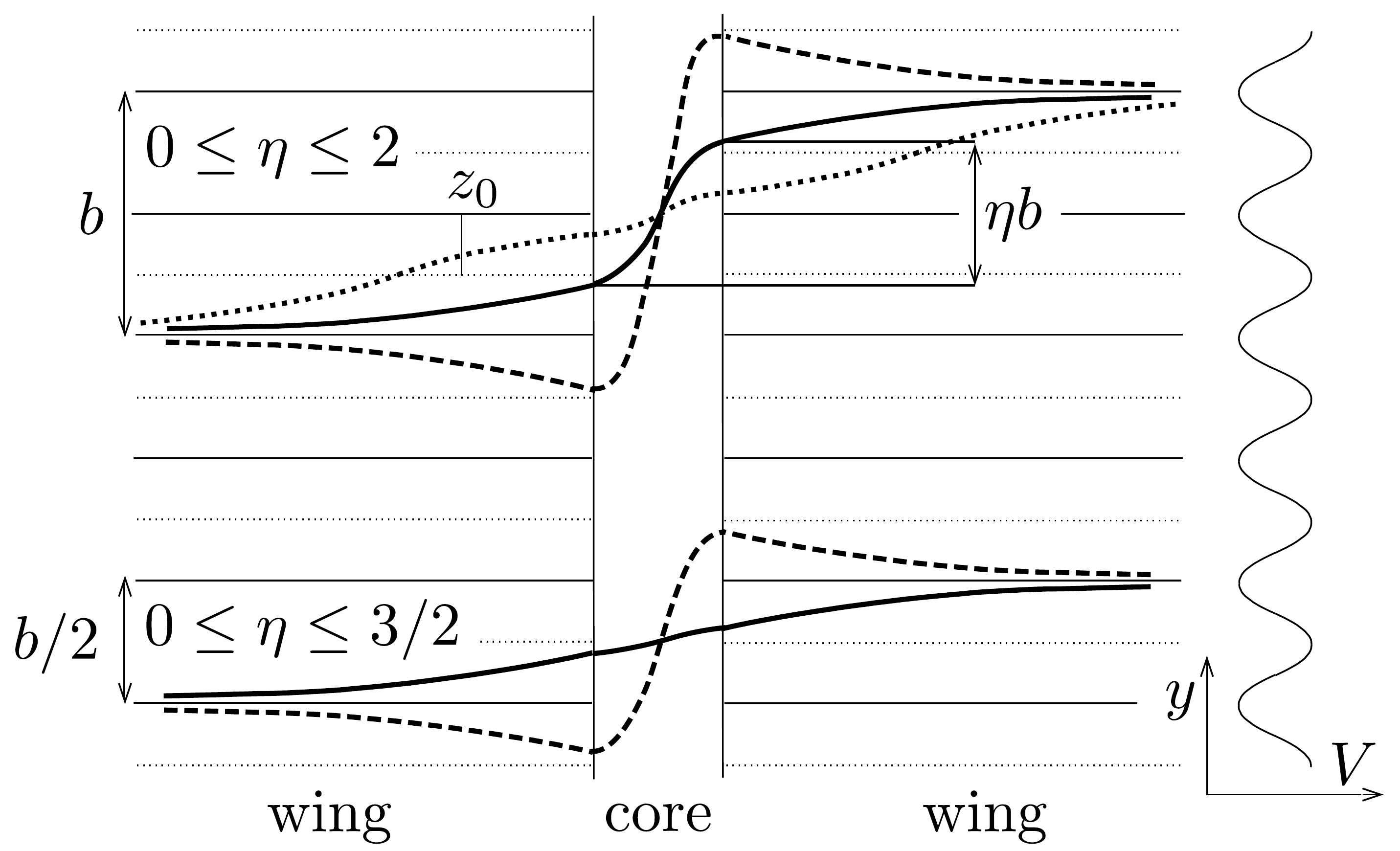}
\end{center}
\caption{\label{fig:wings} Sketch of core ($v_y^{\scriptscriptstyle (0)}$) and
wing ($v_y^{\scriptscriptstyle (1)}$) solutions for the $(1,k)$-solitons for
the cases $k=2$ (top) and $k=1$ (bottom). Solid lines are for
$\eta_{\scriptscriptstyle\rhd} < 1$ (top) and $\eta_{\scriptscriptstyle\rhd} <
1/2$ (bottom); dashed lines are for $\eta_{\scriptscriptstyle\rhd} \geq 1$
(top) and $\eta_{\scriptscriptstyle\rhd} \geq 1/2$ (bottom); the dotted line
(top) is for $\eta_{\scriptscriptstyle\rhd} < 1/2$ where the solution
approaches the shape of two consecutive solitons with shifts $b/2$ each.
Within the core region, the $v_x$-soliton drags the $v_y$-soliton across the
valley thereby binding the two wing solitons into a proper solution. Our
analytic solution ignores the finite width of the core
$v_y^{\scriptscriptstyle (0)}$.}
\end{figure}

Including the potential along $y$, we have to solve the equation
\begin{align}\label{eq:dSGy}
   \alpha^y_{\scriptscriptstyle\rhd} \partial_z^2 (2qv_y)
   &=\sin(2q v_y) - \alpha^y_{\scriptscriptstyle\rhd} 
   \eta_{\scriptscriptstyle\rhd} \partial_z(\partial_z 2qv_x)
\end{align}
with $\alpha^y_{\scriptscriptstyle\rhd}$ given in Eq.\ (\ref{eq:aybb}). Since
typically $\alpha^x_{\scriptscriptstyle \rhd} \ll \alpha^y_{\scriptscriptstyle
\rhd}$ the $v_x$-soliton is rather narrow and the expression $\partial_z v_x$
in (\ref{eq:dSGy}) can be replaced by a $\delta$-function, $\partial_z v_x
\approx -b\, \delta(z)$; the displacement field $v_y$ then can be found as the
solution of the Sine-Gordon equation $\alpha^y_{\scriptscriptstyle\rhd}
\partial_z^2 (2qv_y) =\sin(2q v_y)$ with the additional boundary condition
\begin{align}\label{eq:y-jump}
   v_y(0^+) - v_y(0^-) = \eta_{\scriptscriptstyle\rhd} b.
\end{align}
The solution $v_y$ thus splits into a combination of soliton solutions, the
0-order core part $v_y^{\scriptscriptstyle (0)} = -\eta_{\scriptscriptstyle
\rhd} v_x^{\scriptscriptstyle (0)}$ which we describe as a sharp jump
(\ref{eq:y-jump}) at the origin and a smooth `wing' part
$v_y^{\scriptscriptstyle (1)}$ of extended width
$\sqrt{\alpha^y_{\scriptscriptstyle \rhd}}$ taking the solution to the nearby
potential minimum, see Fig.\ \ref{fig:wings}. The explicit form of the wings
is given by the solutions
\begin{align} \nonumber
   v_y^{\scriptscriptstyle (1)}(z>0)
     &=\frac{kb}{2} \pm \frac{b}{\pi}\arctan[\exp[-(z+z_0)
                  /\sqrt{\alpha^y_{\scriptscriptstyle\rhd}}]],\\
   \label{eq:wings}
   v_y^{\scriptscriptstyle (1)}(z<0)
     &=\mp\frac{b}{\pi}\arctan[\exp[(z-z_0)
                  /\sqrt{\alpha^y_{\scriptscriptstyle\rhd}}]],
\end{align}
with
\begin{align} \label{eq:wings_z0}
   z_0 = -\sqrt{\alpha^y_{\scriptscriptstyle\rhd}}
   \ln\Bigl[\tan\Bigl(\frac{\pi}{2}\big|\eta_{\scriptscriptstyle\rhd}
                                         -k/2\big|\Bigr)\Bigr],
\end{align}
where the upper/lower signs apply to the cases $k/2\leq
\eta_{\scriptscriptstyle\rhd} \leq (k+2)/2$ and $\max[(k-2)/2,0]\leq
\eta_{\scriptscriptstyle\rhd} < k/2$, respectively; two representative cases
for $k=2$ and $k=1$ are illustrated in Fig.\ \ref{fig:wings}. Note that the
two wing solitons can cover at most a shift $\pm 2(b/2)$ along $y$; these
shifts, combined with the jump $\eta_{\scriptscriptstyle\rhd} b$ in the core
has to add up to the total shift $k b/2$ along $y$, what poses some
restrictions on the allowed angles $\theta$ defining the direction of the
$z$-axis.

It remains to determine the total line energy of the $(1,k)$ defect; inserting
the solutions $v_x \approx v_x^{\scriptscriptstyle (0)}$ (we drop the shape
correction $v_x^{\scriptscriptstyle (1)}$ induced by $v_y^{\scriptscriptstyle
(1)}$) and $v_y \approx v_y^{\scriptscriptstyle (0)} +
v_y^{\scriptscriptstyle (1)}$ into the expression for the line energy
$\varepsilon_\mathrm{s}$, we obtain the result
\begin{align} \label{eq:eps1K}
   \varepsilon_\mathrm{s}
   &=4nV\sqrt{\alpha^x_{\scriptscriptstyle \rhd}}
   +\frac{nV^2}{64\Delta}\, 2\int_{0}^{|u_0|} \!\!\!\! du
   \sqrt{2\alpha^y_{\scriptscriptstyle\rhd}} \sqrt{1\!-\!\cos{u}}
\end{align}
with $u_0 = 2 q v_y^{\scriptscriptstyle (0)} (0^-) = \mp 4
\arctan[\exp(-z_0/\sqrt{\alpha^y_{\scriptscriptstyle \rhd}})]$, see Eq.\
(\ref{eq:wings}); the factor 2 before the integral accounts for the two wings
at positive and negative $z$. The final results for the $k=1,2,3$ solitons
then are given by
\begin{align} \label{eq:eps1k}
   \varepsilon^{\scriptscriptstyle (1,k)}_\mathrm{s}
   &=4nV\sqrt{\alpha^x_{\scriptscriptstyle\rhd}}
   +\frac{nV^2}{8\Delta}\sqrt{\alpha^y_{\scriptscriptstyle\rhd}}
   \bigl(1\!-\!\cos[\pi(\eta_{\scriptscriptstyle\rhd}\!-\!k/2)]\bigr).
\end{align}
Obviously, the correction due to the wings vanishes when
$\eta_{\scriptscriptstyle\rhd} = k/2$, i.e., when the jump at the origin
induced by $v_x^{\scriptscriptstyle (0)}$ already matches the imposed boundary
condition set by the shift vector ${\bf d}_{j,k}$.  The critical values for
the substrate potentials $V_c$ and optimal angles $\theta$ are then again
found by (numerically) evaluating the points where the total line energies
$\varepsilon(V,\theta) = \varepsilon_\mathrm{s}(V,\theta) +
\varepsilon_\mathrm{d}(V,\theta)$ go to zero for the first time upon
decreasing $V$ and the results for the $(1,1)$ and $(1,3)$ solitons and for the
$(1,2)$ domain-wall are
\begin{align} \label{eq:vc11}
   V_c^{\scriptscriptstyle (1,1)} &\approx 0.0529\, e_{\scriptscriptstyle D}, \qquad
   \theta^{\scriptscriptstyle (1,1)} \approx 72.45^\circ,\\
   \label{eq:vc12}
   V_c^{\scriptscriptstyle (1,2)} &\approx 0.0536\, e_{\scriptscriptstyle D}, \qquad
   \theta^{\scriptscriptstyle (1,2)} \approx 79.15^\circ,\\
   \label{eq:vc13}
   V_c^{\scriptscriptstyle (1,3)} &\approx 0.0572\, e_{\scriptscriptstyle D}, \qquad
   \theta^{\scriptscriptstyle (1,3)} \approx 54.2^\circ,
\end{align}
with a slight advantage for the $(1,3)$ domain-wall but all values appreciably lower
than the result (\ref{eq:vc01_num}) for the $(0,1)$ domain-wall.
\begin{table}
\caption{\label{table:bb_vs_hex} Analytic (evaluated with the elastic theories for
the $bb$ rhombic and hexagonal (hex) lattices) and precise numerical (num) results for
$V_c^{\scriptscriptstyle (j,k)}$ and optimal angles
$\theta^{\scriptscriptstyle (j,k)}$.}
  \begin{center}
  \begin{tabular}{| c | c | c | c | c | c | c |}
  \hline
  $(j,k)$ & $V_c/e_{\scriptscriptstyle D}$, $bb$ & $\theta$, $bb$
          & $V_c/e_{\scriptscriptstyle D}$, hex & $\theta$, hex 
          & $V_c/e_{\scriptscriptstyle D}$, num & $\theta$ \\
  \hline
  $(0,1)$ & $0.0753$ & $90^\circ$    &  $-$     &   $-$        & $0.0741$ & $45^\circ$\\
  $(1,1)$ & $0.0529$ & $72.45^\circ$ & $0.0309$ & $58.7^\circ$ & $0.0382$ & $63.4^\circ$\\
  $(1,2)$ & $0.0536$ & $79.15^\circ$ & $0.0478$ & $47.8^\circ$ & $0.0501$ & $45^\circ$\\
  $(1,3)$ & $0.0572$ & $54.2^\circ$ & $0.0447$ & $45.3^\circ$    & $0.0544$ & $45^\circ$\\
  \hline
\end{tabular}
\end{center}
\end{table}

Since the soliton core of the $(1,k)$ defects are expected to have a structure
close to the free hexagonal lattice, we have calculated the critical
parameters using the elastic theory for the hexagonal lattice as well (the
core of the $(0,1)$ domain-wall resembles the $bb'$ rhombic lattice, hence trying a
hexagonal elastic theory is not promising). The results are summarized in
Table \ref{table:bb_vs_hex}; the values for $V_c$ calculated for the
hexagonal lattice are systematically smaller and appreciably different from
those obtained via the $bb$ elastic theory.  Once more, we conclude that a
numerical analysis is required in order to faithfully compare the energies of
the various topological defects and determine the type and critical potential
for the best candidate.

\subsection{Numerical analysis}\label{sec:s_sol_num_2}

Our numerical analysis for the $(0,1)$ and $(1,k)$ defects makes use of the
quantization of the topological vector charge ${\bf d}_{j,k}$: the return of
the particles to original lattice points after the passage of one soliton (or
two domain walls) allows us to analyze a periodic array of defects, thereby
reducing the problem of boundary effects and large system size. As a result,
rather than a variational, we will be able to perform a full relaxation of the
soliton shape and thus attain more precise results. In the following, we first
analyze the simplest situation, the $(0,1)$ domain-wall at $\theta = 90^\circ$
and then extend the discussion to other (discrete) angles.  Subsequently, we
study the $(1,k)$ defects for $k = 1,\, 2,\, 3$; the results are summarized in
Table \ref{table:bb_vs_hex} together with the analytic results. New elements
in the analysis will be introduced on the go and not repeated for every case.

\subsubsection{$(0,1)$ domain-wall at $\theta = 90^\circ$}\label{ssec:01}

In order to understand the impact of a finite system size and the interaction
between defects, we first analyze a crude model describing two $(0,1)$
domain-walls in terms of two missing rows of particles separated by
$2y_\mathrm{s}$ in a system of size $2Y$. We start from the $bb$ rhombic
lattice and consider two rows separated by the distance $y$.  Summing the
interaction over the $x$-coordinates (with $N_{\perp}$ the number of unit
cells of size $2b$ along $x$) and using Eq.\ (\ref{eq:sum_l}), we find the
interaction energy between the two rows at the distance $y$
\begin{align}\nonumber
   E_{\scriptscriptstyle\rhd}^\mathrm{int} (y)
   &= N_\perp \, e_{\scriptscriptstyle D}
   \sum_{l} \frac{b^3}{[(2bl)^2 + y^2]^{3/2}}
   \approx N_\perp \,e_{\scriptscriptstyle D} \frac{b^2}{y^2},
\end{align}
where we have ignored corrections due to the sum over $\bar{l}$ in
(\ref{eq:sum_l}) (this approximation, i.e., replacing the sum over $l$ by an
integral, is valid at large $y/b \gg 1$).  With the shift vector ${\bf
d}_{0,1} = (0, b/2)$, we can describe the two domain walls by shifting all
rows with $y > y_\mathrm{s}$ ($y < - y_\mathrm{s}$) up (down) by a distance
$b/2$.  Summing the interaction energies over all rows including these shifts
and substrating the sum without shifts we obtain the interaction part of the
two defect lines in the form (we devide by $N_\perp 2b$ to obtain a line
energy)
\begin{align}\label{eq:e2s_est_1}
   \varepsilon
   \approx
   \frac{e_{\scriptscriptstyle D}}{a_1} \frac{4}{2}
   \Bigl[{\sum_{j\neq j'}}^\mathrm{\scriptscriptstyle 2s}-\sum_{j\neq j'} \Bigr]
   \frac{1}{(j-j')^2},
\end{align}
with $y = j\, b/2$, $b/2$ the distance between rows along the $y$ direction
and the factor $1/2$ avoids double counting of rows. In Eq.\
\eqref{eq:e2s_est_1} the sum $\sum^\mathrm{\scriptscriptstyle 2s}$ has to be
taken between $\pm j_Y=\pm(2Y/b+1)$ but with $j,j' \neq \pm j_\mathrm{s} = \pm
2 y_\mathrm{s}/b$ (we consider a symmetrized situation which provides equal
leading corrections from the two solitons), while the second sum goes over
$j,j'=-(j_Y-1),\dots,(j_Y-1) =- 2Y/b,\dots,2Y/b$.  Then the following terms
survive the cancellation in the difference of sums,
\begin{align}\label{eq:e2s_est_2}
   \varepsilon
   &\approx \frac{4 e_\mathrm{\scriptscriptstyle D}}{a_1}
   \biggl\{\sum_{j\neq \pm j_\mathrm{s}} \biggl[\frac{1}{(j_Y-j)^2}+
                                     \frac{1}{(j_Y+j)^2}\biggr] \\
   & \nonumber
   \qquad
   -\sum_{j}\biggl[\frac{1}{(j_\mathrm{s}-j)^2}+\frac{1}{(j_\mathrm{s}+j)^2}\biggr]
   -\frac{1}{(2j_Y)^2} + \frac{1}{(2j_\mathrm{s})^2}\biggr\},
\end{align}
where the additional factor 2 arises from interchanging the role of $j$ and
$j'$ and the last two terms correct for double counting the interactions
between the `adatoms' at $\pm j_Yb$ and `vacancies' at $\pm j_\mathrm{s} b$;
self-energy terms always have to be dropped from the sums.  Replacing the sums
by integrals, e.g.,
\begin{align}\label{eq:sum_to_int}
   \sum_j\frac{1}{(j_\mathrm{s} \pm j)^2} &\approx
   \biggl[\int_{-j_Y+1}^{\mp j_\mathrm{s}-1} + \int_{\mp j_\mathrm{s}+1}^{j_Y-1}\biggr] \,
   \frac{dx}{(j_\mathrm{s}\pm x)^2}\\
   \nonumber
   &=2-\frac{1}{j_Y-1-j_\mathrm{s}}-\frac{1}{j_Y-1+j_\mathrm{s}},
\end{align}
we can evaluate Eq.\ \eqref{eq:e2s_est_2} and obtain the asymptotic behavior
(we assume $j_\mathrm{s} \ll j_Y$ and drop terms $\propto j_Y^{-2}$)
\begin{align}\label{eq:e_2s_asym}
   \varepsilon
   &\approx \varepsilon^\infty +\frac{e_{\scriptscriptstyle D}}{a_1}
   \biggl(\frac{4b}{Y+y_\mathrm{s}}+\frac{4b}{Y-y_\mathrm{s}}-\frac{2b}{Y}
   +\frac{b^2}{4y_\mathrm{s}^2} \biggr).
\end{align}
The result (\ref{eq:e_2s_asym}) shows that boundary effects decay with inverse
system size $\propto 1/2Y$, while the interaction between defects decays
faster, as the inverse square of the defect separation $2 y_\mathrm{s}$. It is
the long-range interaction $\propto 1/R^3$ between particles that enhances the
defect interaction, from the usual exponential behavior (see
(\ref{eq:g_Vlarge})) to an inverse-square law \cite{HaldaneVillain_81}; the
non-dispersive elastic theories did not catch this effect in our previous
analytic studies. Regarding our numerical studies, we learn that analyzing
periodic systems allows us to avoid boundary effects which decay only slowly
$\propto 1/Y$; furthermore, working with a system size $Y \sim 100\, b$, the
residual interaction between solitons contributes a small error of order
$10^{-4} e_{\scriptscriptstyle D}/a_1$ to the isolated defect energy.  These
small system sizes then allow us to fully relax the defect
shapes.\cite{mathematica}
\begin{figure}[h]
\begin{center}
\includegraphics[width=6cm]{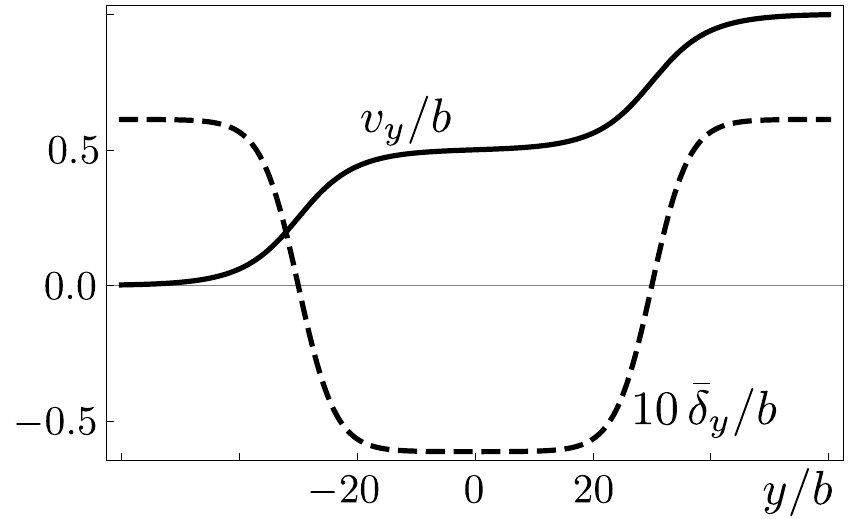}
\end{center}
\caption {\label{fig:horsol} Shape of two $(0,1)$-solitons $v_y(y)/b$ as well
as internal distortion $\bar\delta_y(y)/b$ for $V=0.075\,_{\scriptscriptstyle
D}$. The soliton width is $\sqrt{\alpha_y}\approx 6.1\,b$. The distortion
field $\bar\delta_y(y)$ (dashed) is expanded by a factor 10 for better
visibility.}
\end{figure}

\begin{figure}[h!]
\begin{center}
\includegraphics[width=6cm]{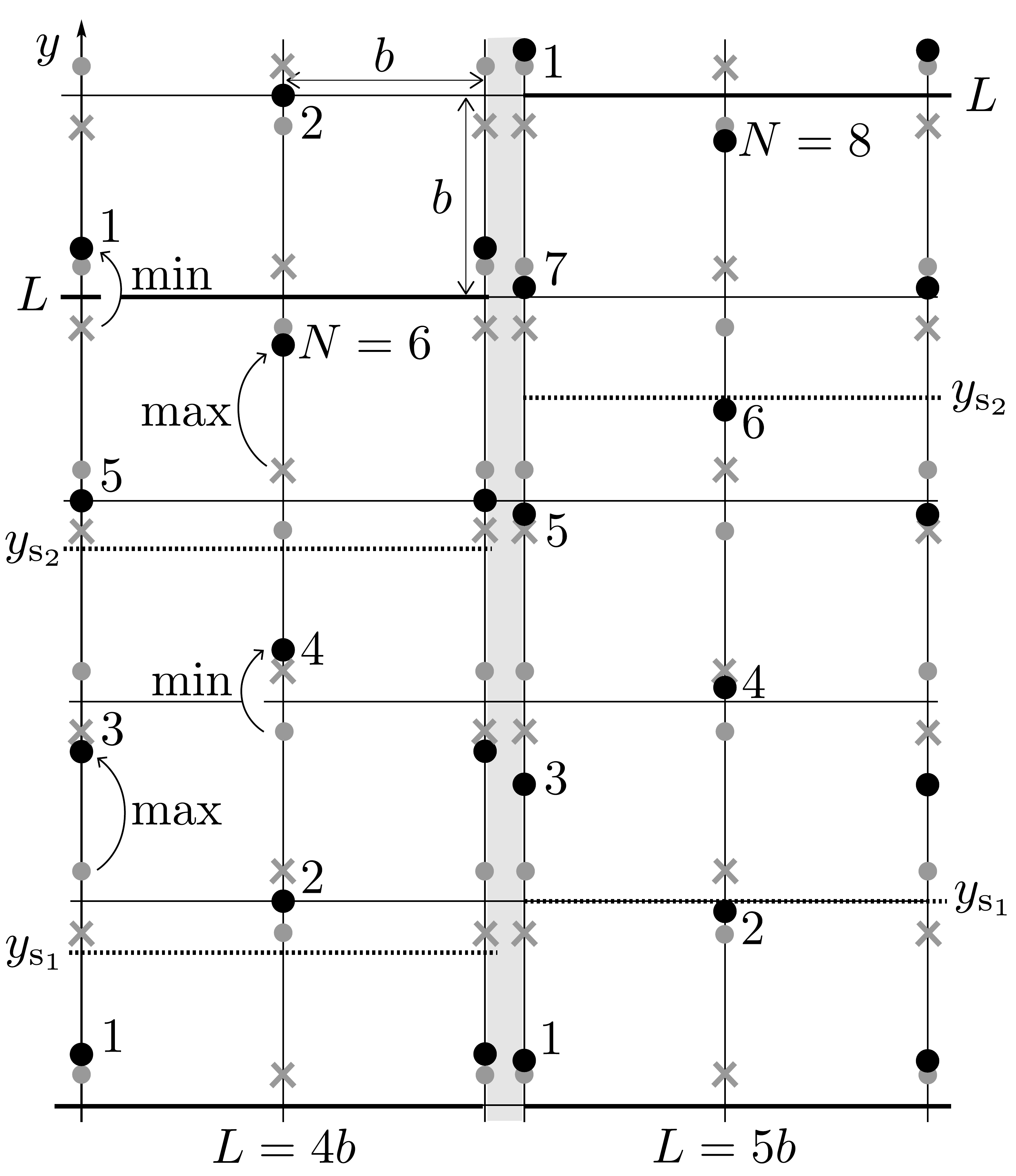}
\end{center}
\caption{\label{fig:period90} Displaced particles (black dots) for a
period-doubled lattice with two domain walls at $y_{s_1}=L/4-b/4$ and
$y_{s_2}=3L/4-b/4$. The domain walls take the period-doubled lattice from the
twin A (grey points) to the twin B phase (grey crosses) and back to the
original lattice. Periodicity is trivially achieved with $L/b \in \mathbb{N}$;
examples are shown for $L = 4b$ with 6 particles (left) and $L = 5b$ (right).
Particles with odd index $j$ first cross a substrate maximum (max) shifting by
$b/2 + \bar\delta_y$ and then a minimum (min, shifting by $b/2 -
\bar\delta_y$) while particles with even $j$'s have the reversed order.}
\end{figure}

The implementation with periodic boundary conditions profits from an
alternative particle enumeration with only one index (at fixed strip index
$l$). The particle positions of the $bb$ rhombic lattice are chosen as
\begin{align} \label{eq:stripe_coord}
   {\bf R}^{\scriptscriptstyle \rhd}_{lj}=\binom{l a_1+x_j}{y_j},
\end{align}
with $x_j/b=(1+(-1)^j)/2$ referring to the alternating columns in the doubled
unit cell and $y_j/b=j/2-1/4$. The coordinates of the period-doubled lattice
are ${\bf R}^\mathrm{pd}_{lj} = {\bf R}^{\scriptscriptstyle \rhd}_{lj}
+(0,(-1)^j \bar\delta_y/2)$ with $\bar\delta_y = (b/\pi)\arcsin(V/8\Delta)$, see
Eq.\ (\ref{eq:pd}), and including two domain-walls at $y_{s_1}$ and $y_{s_2}$
with the displacement and distortion fields $v_y(y)$ and $\bar\delta_y(y)$
(see Fig.\ \ref{fig:horsol})
\begin{align} \label{eq:01dw}
   v_y(y) &= (b/\pi)\bigl\{(\arctan[\exp[(y-y_{\mathrm{s}_1})/\sqrt{\alpha_y}]]
   \\\nonumber &\qquad\qquad
             +\arctan[\exp[(y-y_{\mathrm{s}_2})/\sqrt{\alpha_y}]]\bigr\}, \\
   \label{eq:01d}
   \bar\delta_y(y) &= (b/\pi)\arcsin[(V/8\Delta)\cos[2\pi v_y(y)/b]],
\end{align}
we obtain the coordinates of the particles in the defected lattice ${\bf
R}^s_{lj} = {\bf R}^{\scriptscriptstyle \rhd}_{lj} + (0,v_y(y_j) + (-1)^j
\bar\delta_y(y_j)/2)$.  Working with periodic boundary conditions, the cell
size $L$ has to be chosen such that the boundaries match.  For the $(0,1)$
domain-wall and $\theta=90^\circ$ this is easily satisfied for $L/b \in
\mathbb{N}$ and two domain-walls per period placed at the positions
$y_{s_1}=L/4-b/4$ and $y_{s_2}=3L/4-b/4$, taking the lattice from the twin A
phase to the twin B phase and again back to twin A, see Fig.\
\ref{fig:period90}.

The calculation of the interaction energy \eqref{eq:sum_int_res} is modified
by splitting the sum over particle distances $y_{jj'} = y_j - y_{j'}$ in a sum
over particle distances within one period and then extend the sum over
periodic images. This corresponds to changing the $2/\beta^2$ term in the
interaction (\ref{eq:sum_l}) to $\sum_k 2/(\beta + kL/a_1)^2 = [2(\pi a_1/L) /
\sin(\pi\beta a_1/L)]^2$ and replace the argument in the correction terms by
$\beta^\mathrm{min} = \min [\beta, L/a_1-\beta]$ (due to the exponential decay
of $K_1(y)$ at most the image in the neighboring cell might contribute). The
interaction energy (\ref{eq:sum_int_res}) finally assumes the form (with $\ell
= L/a_1$ and $\alpha^\mathrm{s}_{jj'}$, $\beta^{\mathrm{s}}_{jj'}$ the
relevant scaled difference coordinates, cf.\ Sec.\ \ref{ssec:num_ana})
\begin{align}\label{eq:sum_int_res_01}
   E_\mathrm{s}^\mathrm{int} &=
   \frac{N_{\perp} D}{a_1^3}
   \sum_{j'=1}^N \biggl\{ \frac{4}{\ell^2}\zeta(2)+\!\!\!
   \sum_{j = 1<j'} \biggl[\frac{2 (\pi/\ell)^2}
   {\sin^2(\pi \beta^{\mathrm{s}}_{jj'}/\ell)}\\
   &\nonumber
   +8\pi\sum_{\bar{l}>0}  \bar{l} \,
   \cos(2\pi \bar{l}\alpha^\mathrm{s}_{jj'})
   \frac{K_1(|2\pi \bar{l} \beta^{\mathrm{s,min}}_{jj'}|)}
   {|\beta^{\mathrm{s,min}}_{jj'}|}\biggr] \biggr\},
\end{align}
where the first term accounts for the interaction between a particle and its
periodic images.  The accommodation of the substrate energy
$E^\mathrm{sub}_\mathrm{s}$, see Eq.\ (\ref{eq:sum_sub}), to the new situation
is straightforward and the area change associated with the two domain-walls
with shifts $b/2$ is $\delta A = 2 a_1 b/2 = 2 b^2$.
\begin{figure}[h!]
\begin{center}
\includegraphics[width=5.5cm]{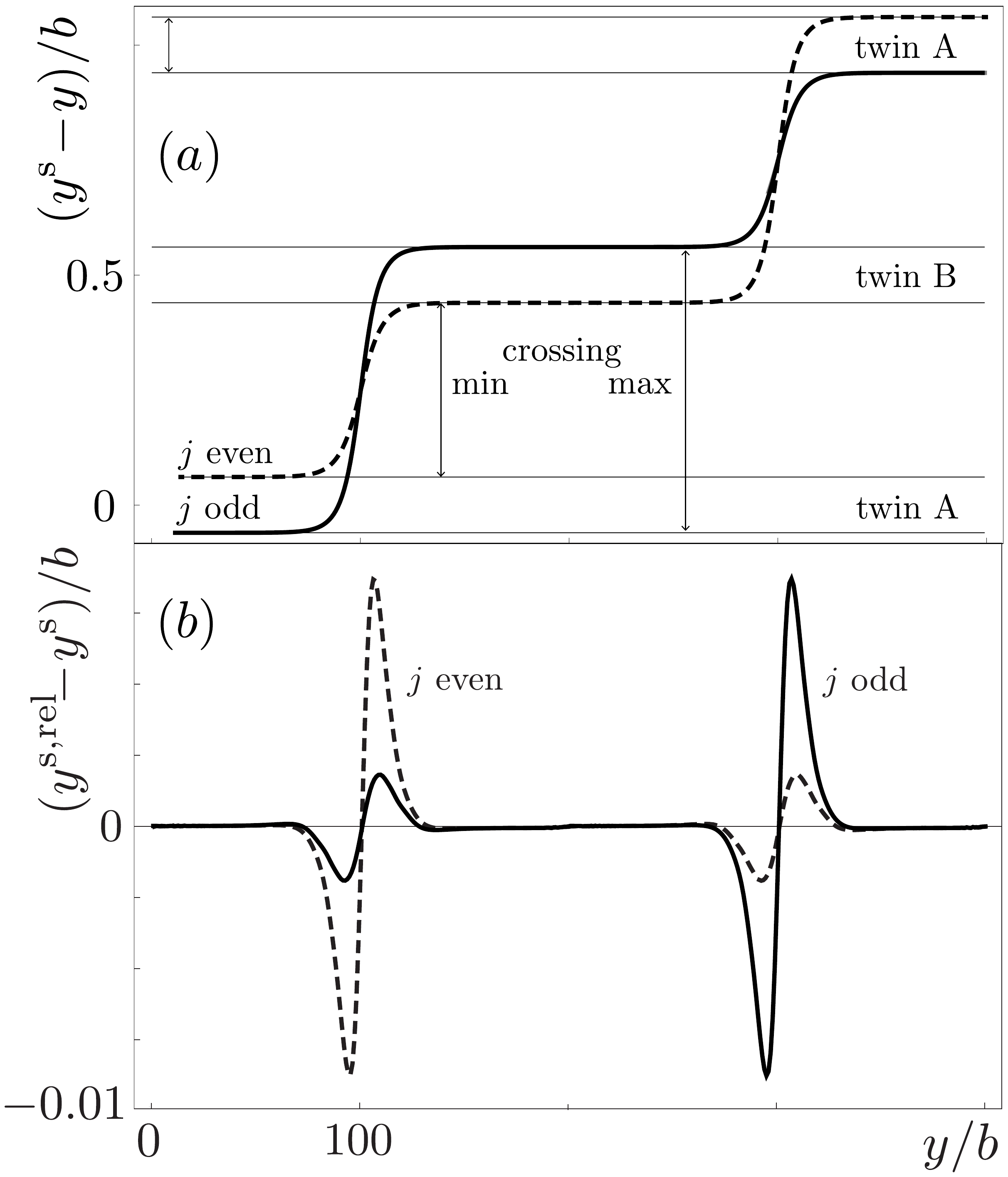}
\end{center}
\caption{\label{fig:ysol90_relax_comb} (a) The displacement $y^\mathrm{s}-y$
associated with two $(0,1)$ domain-walls evaluated at $V =
V_c^{\scriptscriptstyle (0,1)} = 0.0732\, e_{\scriptscriptstyle D}$ after
relaxation. When going from twin A to twin B, particles with odd index $j$
(solid lines, shift by $b/2 +\bar\delta_y$) have to overcome a substrate
maximum while the even $j$'s (dashed lines, shift by $b/2 -\bar\delta_y$)
cross a substrate minimum (and vice versa from twin B back to twin A, see also
Fig.\ \ref{fig:period90}). (b) Relaxation of the two domain-walls during
15 iterations. The maximal shift is on the level of $0.01 \, b$, symmetric
around the defect center and more than 4.5 times larger when particles cross a
substrate minimum than when crossing a substrate maximum.}
\end{figure}

Our numerical study involves a system size $L = 401 \, b$ and 15 relaxational
steps, resulting in a precision of $a_1 \delta \varepsilon /
e_{\scriptscriptstyle D} \sim 10^{-4}$ (note that the precise shift
$\bar\delta_y$ has to be found by numerical relaxation as well). The initial
analytic solution with width $\sqrt{\alpha_{\scriptscriptstyle \rhd}^y}
\approx 6.1\, b$ relaxes only minimally (not visible in Fig.\
\ref{fig:ysol90_relax_comb}(a); the relaxation itself shown in Fig.\
\ref{fig:ysol90_relax_comb}(b) is of the order of $10^{-2} b$ and is larger
when particles cross a minimum). The unrelaxed critical potential
$V_c^{\scriptscriptstyle (0,1),\mathrm{ur}} \approx 0.0730\,
e_{\scriptscriptstyle D}$ increases by a small amount to the relaxed value
\begin{align} \label{eq:Vc^01_90}
   V_c^{\scriptscriptstyle (0,1)} \approx 0.0732\, e_{\scriptscriptstyle D}, 
   \qquad \theta = 90^\circ.
\end{align}

\subsubsection{$(0,1)$ domain-wall at other angles}\label{ssec:01_oa}

\begin{figure}[h!]
\begin{center}
\includegraphics[width=8cm]{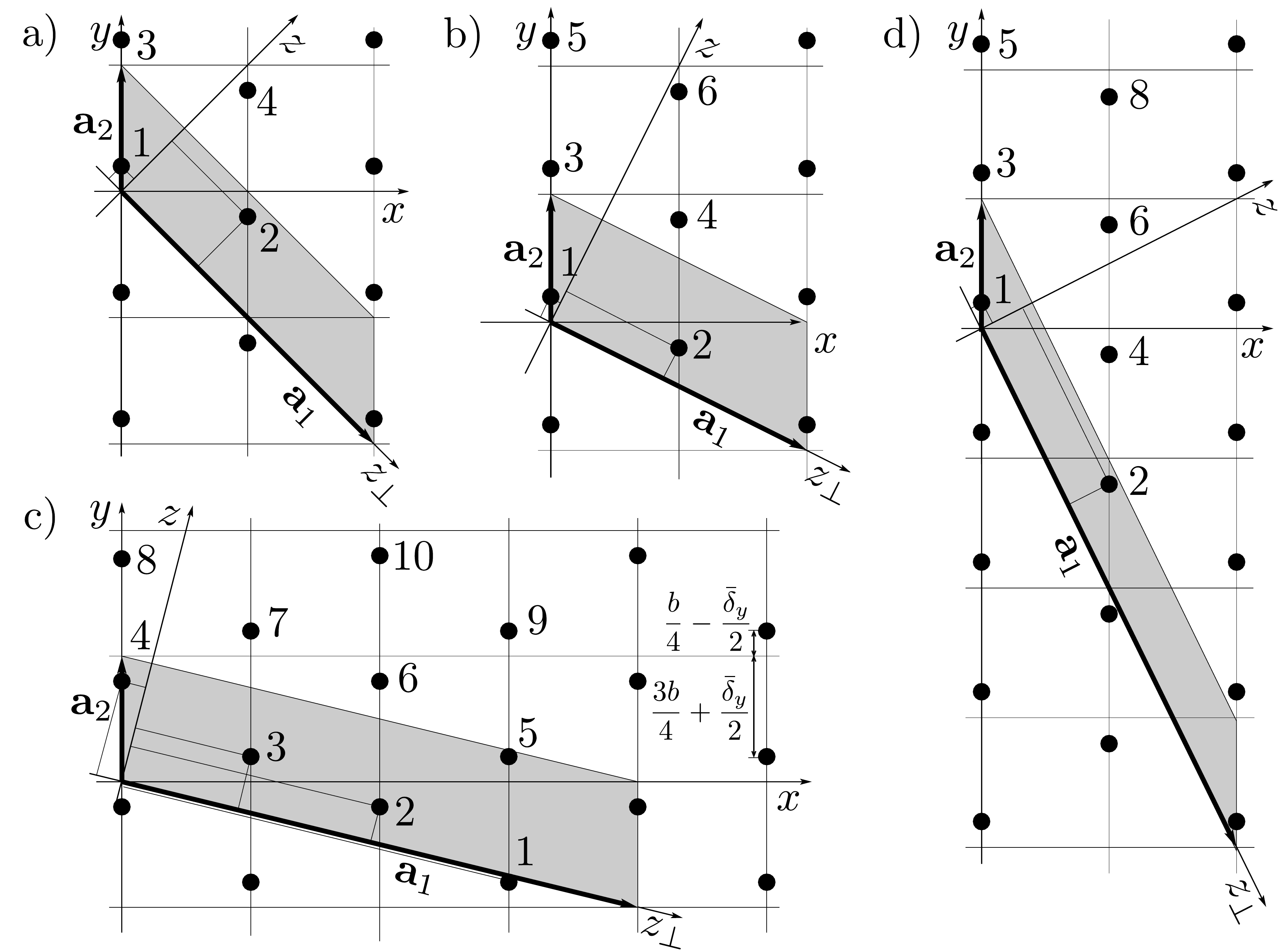}
\end{center}
\caption{\label{fig:ucoverview} Coordinates $z_\perp$ and $z$ for a) $\theta =
45^\circ$ ($(m,n) = (2,2)$), b) $\theta \approx 63.4^\circ$ ($(m,n) = (2,1)$),
c) $\theta \approx 76.0^\circ$ ($(m,n) = (4,1)$), and d) $\theta \approx
26.6^\circ$ ($(m,n) = (2,4)$) describing the period-doubled lattice (shown as
black dots) and an array of domain-walls evolving along $z$. For such uniaxial
displacement fields along the $z$-axis, the structure remains invariant under a
translation by the vector ${\bf a}_1$; choosing the unit cells (grey areas)
with 2(4) particles allows for the summation of the interactions along
$z_\perp$ with period $a_1$.}
\end{figure}

The analytic result Eq.\ (\ref{eq:vc01}) for the critical substrate potential
of the $(0,1)$ domain-wall depends weakly on angle, with a flat maximum at
$\theta = 90^\circ$. Here, we find the angle dependence of
$V_c^{\scriptscriptstyle (0,1)}$ for discrete angles $\theta=\arctan{(m/n)}$
belonging to small Miller indices $(m,n)$, using the methodology in Sec.\
\ref{ssec:num_ana} adapted to the $bb$ rhombic lattice and making use of the
numerical relaxation of the defect shape as in Sec.\ \ref{ssec:01}. The four
cases analyzed below are illustrated in Fig.\ \ref{fig:ucoverview}. The change
in area $\delta A$ (or `charge' $Q = -\delta A/b^2$) associated with a
domain-wall depends on the angle $\theta$,
\begin{align} \label{eq:Amn01}
  \delta A_{m,n}^{\scriptscriptstyle(0,1)}=a_1\,({\bf d}_{0,1}\cdot 
  {\hat{\bf e}_z}(\theta))=(m/2)b^2,
\end{align}
and describes defects diluting the particle lattice (as opposed to the
compressive PT soliton in Sec.\ \ref{ssec:num_ana}).

\begin{figure}[h!]
\begin{center}
\includegraphics[width=7.5cm]{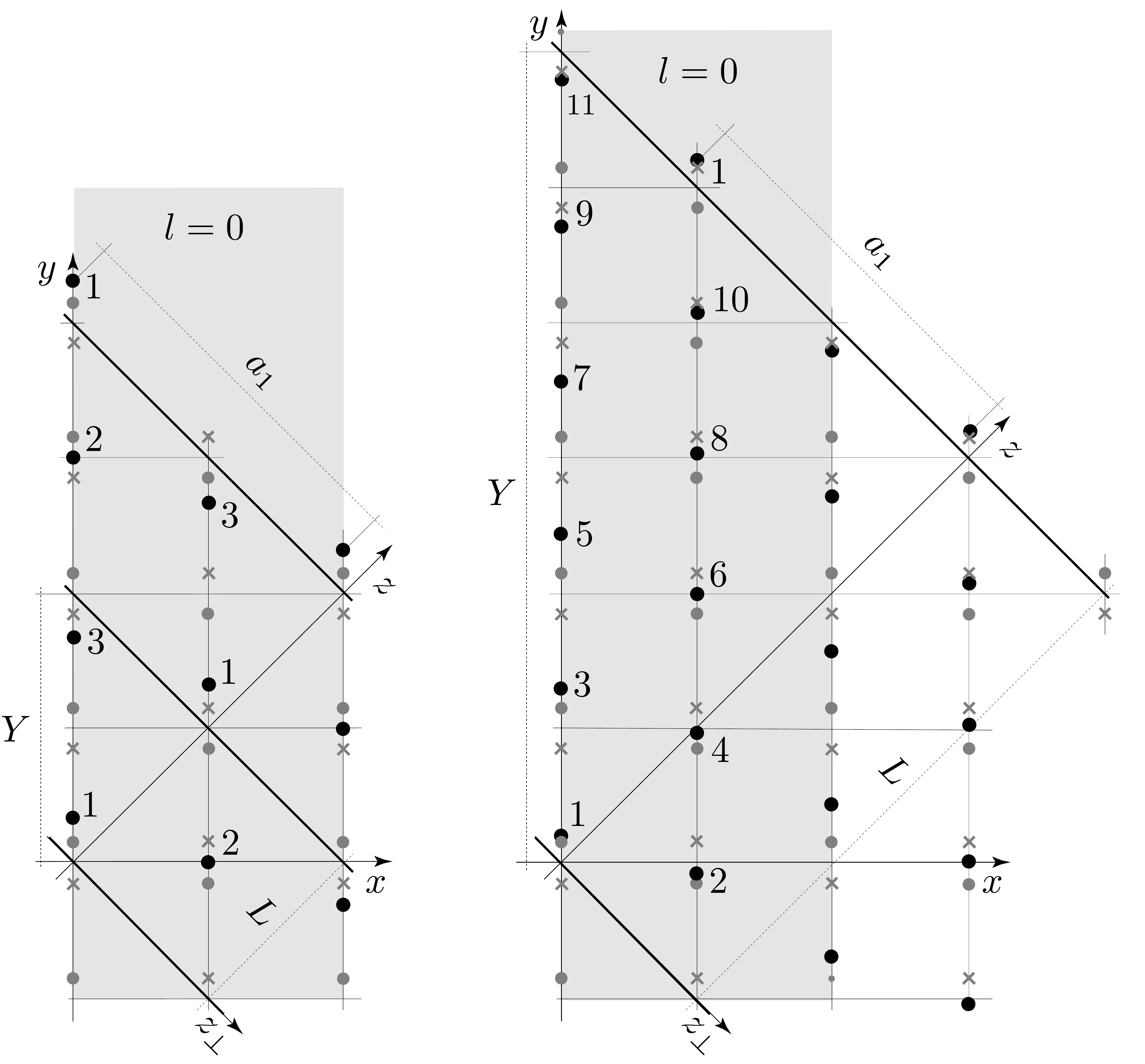}
\end{center}
\caption {\label{fig:01sol45_PBC} Particle positions for the $(0,1)$-soliton
at $\theta=45^\circ$. One soliton ($n_\mathrm{s}=1$) per period $L=\sqrt{2}
(2k+1)\, b$ with $k\in\mathbb{N}$ already ensures correct matching at the
boundaries as illustrated here for the cases $k=0$ (left, shown are two
periods of length $L=\sqrt{2}\,b$ containing $N_\mathrm{s}=3$ particles) and
$k=1$ (right, shown is one period of length $L=3\sqrt{2}\,b$ with
$N_\mathrm{s}=11$ particles).  The shaded area corresponds to the $l=0$-strip.
The particle number $N_\mathrm{s}$ per period is related to $L$ via the
`vertical period' $Y$ (and the charge $Q$ per soliton), $L=Y\sin{\theta}$ and
$N_\mathrm{s}=2Y/b+n_\mathrm{s}Q$. Grey dots and crosses denote twin A and
twin B lattice sites, respectively.}
\end{figure}

An important but not straightforward element is the choice of the periodic
supercell.  Correct matching after the period $L$ (or number of particles $N$)
requires that
\begin{align} \label{eq:PBCzp}
   z_{\perp,N+1}&=z_{\perp,1}+pa_1,\qquad z_{N+1}=z_1+L
\end{align}
for some integer $p$. Note that this condition does not require that equal
twins match up after one period, hence the number of defects per period can be
one or two. Figure \ref{fig:01sol45_PBC} illustrates two cases for $\theta =
45^\circ$ with $N_\mathrm{s} = 3$ and $N_\mathrm{s} = 11$ particles per unit
cell, where $N_\mathrm{s} = N_\mathrm{pd} +n_\mathrm{s} Q$ is the particle
number per supercell in the presence of $n_\mathrm{s}$ defects with `charge'
$Q$ and $N_\mathrm{pd}$ is the particle number in the undistorted supercell.
For the angle $\theta = 45 ^\circ$ the allowed supercell lengths are given by
$L/b = \sqrt{2} (2k+1)$ with $k \in\mathbb{N}$; the allowed values for the
other angles are given in Table \ref{table:01_par_Vc}.

The angle $\theta \approx 63.4^\circ$ associated with the Miller indices
$(2,1)$ involves an additional subtlety: indeed, for this angle the summation
over $l$ in Eq.\ \eqref{eq:sum_l} can lead to (nearly) coinciding particle
rows where $\beta$ becomes small or even vanishes. This is the case when the
internal distortion field $\bar \delta_y$ crosses zero within a domain-wall
and the separation $z^\mathrm{s}_{2q} - z^\mathrm{s}_{2q-1}$ between the
particles $j=2q-1$ and $j'=2q$ vanishes. This spurious divergence can be dealt
with in different ways, e.g., with the help of the Euler-Maclaurin formula
\begin{align}\nonumber
  &\sum_{l=-\infty}^{\infty} \frac{1}{[(l+a)^2 + \beta^2]^{3/2}} \approx
  \sum_{l=-N+1}^{N-1}\frac{1}{[(l+a)^2 + \beta^2]^{3/2}}
   \\ \nonumber
  &{
   +\frac{1}{\beta^2}\bigl(1\!-\!\frac{N+a}{[(N\!+\!a)^2\! +\! \beta^2]^{1/2}}\bigr)
   \!+\!\frac{1}{\beta^2} \bigl(1\!-\!\frac{N-a}{[(N\!-\!a)^2\! +\! \beta^2]^{1/2}}\bigr)}
    \\ \nonumber
 &\quad{
   +\frac{1}{2} \frac{1}{[(N+a)^2 + \beta^2]^{3/2}}
   +\frac{1}{2} \frac{1}{[(N-a)^2 + \beta^2]^{3/2}}}
   \\ \nonumber
  &\quad{
   +\frac{1}{4}\frac{N+a}{[(N+a)^2 + \beta^2]^{5/2}}
   +\frac{1}{4}\frac{N-a}{[(N-a)^2 + \beta^2]^{5/2}}} + \dots
\end{align}
where in the limit $\beta \to 0$ the first two correction terms should be
replaced by $1/[2(N\pm a)^2]$.
\begin{table}
\caption{\label{table:01_par_Vc} Parameters [cell sizes $L$
($k\in\mathbb{N}$), number of defects $n_\mathrm{s}$ per cell, and charge $Q$
per defect (and length $a_1$)] for the numerical analysis of the $(0,1)$
domain-wall for discrete angles $\theta$ and numerical results for the
critical substrate strengths $V_c^{\scriptscriptstyle(0,1)}$ for the $(0,1)$
domain-wall before (superscript `ur = unrelaxed') and after relaxation.  $L$ and
$n_\mathrm{s}$ are not independent quantities and may be chosen differently.}
  \vskip 3 pt
  \begin{center}
  \begin{tabular}{| c || c | c | c | c | c |}
  \hline
$ \theta$ &  $L/b$ & $n_\mathrm{s}$ & $Q$ & $V_c^{\scriptscriptstyle(0,1),\mathrm{ur}}/
               e_{\scriptscriptstyle D}$
          &  $V_c^{\scriptscriptstyle(0,1)}/
               e_{\scriptscriptstyle D}$\\
            \hline
  $26.6^\circ$ & $\sqrt{5}(2k+1)$  & 2 & $-1$ & 0.0684 & 0.0730 \\
  $45^\circ$   & $\sqrt{2}(2k+1)$  & 1 & $-1$ & 0.0718 & 0.0741 \\
  $63.4^\circ$ & $2\sqrt{5}(2k+1)$ & 2 & $-1$ & 0.0727 & 0.0735 \\
  $76.0^\circ$ & $\sqrt{17}(2k+1)$ & 1 & $-2$ & 0.0729 & 0.0733 \\
  $90^\circ$   & $2k+1$            & 2 & $-1$ & 0.0730 & 0.0732 \\
  \hline
\end{tabular}
\end{center}
\end{table}
\begin{figure}[h!]
\begin{center}
\includegraphics[width=7cm]{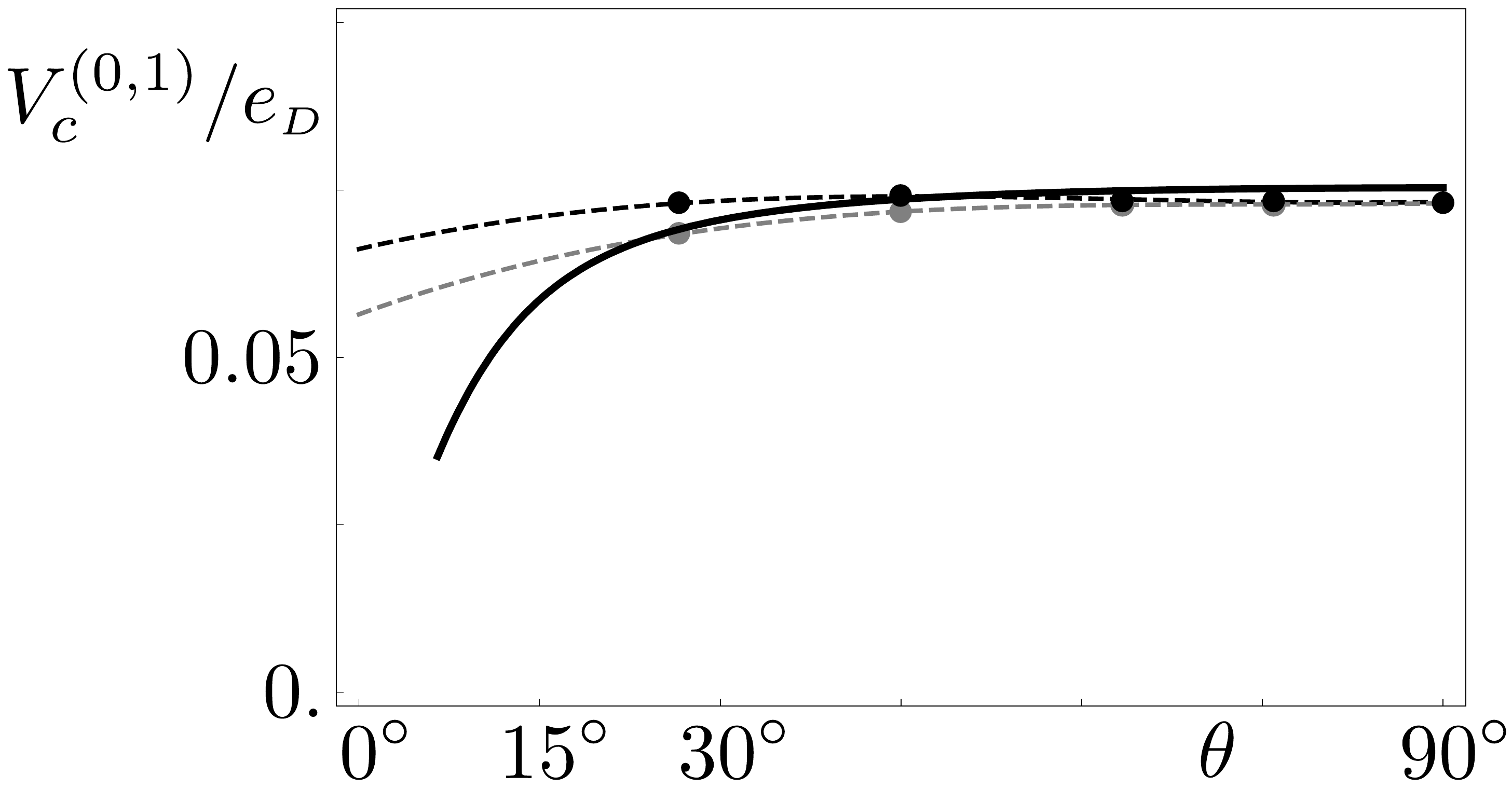}
\end{center}
\caption{\label{fig:01sol_Vc} The critical substrate strength
$V_c^{\scriptscriptstyle (0,1)}$ for the $(0,1)$ domain-wall as a function of
$\theta$. Upon relaxation the maximum shifts from $\theta=90^\circ$ (grey
points are numerical results using the analytic soliton shape) to
$\theta=45^\circ$; the angle dependence of the relaxed configuration (black
dots) is very flat. The dashed lines are guides to the eye. The solid line
shows the analytic result from the continuum elastic description, see Eq.\
\eqref{eq:vc01}.}
\end{figure}
Accounting for all these measures, the optimal domain-wall shapes are found
numerically and the critical substrate potentials can be determined.  The
results are summarized in Table \ref{table:01_par_Vc} and are illustrated in
Fig.\ \ref{fig:01sol_Vc}. Quite surprisingly, the optimal domain-wall does not
appear at the symmetric angle $\theta = 90^\circ$ but rather far away near
$\theta \approx 45 ^\circ$, an angle that is unrelated to the symmetry axes
of the parent crystal.

\subsubsection{$(1,k)$ defects}\label{ssec:1k}

\begin{figure}[h]
\begin{center}
\includegraphics[width=6cm]{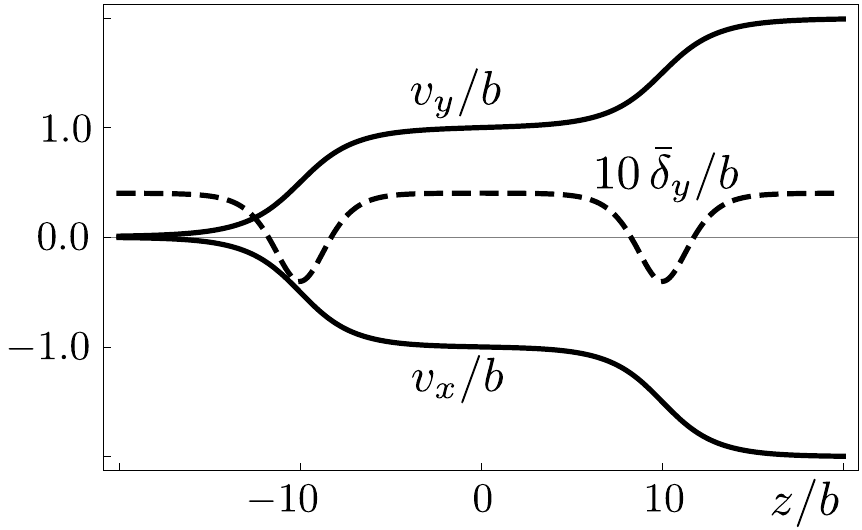}
\end{center}
\caption{\label{fig:diagsol} Displacements $v_x(z)/b$ and $v_y(z)/b$
associated with the $(1,2)$-soliton and the corresponding internal distortion
field $\bar{\delta}_y(z)/b$ for $V=0.05\,e_{\scriptscriptstyle D}$. The
soliton widths are $\sqrt{\alpha_x} \approx 1.9\,b$ and
$\sqrt{\alpha_y}\approx 6.6\,b$; within the soliton core, $v_y$ rapidly
changes by $\eta b\approx 0.94\,b$, leaving only small wing amplitudes. The
distortion field $\bar{\delta_y}$ (dashed) is expanded by a factor 10 for
better visibility.}
\end{figure}

Next, we analyze defects with a displacement field that includes a component
along the $x$-direction. We start out with the $(1,2)$ domain-wall shifting
the particles by a vector $(-b,b)$. The analytic expressions for the
displacements describing the body (\ref{eq:body}) and wings (\ref{eq:wings})
of the defect are illustrated in Fig.\ \ref{fig:diagsol}; furthermore, the
intracell distortion $\bar\delta_y(z)$ as given by Eq.\ (\ref{eq:01d}) is
quite different from the one of the $(0,1)$ domain wall as $v_y$ increases by
$b$ across one domain wall (instead of $b/2$ in the $(0,1)$ defect). As a
result, the deformation is maximal and of opposite sign in the core and
returns to its original value behind the defect, see Fig.\ \ref{fig:diagsol},
while for the $(0,1)$ domain-wall this deformation vanished in the defect
center and finally changed sign across the defect, thereby taking the
particles to a different twin, see Fig.\ \ref{fig:horsol}.  On the contrary,
it is the shift along $x$ which leads to the different twin after crossing the
$(1,2)$ defect.

Another peculiarity of the $(1,2)$ domain-wall is its areal change $\delta
A_{m,n}^{\scriptscriptstyle(1,2)}=a_1\,({\bf d}_{1,2}\cdot {\hat{\bf
e}_z}(\theta))=(m-n)b^2$ or `charge' $Q = n-m$ which changes sign at $\theta =
45^\circ$---the $45^\circ$ domain-wall then is uncharged, while the one at
$\theta \approx 26.6^\circ$ with $(m,n) = (2,4)$ is a compression defect with
$Q = 2$.
\begin{figure}[h!]
\begin{center}
\includegraphics[width=7cm]{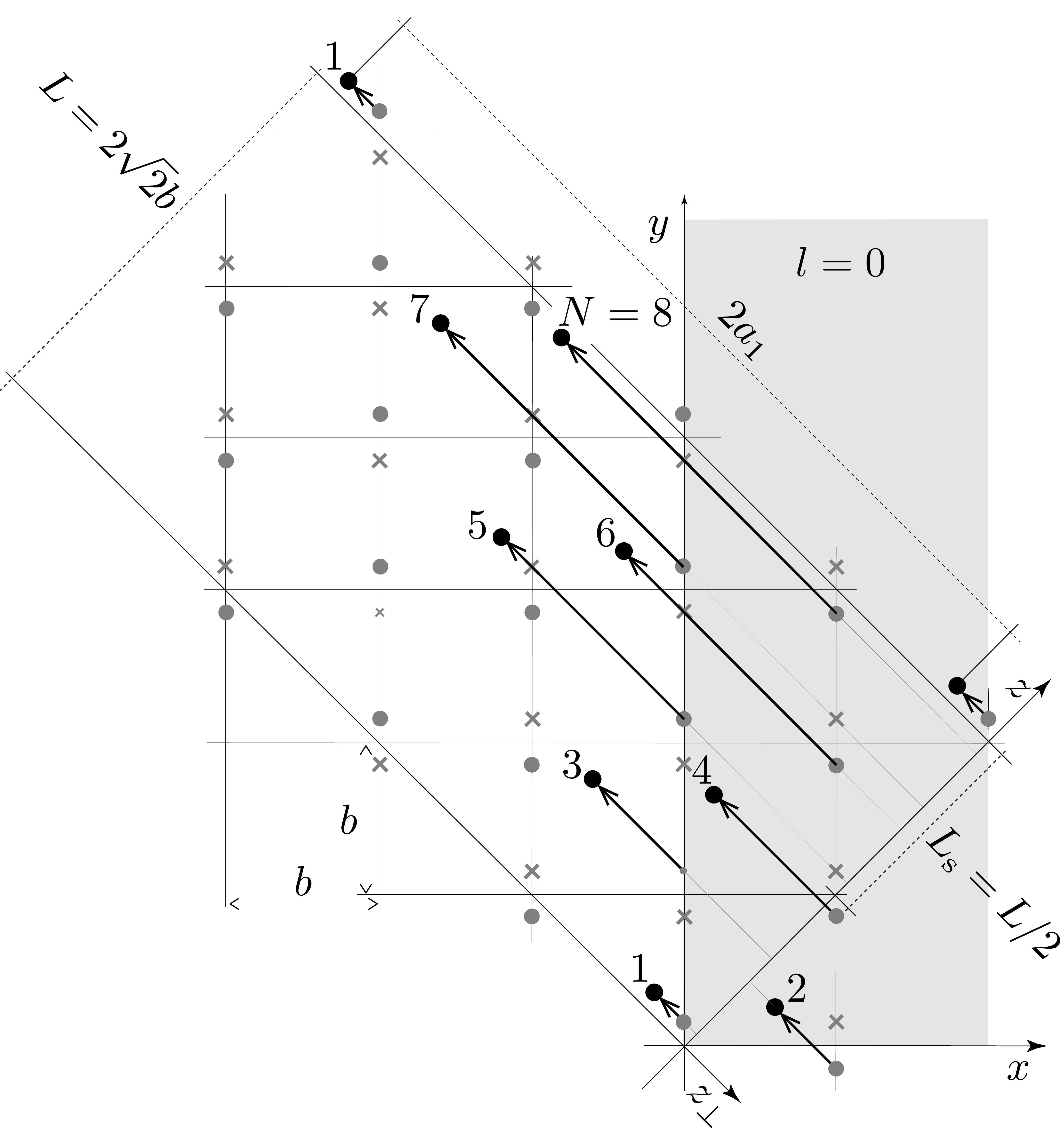}
\end{center}
\caption {\label{fig:period45} Sketch of two $(1,2)$-solitons (of width $L_s$)
within the period $L/b=2\sqrt{2}$ ($N=8$, i.e., four particles per soliton)
going from twin A (grey dots) to twin B (grey crosses) and back to twin A. The
shaded area corresponds to the $l=0$-strip. With $(z_{\perp,N+1} -
z_{\perp,1})=-2 a_1$ the boundary condition \eqref{eq:PBCzp} is properly
satisfied. Grey dots and crosses denote twin A and twin B lattice sites,
respectively.}
\end{figure}

\begin{figure}[t]
\begin{center}
\includegraphics[width=7cm]{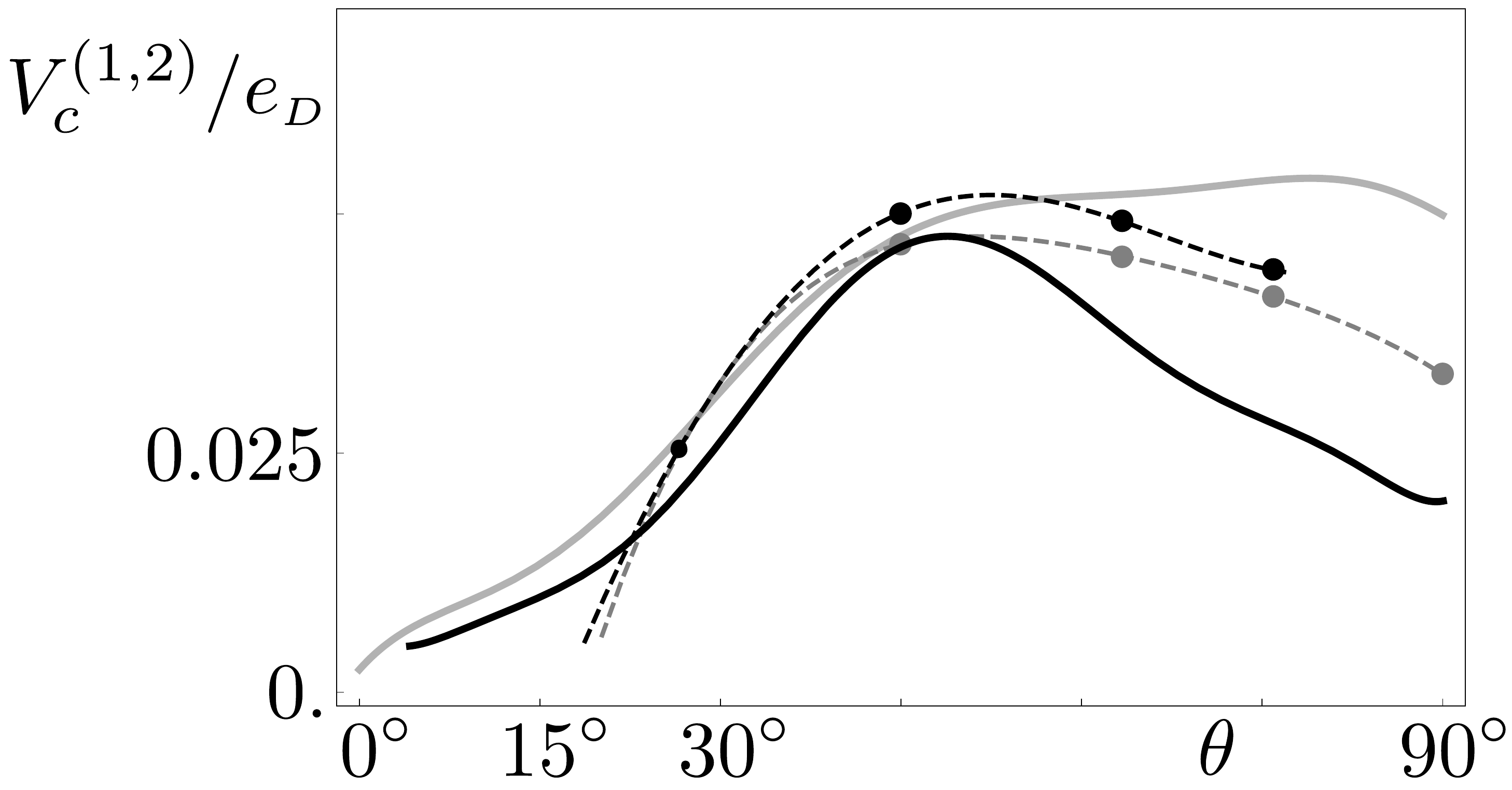}
\end{center}
\caption{\label{fig:12sol_Vc} The critical substrate strength
$V_c^{\scriptscriptstyle (1,2)}$ of the $(1,2)$ domain-wall as a function of
$\theta$. Grey points show the numerical results using the (unrelaxed) defect
shape obtained analytically, black points are the values after relaxation. The
dashed lines are guides to the eye. The black solid line is the analytic
result obtained with the elastic theory for the hexagonal lattice, the grey
solid line is the analytic result using the rhombic elasticity theory, see
Sec.\ \ref{sec:sol_dw}.}
\end{figure}

We make again use of a periodic arrangement of defects with the period $L$ of
the supercell chosen appropriately, see Fig.\ \ref{fig:period45} for a sketch
of two $(1,2)$ domain-walls at $\theta = 45^\circ$ with $L = 2\sqrt{2}\, b$
and $N_\mathrm{s} = 8$ particles and Table \ref{table:12sol_L} for a summary
of suitable sizes $L$.  Next, we determine the critical substrate potentials
at the various discrete angles $\theta$. We optimize the domain-wall shapes by
numerical relaxation of the initial analytical solution and obtain the results
listed in Table \ref{table:12sol_L}; figure \ref{fig:12sol_Vc} compares the
results from the analytic solution with those obtained numerically without and
with relaxation. We find that the best $(1,2)$ defect is directed close to
$\theta = 45^\circ$ with $V_c^{\scriptscriptstyle (1,2)} \approx 0.0501 \,
e_{\scriptscriptstyle D}$.

An interesting feature of the $(1,2)$ domain wall reveals itself for the
angles $\theta \approx 76^\circ$ and $\theta = 90^\circ$. Indeed, for the
$76^\circ$ angle, the relaxation process, although still converging, lasts
much longer. While the initial displacement $v_x$ along $x$ remains nearly
unchanged, the displacement $v_y$ along $y$ changes quite appreciably. In
fact, the relaxation tends to dissolve the $(1,2)$ defect into a $(1,0)$ and a
$(0,2)$ part where the latter one tends to split into a domain-wall pair
$(0,1) + (0,1)$, resembling the sketch in figure \ref{fig:wings} (dotted
line).  However, at $76^\circ$ the three parts still remain bounded and the
relaxation converges. This is no longer the case at $90^\circ$ where the
relaxation never converged (explaining for the missing entry of a value for
$V_c^{\scriptscriptstyle (1,2)}$ in table \ref{table:12sol_L}).  We note that
the $(1,0)$ defect is compressing the lattice and hence involves a positive
`charge', while the $(0,1)$ defects are diluting the lattice and hence are
negatively charged. These oppositely charge defects then tend to bind into a
cluster. However, the $(1,0)$ defect becomes pure shear when $\theta$
approaches $90^\circ$, supporting the interpretation for a complete
dissolution of the $(1,2)$ soliton into a regular array of $(0,1) + (1,0) +
(0,1)$ defects at $90^\circ$.
\begin{table}[h!]
\caption{\label{table:12sol_L} Parameters [cell sizes $L$ ($k\in\mathbb{N}$),
number of defects $n_\mathrm{s}$ per cell, and charge $Q$ per defect (and
length $a_1$)] for the numerical analysis of the $(1,2)$ domain-wall for
discrete angles $\theta$ and numerical results for the critical substrate
strengths $V_c^{\scriptscriptstyle(1,2)}$ for the $(1,2)$ domain-walls before
(superscript `ur = unrelaxed') and after relaxation.  $L$ and $n_\mathrm{s}$
are not independent quantities and may be chosen differently.}
  \vskip 3 pt
  \begin{center}
  \begin{tabular}{| c || c | c | c | c | c |}
  \hline
$ \theta$ &  $L/b$ & $n_\mathrm{s}$ & $Q$ &
             $V_c^{\scriptscriptstyle(1,2),\mathrm{ur}}/
               e_{\scriptscriptstyle D}$
          &  $V_c^{\scriptscriptstyle(1,2)}/
               e_{\scriptscriptstyle D}$\\
            \hline
  $26.6^\circ$ & $\sqrt{5}(2k+1)$  &2 & $2$   & 0.0253 & 0.0255 \\
  $45^\circ$   & $2\sqrt{2}(2k+1)$ &2 & $0$   & 0.0468 & 0.0501 \\
  $63.4^\circ$ & $\sqrt{5}(2k+1)$  &1 & $-1$  & 0.0456 & 0.0492 \\
  $76.0^\circ$ & $\sqrt{17}(2k+1)$ &1 & $-3$  & 0.0413 & 0.0441 \\
  $90^\circ$   &  $2k+1$           &2 & $-2$  & 0.0332 & -- \\
  \hline
\end{tabular}
\end{center}
\end{table}

Finally, we briefly report on our study of the $(1,1)$ and $(1,3)$ solitons.
The $(1,3)$ soliton extends over a large distance $3b/2$ along the
$y$-direction, of which a distance $b/2$ has to be covered by the core,
requiring that $\eta > 1/2$ and thereby restricting the allowed angles
$\theta$, see the discussion in Sec.\ \ref{sec:sol_dw}. Another special case
is the $(1,1)$ soliton at $\theta \approx 26.6^\circ$, where $\eta \approx
1.4$ is large and nearly fully developed wing-solitons are needed to bring the
$v_y$-overshoot in the core back to the imposed shift $b/2$.  Such extensive
wings require very large periods (soliton separations) in order to minimize
the soliton-soliton interaction. Relaxing the soliton shape then necessitates
a lot of computing time and we have abstained from its detailed study as this
direction is not favorable anyway.
\begin{table}[h]
\caption{\label{table:113sol_L} Parameters [cell sizes $L$ ($k\in\mathbb{N}$),
number of defects $n_\mathrm{s}$ per cell, and charge $Q$ per defect (and
length $a_1$)] for the numerical analysis of the $(1,1)$ and $(1,3)$ solitons
at discrete angles $\theta$.}
  \vskip 3 pt
  \begin{center}
  \begin{tabular}{| c || c | c | c || c | c | c |}
  \hline
  \multicolumn{1}{| c ||}{ } &\multicolumn{3}{c||}{(1,1)}
                             &\multicolumn{3}{c |}{(1,3)} \\
  \hline
$ \theta$ &  $L/b$ & $n_\mathrm{s}$ & $Q$&  $L/b$ & $n_\mathrm{s}$ & $Q$\\
            \hline
  $26.6^\circ$ & $\sqrt{5}k$ i          & 1 & $3$ & $\sqrt{5}k$  & 1 & 1\\
  $45^\circ$   & $2\sqrt{2}k$           & 1 & $1$ & $2\sqrt{2}k$ & 1 &$-1$\\
  $63.4^\circ$ & $4\sqrt{5}k$           & 2 & $0$ & $2\sqrt{5}k$ & 1 &$-2$\\
  $76.0^\circ$ & $2\sqrt{17}(2k\!+\!1)$ & 1 &$-1$ & $2\sqrt{17}(2k\!+\!1)$ &1&$-5$\\
  $90^\circ$   & $k$                    & 1 &$-1$ & $k$          & 1 & $-3$\\
  \hline
\end{tabular}
\end{center}
\end{table}

The supercell lengths $L$ for the periodic arrays used in the numerical
relaxation are summarized in Table \ref{table:113sol_L} and the final results
for the critical substrate potentials are presented in Table
\ref{table:113sol_Vc}. Similar to the $(1,2)$ domain-wall, the $(1,3)$ soliton
dissociates into elementary solitons $(1,3) \to (0,1) + (1,1) + (0,1)$ for large
angles (the $(1,3)$ defect is pure shear type at $\theta \approx 63.4^\circ$
and thus has zero `charge').

\begin{table}
\caption{\label{table:113sol_Vc} Numerical results for the critical substrate
amplitudes before (`ur = unrelaxed') and after relaxation of the soliton shape
for the $(1,1)$ and the $(1,3)$ solitons.}
  \vskip 3 pt
  \begin{center}
  \begin{tabular}{| c || c | c || c | c |}
  \hline
  $\theta$
  &$V_c^{\scriptscriptstyle (1,1),\mathrm{ur}}/e_{\scriptscriptstyle D}$
  &$V_c^{\scriptscriptstyle (1,1)}/e_{\scriptscriptstyle D}$
  &$V_c^{\scriptscriptstyle (1,3),\mathrm{ur}}/e_{\scriptscriptstyle D}$
  &$V_c^{\scriptscriptstyle (1,3)}/e_{\scriptscriptstyle D}$\\
            \hline
   $26.6^\circ$ & 0.0140 &  & 0.0328& 0.0337\\
   $45^\circ$ & 0.0267 & 0.0287 & 0.0493 & 0.0544\\
   $63.4^\circ$ & 0.0357 & 0.0382 & 0.0544 &--\\
   $76.0^\circ$ & 0.0332 & 0.0350 & -- & -- \\
   $90^\circ$ & 0.0249 & 0.0266 &-- & --\\
  \hline
\end{tabular}
\end{center}
\end{table}

Quite surprizingly, it is still the rather large $(1,3)$ soliton that turns
out as the best $(1,k)$ defect with the highest critical substrate potential
$V_c^{\scriptscriptstyle (1,3)} \approx 0.0544 \, e_{\scriptscriptstyle D}$ at
$\theta = 45^\circ$. However, this value is appreciably below the critical
potential $V_c^{\scriptscriptstyle (0,1)} \approx 0.0741 \,
e_{\scriptscriptstyle D}$ for the $(0,1)$ domain-wall at $\theta = 45^\circ$.

\section{Transformation Pathway from Square to Triangular}\label{sec:pathway}

The following scenario then describes the transition from the square lattice
to the hexagonal phase with decreasing substrate potential, see Fig.\
\ref{fig:phase_dia}: Starting out at large substrate potential, the square
lattice first undergoes a smooth transition at $V_{\scriptscriptstyle \square}
\approx 0.2\, e_{\scriptscriptstyle D}$ to a period-doubled zig-zag phase,
thereby spontaneously breaking the $x$-$y$ symmetry and selecting a strongly
modulated direction (in this paper always chosen along $x$), leaving a
weakly modulated double-periodic effective potential along the other direction
(here, along $y$). This period-doubled lattice appears in two twin versions,
where one twin transforms into the other by a shift $b$ along $x$.

\begin{figure}[h]
\begin{center}
\includegraphics[width=6cm]{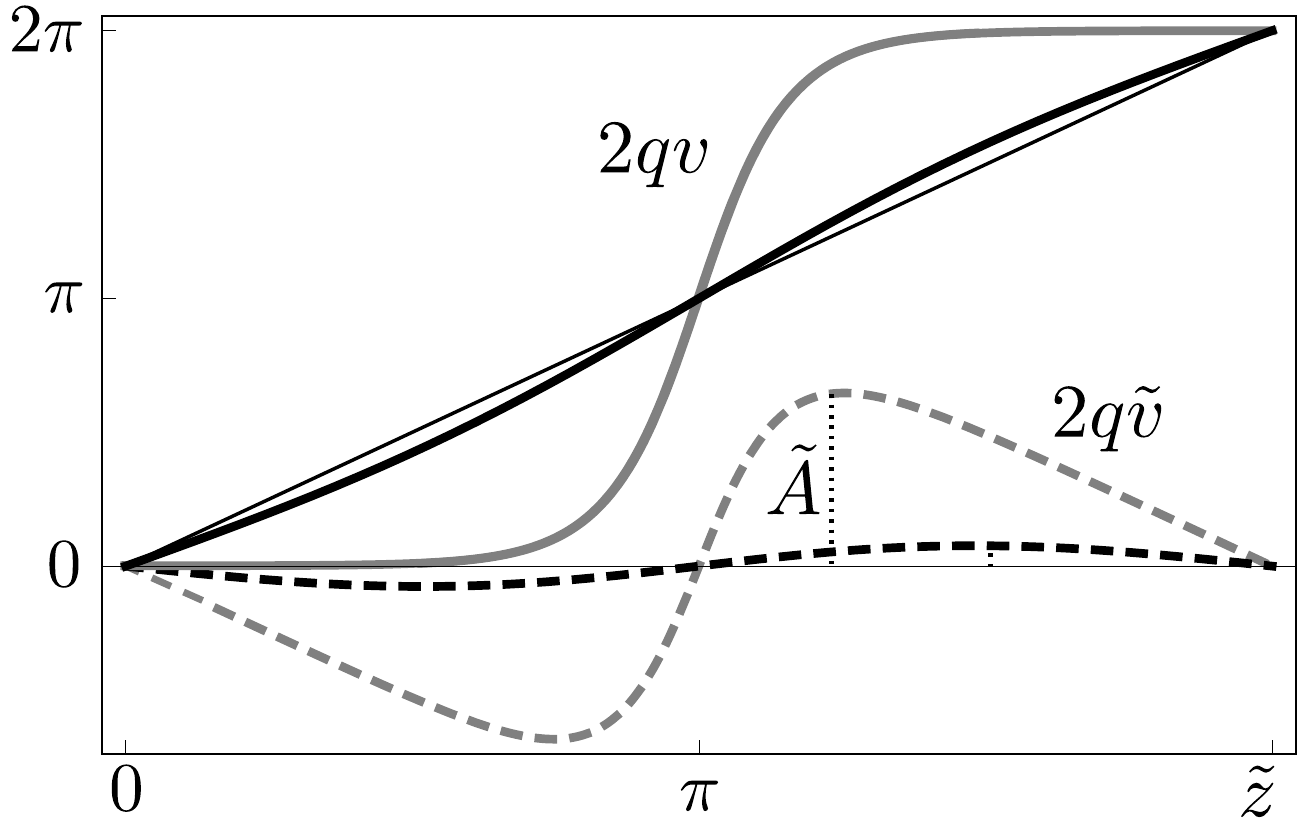}
\end{center}
\caption {\label{fig:washing_out} Displacements $v$ and $\tilde{v}$ for the
$(0,1)$ defect array as a function of $\tilde{z}$ at different substrate
potential strengths.  Shown here are the total (dimensionless) displacement
field $2qv = \tilde{z} + 2q\tilde{v}$ (solid lines) and its periodic part
$2q\tilde{v}$ (dashed lines, with amplitudes $\tilde{A}$) for the
$(0,1)$ soliton phase at $V \approx 0.046\, e_{\scriptscriptstyle D}$ (black
curves) and $V \approx 0.075\,e_{\scriptscriptstyle D}$ (grey curves).  The
shapes are obtained from the analytic calculation using the $bb$ rhombic
elasticity. The large decrease of $\tilde{A}$ with decreasing $V$ shows that at
$V_c^{\rm\scriptscriptstyle PT}\approx 0.046\,e_{\scriptscriptstyle D}$ the
$bb'$ rhombic lattice is already established to within $10\%$.}
\end{figure}

At $V_c^{\scriptscriptstyle (0,1)} \approx 0.0741 \, e_{\scriptscriptstyle D}$
the $(0,1)$ domain walls directed along $\theta \approx 45^\circ$ enter the
period-doubled lattice. As these dilution defects start overlapping, they wash
out the flat double-periodic substrate potential along $y$, giving way to the
$bb'$ rhombic phase. The $bb'$ rhombic lattice then provides the proper parent
lattice for the appearance of the Pokrovsky-Talapov solitons at $V_c^{\rm
\scriptscriptstyle PT} \approx 0.046 \, e_{\scriptscriptstyle D}$ near the
angle $\theta \approx 44.5^\circ$. At this value of the substrate potential,
the $(0,1)$ domain-wall phase has approached the $bb'$-lattice to within
$\approx 10$ \%, as measured by the ratio of amplitudes $\tilde{A}$ of the
periodic displacement $\tilde{\bf v}$ generated by the $(0,1)$ domain wall
array at different substrate potentials, $\tilde{A} (V_c^{\rm
\scriptscriptstyle PT})/ \tilde{A} (V_c^{\scriptscriptstyle (0,1)}) =
0.019/0.25 \approx 0.08$ (the average misfit $(\langle b'\rangle-b)/b$ differs
by $\approx 3$ \% from the asymptotic value $(b'-b)/b$, $(\langle b'\rangle
-b')/(b'-b) \approx 0.03$, see Fig.\ \ref{fig:washing_out}). The proliferation
of PT solitons then smoothly eliminates the $x$-harmonic and completes the
transition to the distorted and rotated hexagonal lattice at small $V$.

\section{Summary and Conclusion}\label{sec:con}

We have studied the competition between different lattice structures in a
two-dimensional particle system with long-range dipolar interaction. Assuming
an underlying substrate potential with square symmetry, the latter is in
competition with the hexagonal lattice favored by the isotropic repulsion
between particles.  This setup generalizes the famous Frenkel-Kontorova model
\cite{FrenkelKontorova_39} in one dimension where the competition is between
two incommensurate lattice constants $a$ for the particle system and $b$ for
the periodic substrate potential. In two dimensions, besides different lattice
constants for the particle- and substrate lattice, the two systems also may
involve different lattice symmetries---this is the case in the present study.

An important degree of freedom is the applied pressure $p$ (or chemical
potential $\mu$) determining the particle density; here, we have chosen a
situation with a commensurate density, i.e., the same density $n$ of free
particles and density of minima in the substrate potential $n = 1/b^2$,
defining an appreciable misfit $s = b/h - 1 = 0.0746$ between the two
lattices. We have studied the purely classical system free of any
fluctuations, either quantum or thermal, thus providing the starting point for
later studies of the full dynamical phase diagram including fluctuations and
external drive. The regime of validity of our results has to be checked case
by case by comparing the interaction energy $e_{\scriptscriptstyle D}$ with
the quantum recoil ($e_r = \hbar^2 \pi^2/2 m b^2$) and thermal energies
($k_{\rm\scriptscriptstyle B} T$). Critical values for the
quantum\cite{Buechler_07} and classical\cite{Kalia_81} phase transitions
between lattice and fluid are $r_{\scriptscriptstyle Q} =
e_{\scriptscriptstyle D}/e_r \approx 18$ and $r_T = e_{\scriptscriptstyle D} /
k_{\rm\scriptscriptstyle B} T \approx 11$.

We have found the complete pathway taking the locked square lattice at large
substrate potential $V$ to the floating hexagonal phase at zero $V$. This
includes a first transition at $V_{\scriptscriptstyle \square} \approx 0.2\,
e_{\scriptscriptstyle D}$ to a period-doubled zig-zag phase, a transition to a
non-uniform phase with $(0,1)$ domain walls at $V_c^{\scriptscriptstyle (0,1)}
\approx 0.0741 \, e_{\scriptscriptstyle D}$ approaching the $bb'$ rhombic
phase, and a second solitonic transition at $V_c^{\rm \scriptscriptstyle PT}
\approx 0.046 \, e_{\scriptscriptstyle D}$ with $(-1, \sqrt{\nu})$ solitons
that transform the particle system to the rotated and deformed hexagonal phase
at small $V$. This orientationally locked phase then approaches the free
floating hexagonal lattice as the substrate potential $V$ vanishes.  Quite
unexpectedly, we have found that the optimal orientation of the $(0,1)$ domain
wall does not follow a common symmetry axis of the substrate and the parent
crystal, although such a special symmetry has often been considered as natural
in the literature \cite{Chaikin_95}. Furthermore, the geometrical constraints
associated with the transformation between lattices are incompatible with the
occurrence of just one solitonic transition involving simple line defects as
it appears within the framework of the resonance approximation
\cite{PT_pap,PT_book}. An interesting scenarium alternative to the one we have
found in the present problem is a transformation involving the formation of a
network of crossing solitons. If the most favorable solitons have close
critical potentials and a negative intersection energy, the two smooth
transitions may merge to a single first-order one. In our analysis of the
square-to-hex transition in the dipolar system, we have found far separated
values for the two transitions at $V_c^{\scriptscriptstyle (0,1)}$ and
$V_c^{\rm\scriptscriptstyle PT}$, favoring our scenario with two subsequent
transitions and discouraging an alternative scenario involving a soliton
network.

Besides a direct structural observation (in direct or reciprocal space) of the
different phases appearing along the transformation pathway, an alternative
way of observing the various transitions is via the system's dynamical
response under an applied force field. Indeed, it turns out, that each of the
phases reacts to a force field with its specific dynamical characteristic.
The ordered lattices, square and period-doubled, are pinned to the optical
lattice, one symmetrically along the $x$ and $y$ axes, the other
asymmetrically with a reduced pinning along $y$, e.g., with a suppression
factor $V/32 \Delta \sim 1/8$ at $V = V_{\square}/2 = 4 \Delta$.  The
domain-wall and soliton phases exhibit a very interesting dynamical response:
the domain walls/solitons are (exponentially in $\sqrt{\alpha}/b$) weakly
pinned by the lattice (unpinned within the continuum elastic description). A
force (density) field ${\bf f}$ acting on the particles will act with the line
force ${\bf f} \cdot {\bf d}_{j,k}$ directed along $z$ on the defects. Their
motion along $z$ then produces a mass flow along the displacement field ${\bf
d}_{j,k}$ which is longitudinal for a pure dilution defect and transverse for
a pure shear defect (our domain walls and solitons are neither pure dilution
nor pure shear); the observation of this characterisitc flow allows for the
identification of the two non-uniform phases.  Finally, our analysis provides
the starting point for further studies, including other pressures or densities
and hence misfits, substrate lattices with different symmetries, alternative
transformation scenarios, and effects due to quantum and thermal fluctuations.

We thank Matthias Troyer for helpful discussions and acknowledge financial
support of the Fonds National Suisse through the NCCR MaNEP; one of us (SEK)
thanks the Pauli Center for Theoretical Physics for its generous hospitality.

\appendix

\section{Ewald summation}
\label{app:ewald}

The Ewald summation method \cite{Ewald_21} allows to sum up lattice energies
for long-range interacting particles by splitting the interaction into two
parts describing distant and nearby particles,
\begin{equation} \label{eq:ewaldsplitting}
   \frac{1}{R^{\eta}} = C_\eta \Bigr\{ \int_0^\epsilon \!\! dt\, t^{\eta/2-1}
   e^{-tR^2} \!\! + \!\! \int_\epsilon^\infty \!\!\! dt \, t^{\eta/2-1}
   e^{-tR^2} \Bigl\},
\end{equation}
with $\epsilon > 0$ a parameter and the constant $C_\eta$ is given by
$C_\eta^{-1}=\Gamma(\eta/2)$ with the Gamma function
$\Gamma(x)=\int_0^\infty\!du\,u^{x-1}e^{-u}$.  We make use of Poisson's
summation formula (the set $\{K_j\}$ denote the reciprocal lattice sites,
$\Omega = 1/n$ denotes the unit cell area)
\begin{equation} \label{eq:psr}
   \sum_i f({\bf R}_i)=\frac{1}{\Omega}
   \sum_j \hat{f}({\bf K}_j),
\end{equation}
with the Fourier transform 
\begin{equation} \label{eq:FT}
  \hat{f}({\bf K})=\int d^2R\,f({\bf R})\exp(-i{\bf R}\cdot{\bf K}),
\end{equation}
and treat the first (long distance) term in (\ref{eq:ewaldsplitting}) in
Fourier space. Using the Fourier transform $\hat{f}({\bf K}_j) = (\pi/t)
\exp(-K_j^2/4t)$, the interaction energy per particle assumes the form
\begin{align} \label{eq:ewald}
   \frac{2e^\textrm{int}}{DC_\eta}
   &=\sum_{j\neq0}\int_\epsilon^\infty\!\!\!\!dt\,t^{\eta/2-1}e^{-tR_j^2}
   \nonumber\\
   &\qquad +\frac{\pi}{\Omega}\sum_{j} \int_0^\epsilon\!\!dt\,t^{\eta/2-2}e^{-K_j^2/4t}
   -\frac{2}{\eta}\epsilon^{\eta/2}
   \nonumber\\
   &=\, \Bigr\{\epsilon^{\eta/2}\Bigr[\sum_{j\neq0}\Psi_{\frac{\eta-2}{2}}(\epsilon R_j^2)
   -\frac{2}{\eta}\Bigl]\\
   &\qquad+\frac{\pi}{\Omega} \epsilon^{\eta/2-1}
   \Bigr[\frac{2}{\eta-2}+\sum_{j\neq0}\Psi_{-\frac{\eta}{2}}(K_j^2/4\epsilon)\Bigl]\Bigl\}.
   \nonumber
\end{align}
Note that the $j=0$ term has to be separately handled (as it is not present in
the energy but contributes to the Poisson formula) and we have substituted
$t=\epsilon u$ and $t=\epsilon/u$ in the integrals of the first and second
sum, respectively.  The function $\Psi_{x}(\beta)$ is related to the
Incomplete Gamma function $\Gamma(x,\beta) = \int_\beta^\infty \! du \,
u^{x-1} e^{-u}$ via $\Psi_{x}(\beta)=\beta^{-(x+1)}\Gamma(x+1,\beta)$.

The choice $\epsilon=\pi/\Omega$ simplifies the formula
(\ref{eq:ewald}) substantially, as for any integers $p$ and $q$ the real- and
$K$-space lattice vectors ${\bf R}_{p,q}=p{\bf R}_1+q{\bf R}_2$ and ${\bf
K}_{p,q} = p{\bf K}_1 + q{\bf K}_2 = (2\pi/\Omega)[(pR_{2y} - qR_{1y}),
(-pR_{2x} + qR_{1x})]$ [note that ${\bf K}_1 = (2\pi/\Omega) ({\bf R}_2 \times {\bf
R}_3)$, and cyclic with ${\bf R}_3 = (0,0,1)$] are related through
\begin{equation} \label{eq:RrelK}
   K_{p,q}^2 = (2\pi/\Omega)^2 R_{-q,p}^2.
\end{equation}
We thus can reexpress the sum in (\ref{eq:ewald}) over the reciprocal space as
a sum over real space and find ($C_3 = 2/\sqrt{\pi}$)
\begin{align} \label{eq:ewald2}
   e^\textrm{int} &= \, C_\eta\pi^{\eta/2}D \Omega^{-\eta/2}
   \Bigr\{\frac{2}{\eta(\eta-2)}\\
   +\frac{1}{2}&\sum_{(p,q)\neq0}\Bigr[
    \Psi_{\frac{\eta-2}{2}} \bigl(\pi R_{pq}^2/\Omega\bigr)
   +\Psi_{-\frac{\eta}{2}}\bigl(\pi R_{-q,p}^2/\Omega\bigr)
   \Bigl]\Bigl\},\nonumber
\end{align}
where the specific lattice type enters via the parameterization of the
primitive lattice vectors ${\bf R}_1$ and ${\bf R}_2$.  The functions
$\Psi_x(\beta)$ die off exponentially with $\beta$ and the first few shells of
lattice sites already give a significant contribution to the total sum,
allowing for a very fast determination of the interaction energy for particles
on a lattice. For any exponent $\eta>1$ the expression (\ref{eq:ewald2})
assumes its global minimum for the hexagonal lattice.

\section{Effective double-periodic potential}
\label{app:eff_pot}

For the derivation of the effective second-mode substrate potential
\eqref{eq:edsapp_iso} we divide the particle lattice into two sublattices, each
forming a rectangular Bravais lattice spanned by the vectors ${\bf
a}_1=(2b,0)$ and ${\bf a}_2=(0,b)$ and shifted with respect to one another by
the vector ${\bf c}=(b,\delta)$.  With ${\bf R}^{\rm \scriptscriptstyle R}_j$
denoting the sites of the rectangular lattice, the interaction energy per
particle can be written as
\begin{align}
\label{eq:eintPD}
   e^\mathrm{int}_\mathrm{pd}(\delta)
   &=\frac{1}{2}\sum_{j=1}^{N/2}\frac{D}{(R^{\rm\scriptscriptstyle R}_j)^3}
   +\frac{1}{2}\sum_{j=1}^{N/2}\frac{D}{\vert{\bf R}^{\rm\scriptscriptstyle R}_j
   +{\bf c}\vert^3}\\ \nonumber
   &=e^\mathrm{int}_{\rm\scriptscriptstyle R} + \sum_{m>0}\sum_l
   \frac{D}{\bigl[ (2m-1)^2b^2+(lb+\delta)^2\bigr]^{3/2}},
\end{align}
where $m$ runs over columns, $l$ over rows, and $e^\mathrm{int}_{\rm
\scriptscriptstyle R} = 2.025\,e_{\scriptscriptstyle D}$ is the interaction
energy per particle of the rectangular lattice (given by the first sum in the
first line and obtained using Ewald summation).  Making use of the Poisson
summation formula (\ref{eq:psr}) (with the Fourier transform (\ref{eq:FT}) and
the inverse $f({\bf R})=\int [d^2 K/(2\pi)^2] \,\hat{f}({\bf K})\exp(i{\bf
K}\cdot{\bf R})$), the sum over $l$ takes the form
\begin{align} \label{eq:eintPDrows}
   &\sum_l\frac{D}{\bigl[ (2m-1)^2b^2+(lb+\delta)^2\bigr]^{3/2}}\nonumber\\
   &~~ =\sum_{l'}\int\!\frac{dy}{b}\,
   \frac{D e^{-2\pi i\, l'y/b}}{[(2m\!-\!1)^2b^2\!+\!(y\!+\!\delta)^2]^{3/2}}
   \nonumber\\
   &~~ = \int\!\frac{dy'}{b}\,
   \frac{D}{[(2m\!-\!1)^2b^2\!+\!y'^2]^{3/2}}\nonumber\\
   &~~~~~~ +\sum_{l'>0}\int\!\frac{dy'}{b}\,
   \frac{D(e^{-2\pi i\, l'(y'-\delta)/b} + e^{2\pi i\, l'(y'-\delta)/b})}
   {[(2m\!-\!1)^2b^2\!+\!y'^2]^{3/2}}
   \nonumber\\
   &~~=\frac{2D}{(2m\!-\!1)^2b^3}\\
   &~~~~~~ +\frac{8\pi\, D}{(2m\!-\!1)b^3}
   \sum_{l'>0} l'\cos{(2\pi l' \delta/b)}K_1[2\pi l'(2m-1)],\nonumber
\end{align}
with $K_1$ the modified Bessel function of the second kind (see Ref.\
\onlinecite{abramowitz_72}).  Inserting \eqref{eq:eintPDrows} into
\eqref{eq:eintPD}, the first term in Eq.\ \eqref{eq:eintPDrows}, corresponding
to $l'=0$, yields $2 e_{\scriptscriptstyle D} \sum_{m=1}^\infty
(2m-1)^{-2}=\pi^2 e_{\scriptscriptstyle D} /4$ and the interaction energy
\eqref{eq:eintPD} reads
\begin{align} \label{eq:eintPD2}
   e^\textrm{int}(\delta)
   &= e_{\rm\scriptscriptstyle R}^\textrm{int}+\frac{\pi^2}{4}\,
   e_{\scriptscriptstyle D}\\
   &+8\pi e_{\scriptscriptstyle D}\, \sum_{m>0}\sum_{l'>0}
   \frac{l'K_1[2\pi l'(2m-1)]}{2m-1}\cos(ql'\delta).\nonumber
\end{align}
Due to the exponential decay of $K_1(z) \propto e^{-z}$, we neglect terms with
$m>1$ and $l'>1$ in the second line and arrive at the approximative formula
\eqref{eq:edsapp} for the interaction energy in the period-doubled phase.

\section{Elastic constants}
\label{app:el_const}

Usual pair potentials $\Phi({\bf R})$ in solids involve both repulsive and
attractive components at small and large distances, respectively. Such
two-body potentials exhibit a minimum at a distance $R_0$ defining the
approximate location of the equilibrium particle spacing and stabilizing the
system at a specific equilibrium density.  A deformation of the bulk material
away from its equilibrium state contributes the elastic energy which in the
continuum limit takes the form,
\begin{equation} \label{eq:intcont1}
   E^\textrm{el}=\frac{1}{2}\int\!d^dr\,
   \lambda_{\mu\nu\sigma\rho}u_{\mu\nu}u_{\sigma\rho},
\end{equation}
where the linearized strain tensor $u_{\mu\nu}$ and the elastic moduli
$\lambda_{\mu\nu\sigma\rho}$ are given by (see standard solid state physics
text books, e.g., Ref.\ \onlinecite{AshcroftMermin_67})
\begin{align}
   u_{\mu\nu}&= (\partial_\mu u_\nu
   +\partial_\nu u_\mu)/2,\label{eq:straintensor}\\
   \lambda_{\mu\nu\sigma\rho}&= \frac{1}{8\Omega}
   \sum_i\bigl\{ \Phi_{\mu\sigma}(R_i)R_{i\nu}R_{i\rho}
   + \Phi_{\nu\sigma}(R_i)R_{i\mu}R_{i\rho}\nonumber\\
   &+ \Phi_{\mu\rho}(R_i)R_{i\nu}R_{i\sigma}
   + \Phi_{\nu\rho}(R_i)R_{i\mu}R_{i\sigma}\bigr\},\label{eq:elasticmoduli}
\end{align}
with the derivatives $\Phi_{\mu\nu} = \partial_\mu\partial_\nu \Phi$ and the
unit cell area $\Omega = A/N$.  These expressions implicitly assume that the
system is situated in a homogeneous and isotropic `background' such that rigid
rotations do not cost any energy.  Consequently, only the symmetric part
$u_{\mu\nu}$ of the derivatives $\partial_\mu u_\nu$ enter in the formula
\eqref{eq:elasticmoduli}.  Another problem of direct relevance in the present
context is the purely repulsive two-body potential $\Phi({\bf
R}_{ij})=\Phi(R_{ij})$ requiring an additional external stabilization, e.g.,
by adding a pressure term $pA$; otherwise the repulsive particles would move
apart and attain a state of vanishing density. In this situation, one should
minimize the Gibbs free energy density $g$ rather than the internal energy
$e$. The continuum limit of the elastic energy density then reads
\begin{align} \label{eq:gelBL}
   g^\mathrm{el}&=e^\textrm{el}+p\,\frac{\delta A}{A}
   =(\gamma_x\!+\!p)(\partial_xu_x)+(\gamma_y\!+\!p)(\partial_yu_y)\\
   &\,+\frac{\lambda_1}{2}(\partial_xu_x)^2+\frac{\lambda_2}{2}(\partial_yu_y)^2
   +(\lambda_3\!+\!p)(\partial_xu_x)(\partial_yu_y)\nonumber\\
   &\,+\frac{\lambda_4}{2}(\partial_yu_x)^2+\frac{\lambda_5}{2}(\partial_xu_y)^2
   +(\lambda_6\!-\!p)(\partial_yu_x)(\partial_xu_y),\nonumber
\end{align}
with
\begin{align}
   \gamma_x&=\frac{1}{2\Omega}
   \sum_{j\neq0}\Phi'_j\frac{x_j^2}{R_j},\quad
   \gamma_y=\frac{1}{2\Omega}
   \sum_{j\neq0}\Phi'_j\frac{y_j^2}{R_j},
   \label{eq:linterm}\\
   \lambda_1&=\frac{1}{2\Omega}\sum_{j\neq0}\Bigl[\Phi''_j\!-\!
     \frac{1}{R_j}\Phi'_j\Bigr]\frac{x_j^4}{R_j^2}+\gamma_x,
   \label{eq:l1}\\
   \lambda_2&=\frac{1}{2\Omega}\sum_{j\neq0}\Bigl[\Phi''_j\!-\!
     \frac{1}{R_j}\Phi'_j\Bigr]\frac{y_j^4}{R_j^2}+\gamma_y,
   \label{eq:l2}\\
   \lambda_3&=\frac{1}{2\Omega}\sum_{j\neq0}\Bigl[\Phi''_j\!-\!
     \frac{1}{R_j}\Phi'_j\Bigr]\frac{x_j^2y_j^2}{R_j^2},
   \label{eq:l3} \\
   \lambda_4&=\frac{1}{2\Omega}\sum_{j\neq0}\Bigl[\Phi''_j\!-\!
     \frac{1}{R_j}\Phi'_j\Bigr]\frac{x_j^2y_j^2}{R_j^2}+\gamma_y,
   \label{eq:l4}\\
   \lambda_5&=\frac{1}{2\Omega}\sum_{j\neq0}\Bigl[\Phi''_j\!-\!
    \frac{1}{R_j}\Phi'_j\Bigr]\frac{x_j^2y_j^2}{R_j^2}+\gamma_x,
   \label{eq:l5}\\
   \lambda_6&=\frac{1}{2\Omega}\sum_{j\neq0}\Bigl[\Phi''_j\!-\!
     \frac{1}{R_j}\Phi'_j\Bigr]\frac{x_j^2y_j^2}{R_j^2},
   \label{eq:l6}
\end{align}
where we have used the abbreviations $\Phi_j\equiv\Phi(R_j)$,
$\Phi'_j=\mathrm{d}\Phi(R_j)/\mathrm{d}R_j$, and
$\Phi''_j=\mathrm{d}^2\Phi(R_j)/\mathrm{d}R_j^2$.  Note that in Eq.\
\eqref{eq:gelBL} we have assumed that the lattice possesses mirror symmetry
along both the $x$- and the $y$-axis. Otherwise, the expression would also
depend on the linear terms $(\partial_yu_x)$ and $(\partial_xu_y)$ and on the
quadratic terms $(\partial_xu_x)(\partial_yu_x)$,
$(\partial_xu_x)(\partial_xu_y)$, $(\partial_yu_y)(\partial_yu_x)$, and
$(\partial_yu_y)(\partial_xu_y)$.  

For an isotropic repulsion, the energetically most favorable configuration is
a hexagonal lattice; the linear terms in Eq.\ \eqref{eq:gelBL} have to vanish
and hence $\gamma_x=\gamma_y=-p$. The pressure then is balanced by the
repulsive forces via (we use $p = -(\gamma_x+\gamma_y)/2$)
\begin{align} \label{eq:platticeconst}
   p=- \frac{1}{4\Omega}\sum_{j\neq 0} \Phi'(R_{j})R_j = \frac{\eta}{2\Omega} 
   e_{\scriptscriptstyle \triangle},
\end{align}
where we have used that $e_{\scriptscriptstyle \triangle} = (1/2)\sum_j
D/R_j^\eta$.  As the right hand side of Eq.\ \eqref{eq:platticeconst} is a
function of the unit cell area $\Omega$, we obtain a relation between the
applied pressure $p$ and the area $\Omega$ or the density $n=1/\Omega$; for
$\eta = 3$, $p = 6.670\,e_{\scriptscriptstyle D} n$, in agreement with Eq.\
\eqref{eq:p}.

\subsection{Hexagonal Lattice}
\label{app:eltr}

Due to the high symmetry of the hexagonal lattice, see the
relations\,\eqref{eq:symprop}, the continuum elastic energy density of Eq.\
\eqref{eq:gelBL} simplifies to the standard form describing a homogeneous and
isotropic system \cite{Landau_70}
\begin{align} \label{eq:fhex}
  g^\mathrm{el}_{\scriptscriptstyle \triangle}
   &=\frac{\kappa}{2}(\partial_xu_x+\partial_yu_y)^2
   +\frac{\mu}{2}[(\partial_xu_x-\partial_yu_y)^2\\ \nonumber
   &\qquad\qquad\qquad\qquad\qquad\quad   +(\partial_yu_x+\partial_xu_y)^2],
\end{align}
where the shear and compression moduli $\mu$ and $\kappa$ are linear
combinations of the $\lambda_j$'s, 
\begin{align}\label{eq:kappa}
 \kappa&=\frac{\lambda_1+\lambda_3+p}{2},
         \qquad \lambda_1 = \lambda_2,\\
 \mu&=\frac{\lambda_{1}-\lambda_3-p}{2}=\lambda_4=\lambda_5=\lambda_6-p.
 \label{eq:mu}
\end{align}
The evaluation of the infinite sums is simplified considerably by first adding
terms over sites ${\bf R}_i$ arranged in shells of radius $R_i$,
\begin{align} \label{eq:symprop}
   x_j^2&=\frac{1}{6}\sum_{s=0}^5R_j^2\cos^2{(\vartheta_j
   +\frac{\pi}{3}s)}=\frac{R_j^2}{2}=y_j^2,\nonumber\\
   x_j^2y_j^2&=\frac{1}{6}\sum_{s=0}^5R_j^4\cos^2{(\vartheta_j
   +\frac{\pi}{3}s)}\sin^2{(\vartheta_j+\frac{\pi}{3}s)}
   =\frac{R_j^4}{8},\nonumber\\
   x_j^4&=\frac{1}{6}\sum_{s=0}^5R_j^4\cos^4{(\vartheta_j
   +\frac{\pi}{3}s)}=\frac{3}{8}R_j^4=y_j^4,
\end{align}
where the dependence on the angle $\vartheta_j$ (with ${\bf R}_j = R_j
\exp{i\vartheta_j}$) drops out due to averaging. Using these angular averages
in the expressions for the elastic coefficients \eqref{eq:l1} to \eqref{eq:l6}
and combining these with the pressure in \eqref{eq:platticeconst} to the
elastic moduli $\kappa$ and $\mu$ as given by Eqs.\ \eqref{eq:kappa} and
\eqref{eq:mu}, we obtain the intermediate results
\begin{align} \label{eq:kap-mu-Phi}
   \kappa &=\frac{1}{8\Omega}\sum_{j\neq 0}
          \bigl[ \Phi''_j\, R_j^2 - \Phi'_j\, R_j\bigr], \\
   \nonumber
   \mu &= \frac{1}{16\Omega}\sum_{j\neq 0}
          \bigl[ \Phi''_j\, R_j^2 + 3\Phi'_j\, R_j\bigr].
\end{align}
Assuming an interaction potential of the form $\Phi(R)=D/R^\eta$ with
$\eta >2$, these moduli can be expressed in terms of the interaction energy
$e_{\scriptscriptstyle \triangle}= (1/2) \sum_{j\neq 0} \Phi_j$ of the
hexagonal lattice and we arrive at the final results
\begin{align}
  \kappa&=\frac{\eta(\eta+2)}{4}
   e_{\scriptscriptstyle \triangle} n \quad \mathrm{and} \quad
   \mu=\frac{\eta(\eta-2)}{8}
   e_{\scriptscriptstyle \triangle} n.
   \label{eq:mukap}
\end{align}
The expressions (\ref{eq:mukap}) scale with $n^{(\eta+2)/2}$ in particle
density $n$; their ratio is only determined by the exponent $\eta$ of the
power-law interaction potential,
\begin{equation} \label{eq:kmratio}
   \frac{\kappa}{\mu}=\frac{2(\eta+2)}{\eta-2},
\end{equation}
leading to a Poisson ratio
\begin{equation}
\label{eq:poissonratio}
\nu=\frac{\kappa+\mu}{\kappa-\mu}=\frac{\eta+6}{3\eta+2}.
\end{equation}
In our particular case with $\eta=3$, we have $\kappa = 16.674\,
e_{\scriptscriptstyle D} n$, $\mu = 1.667\, e_{\scriptscriptstyle D} n$,
$\kappa=10\, \mu$ and $\nu=9/11$. 

The analysis of long-range interactions with $\eta \leq 2$ requires a more
careful study. For $\eta = 2$ the sum $e_{\scriptscriptstyle \triangle}$
diverges and the compression modulus becomes dispersive with $\kappa (K \to 0)
\sim e_{\scriptscriptstyle D} n \ln (Kb)$, where $e_{\scriptscriptstyle D} =
D/b^\eta$ is the interaction energy scale. On the other hand, for the shear
modulus, the divergence of $e_{\scriptscriptstyle \triangle}$ is compensated
by the factor $\eta-2$, producing a finite result $\mu \sim
e_{\scriptscriptstyle D} n$ when $\eta = 2$. Similar results have been
obtained for the Wigner crystal \cite{Bonsall_77} with $\eta = 1$ or a 2D
superfluid vortex lattice with $\eta = 0$, i.e., a logarithmic interaction:
the compression modulus is dispersive, $\kappa (K \to 0) \sim
e_{\scriptscriptstyle D} n/(Kb)^{2-\eta}$, while the shear modulus remains
finite $\mu \sim e_{\scriptscriptstyle D} n$. The calculation of the shear
modulus is particularly subtle and requires an analysis with a finite
screening length $\lambda$ or at finite wavevector $K$. Interestingly, the
final result turns out not to depend on $\lambda$ or $K$ and involves only
short scales of order of the lattice constant.

\subsection{Rhombic Lattices}
\label{sec:eliso}

Due to the anisotropic character of a rhombic (or isosceles triangular)
lattice one finds that $\gamma_x \neq \gamma_y$ and that the linear term in
the harmonic expansion Eq.\ \eqref{eq:gelBL} does not vanish.  Without further
stabilization by an additional potential, an external boundary condition,
etc., the system will not remain in this structure.  Even though this
configuration is not stable by itself in homogeneous space, it has to be
invariant under global rotations. This may be checked by inserting the
displacement field of a rotation by the angle $\varphi$ (up to order
$\varphi^2$), $u_x({\bf r})=-\varphi^2 x/2+\varphi y$, $u_y({\bf
r})=-\varphi x-s\varphi^2 y / 2$ into the expression \eqref{eq:gelBL}. The
energy change reads $\delta g^\textrm{rot}=\bigl(-\gamma_x-\gamma_y+\lambda_4
+\lambda_5-2\lambda_6\bigr) {\varphi^2}/{2}$ which vanishes, as easily
verified using the formulas \eqref{eq:linterm} through \eqref{eq:l6}.

Applying the Ewald summation technique for $\eta=3$ (the factors $x_j^n y_j^m$
in the expressions for the coefficients $\gamma_{x,y}$ and $\lambda_{1,\dots,
6}$ are written as derivatives $\partial_{K_{xj}}^n\partial_{K_{yj}}^m
\exp(-K^2/4t)$), the elastic moduli for the rhombic lattice can be combined
from the expressions (the coefficients $\gamma_y$ and $\lambda_y$ are obtained
by replacing $x^2 \to y^2$, $K_x^2 \to K_y^2$, etc., with $y^2 =
(pR_{1y}+qR_{2y})^2$ and $K_y^2 = (2\pi/\Omega)^2 (-p
R_{2x}+qR_{1x})^2$)
\begin{align} \label{eq:gx_ew}
   &\gamma_x = -\frac{2\pi D}{\Omega^{5/2}}
   \Bigl[1+\frac{1}{2}\sum_{p,q}
      \Psi_{-\frac32}(\pi R_{pq}^2/\Omega)\\
   &\quad-\frac{\pi}{\Omega}\sum_{p,q} (pR_{2y}-qR_{1y})^2
      \Psi_{-\frac12}(\pi R_{-qp}^2/\Omega)\nonumber\\
   &\quad+\frac{\pi}{\Omega}\sum_{p,q} (pR_{1x}+qR_{2x})^2
      \Psi_{\frac32}(\pi R_{pq}^2/\Omega)\Bigr],\nonumber\\
   \label{eq:lx_ew}
   &\lambda_x =\frac{1}{2\Omega}\sum_{j\neq0}\Bigl[\Phi''_j\!-\!
     \frac{1}{R_j}\Phi'_j\Bigr]\frac{x_j^4}{R_j^2}\\ \nonumber
     & \quad = \frac{4\pi D}{\Omega^{5/2}}
   \Bigl[\frac{3}{2}
   +\frac{3}{4}\sum_{p,q} \Psi_{-\frac32}(\pi R_{-qp}^2/\Omega)\\
   &\quad+\frac{\pi^2}{\Omega^2}\sum_{p,q} (pR_{1x}+qR_{2x})^4
      \Psi_{\frac52}(\pi R_{pq}^2/\Omega) \nonumber\\
   &\quad-\frac{3\pi}{\Omega}\sum_{p,q} (pR_{2y}-qR_{1y})^2
      \Psi_{-\frac12}(\pi R_{-qp}^2/\Omega)\nonumber\\
   &\quad+\frac{\pi^2}{\Omega^2}\sum_{p,q} (pR_{2y}-qR_{1y})^4
      \Psi_{\frac12}(\pi R_{-qp}^2/\Omega) \Bigr],\nonumber \\
   \label{eq:lxy_ew}
   &\lambda_{xy} =\frac{1}{2\Omega}\sum_{j\neq0}\Bigl[\Phi''_j\!-\!
     \frac{1}{R_j}\Phi'_j\Bigr]\frac{x_j^2y_j^2}{R_j^2}\\ \nonumber
     & \quad = \frac{4\pi D}{\Omega^{5/2}}
   \Bigl[\frac{1}{2}
   +\frac{1}{4}\sum_{p,q} \Psi_{-\frac32}(\pi R_{-qp}^2/\Omega)\\
   &\quad-\frac{\pi}{2\Omega}\sum_{p,q} R_{-qp}^2
      \Psi_{-\frac12}(\pi R_{-qp}^2/\Omega)\nonumber\\ \nonumber
   &+\!\frac{\pi^2}{\Omega^2}\!\sum_{p,q} (pR_{2y}\!-\!qR_{1y})^2
       (-pR_{2x}\!+\!qR_{1x})^2 
   \Psi_{\frac12}(\pi R_{-qp}^2/\Omega) \\ \nonumber
   &+\!\frac{\pi^2}{\Omega^2}\!\sum_{p,q} (pR_{1x}\!+\!qR_{2x})^2
   (pR_{1y}\!+\!qR_{2y})^2 \Psi_{\frac52}(\pi R_{pq}^2/\Omega) \Bigr],
\end{align}
where the terms with $R_{-qp}^2$ arise from the $K$-transformed part in the
Ewald summation.

\subsubsection{The $bb$ rhombic lattice}
\label{subsec:elisobb}

The expansion coefficients for a rhombic lattice with height and base equal
to $b$  take the values (for convenience, we include the `correction' terms
$\pm p$ with $p=6.670\,e_{\scriptscriptstyle D} n$ with the moduli $\lambda_3$
and $\lambda_6$)
\begin{align}
\gamma_x&=-6.387\,e_{\scriptscriptstyle D} n,\label{eq:gxiso}\\ 
\gamma_y&=-7.015\,e_{\scriptscriptstyle D} n,\label{eq:gyiso}\\
\kappa_x=\lambda_1&=18.193\,e_{\scriptscriptstyle D} n,
\label{eq:xxxxiso}\\
\kappa_y=\lambda_2&=20.707\,e_{\scriptscriptstyle D} n,
\label{eq:yyyyiso}\\
\kappa_{xy}=\lambda_3+p&=14.023\,e_{\scriptscriptstyle D} n,
\label{eq:xxyyiso}\\
\mu_x=\lambda_4&=\phantom{1}0.338\,e_{\scriptscriptstyle D} n,
\label{eq:yxyxiso}\\
\mu_y=\lambda_5&=\phantom{1}0.967\,e_{\scriptscriptstyle D} n,
\label{eq:xyxyiso}\\
\mu_{xy}=\lambda_6-p&=\phantom{1}0.684\,e_{\scriptscriptstyle D} n.
\label{eq:yxxyiso}
\end{align}

\subsubsection{The $bb'$ rhombic lattice}
\label{subsec:elisobb'}

Locking the particles to the period $b$ along $x$, these form a $bb'$-lattice
where $b'$ adjusts itself such that the drive along $y$, $\gamma_y+p$ (see
Eq.\ \ref{eq:gelBL}), vanishes; using the Ewald technique, we find that this
is the case for $b'=1.0173\,b$ (alternatively, $b'$ can be found by minimizing
the Gibbs free energy $g_{\scriptscriptstyle \rhd'} (b')$ at given $p$ with
respect to $b'$ as done in Sec.\ \ref{sec:sol_ra}).  The elastic moduli for
this $bb'$ rhombic lattice are
\begin{align}
\gamma'_x&=-6.155\,e_{\scriptscriptstyle D} n,\label{eq:gxb'}\\
\gamma'_y&=-p=-6.670\,e_{\scriptscriptstyle D} n,\label{eq:gyb'}\\
\kappa'_x=\lambda_1'&=17.469\,e_{\scriptscriptstyle D} n,\label{eq:xxxxb'}\\
\kappa'_y=\lambda_2'&=19.531\,e_{\scriptscriptstyle D} n,\label{eq:yyyyb'}\\
\kappa'_{xy}=\lambda_3'+p&=13.820\,e_{\scriptscriptstyle D} n,\label{eq:xxyyb'}\\
\mu'_x=\lambda_4'&=\phantom{1}0.480\,e_{\scriptscriptstyle D} n,\label{eq:yxyxb'}\\
\mu'_y=\lambda_5'&=\phantom{1}0.995\,e_{\scriptscriptstyle D} n,\label{eq:xyxyb'}\\
\mu'_{xy}=\lambda_6'-p&=\phantom{1}0.480\,e_{\scriptscriptstyle D} n,\label{eq:yxxyb'}
\end{align}
where we have used $b$ as our length unit and have expressed our energy
densities through $D n^{5/2} = e_{\scriptscriptstyle D} n$.

\hfill


\begin{thebibliography}{10}
\bibitem{remark_mu} Fixing the chemical potential and minimizing the grand
potential $\omega (\mu,n) = n e_{\scriptscriptstyle \triangle}(n) - \mu n$
instead, we find that $\mu = (5/2) e_{\scriptscriptstyle \triangle}$.
\bibitem{remark_Kron} With a discrete reciprocal space, the Kronecker symbol
has to be understood as the characteristic function over the mode volume
$(2\pi)^2/A$.
\bibitem{remark:KE} The complete elliptic integrals are approximated as
%
\begin{align}
   K\Bigl[\!\sqrt{\frac{2}{2+\beta}}\Bigr]
   &=\frac{5}{2}\log2\!-\!\frac{1}{2}\log\beta\!
   +\!\frac{\beta}{16}(2\!+\!5\log2\!-\!\log\beta),\nonumber\\
   E\Bigl[\!\sqrt{\frac{2}{2+\beta}}\Bigr]
   &=1\!-\!\frac{\beta}{8}(1\!-\!5\log2)\!
   -\!\frac{\beta}{8}\log\beta,\nonumber
\end{align}
%
where terms of order $\mathcal{O}(\beta^2\log\beta,\beta^2)$ have been ignored.
\bibitem{mistake} In Ref.\ \onlinecite{PT_pap} the Poisson ratio was
mistakenly evaluated to $\nu = 5/11$ producing an angle $\theta \approx
24^\circ$ (corrected in Ref.\ \onlinecite{PT_book}).
\bibitem{mathematica} We make use of a standard minimizer algorithm, specifically
the FindMinimum routine in Wolfram's Mathematica.
\end{thebibliography}
\end{document}